\documentclass[12pt]{article}
\usepackage{amsmath,amssymb,amsfonts,graphicx,verbatim, cite}
\usepackage{wrapfig}
\pdfoutput=1

\usepackage[usenames,dvipsnames]{xcolor}

\usepackage[pdftex, bookmarks={false},
pdfauthor={Brian Swingle, John McGreevy}, pdftitle={s-sourcery}]{hyperref}
\hypersetup{colorlinks=true, linkcolor=BrickRed, citecolor=Violet, filecolor=OliveGreen, urlcolor=RoyalBlue, filebordercolor={.8 .8 1}, urlbordercolor={.8 .8 0}}

\definecolor{palatinatepurple}{rgb}{0.41, 0.16, 0.38}

\definecolor{uglybrown}{rgb}{0.8,  0.7,  0.5}

\newcommand{\preprint}[1]{\begin{table}[t]  
             \begin{flushright}               
             {#1}                             
             \end{flushright}                 
             \end{table}}                     
\renewcommand{\title}[1]{\vbox{\center\LARGE{#1}}\vspace{5mm}}
\renewcommand{\author}[1]{\vbox{\center#1}\vspace{5mm}}
\newcommand{\address}[1]{\vbox{\center\em#1}}

\numberwithin{equation}{section}

\renewcommand{\theequation}{\arabic{section}.\arabic{equation}}

\newcommand{\beq}{\begin{equation}}
\newcommand{\eeq}{\end{equation}}
\def\bea{\begin{eqnarray}}
\def\eea{\end{eqnarray}}

\newcommand{\tr}{\text{tr}}

\newcommand{\IZ}{{\mathbb Z}}

\def\half{{1\over 2 }}

\newtheorem{thm}{Theorem}
\newtheorem{lem}{Lemma}

\newtheorem{DEF}{Definition}
\newtheorem{conj}{Conjecture}
\newtheorem{physth}{``Physics" Theorem}

\def\({\left(}
\def\){\right)}

\def\CC{{\cal C}}

\def\CF{{\cal F}}

\def\CQ{{\cal Q}}

\begin{document}

\begin{titlepage}

\title{Renormalization group constructions\\ of  topological quantum liquids\\ and beyond \\}

\preprint{UCSD/PTH 14-05}

\author{Brian Swingle${}^a$ and John McGreevy${}^b$}

\address{${}^a$ Department of Physics, Harvard University, Cambridge, MA 02138, USA}

\address{${}^b$ Department of Physics, University of California at San Diego, La Jolla, CA 92093, USA}

\begin{abstract}

We give a detailed physical argument for the area law for entanglement entropy in gapped phases of matter arising from local Hamiltonians. Our approach is based on renormalization group (RG) ideas and takes a resource oriented perspective. We report four main results. First, we argue for the ``weak area law": any gapped phase with a unique ground state on every closed manifold obeys the area law. Second, we introduce an RG based classification scheme and give a detailed argument that all phases within the classification scheme obey the area law. Third, we define a special sub-class of gapped phases, \textit{topological quantum liquids}, which captures all examples of current physical relevance, and we rigorously show that TQLs obey an area law. Fourth, we show that all topological quantum liquids have MERA representations which achieve unit overlap with the ground state in the thermodynamic limit and which have a bond dimension scaling with system size $L$ as $e^{c \log^{d(1+\delta)}(L)}$ for all $\delta >0$. For example, we show that chiral phases in $d=2$ dimensions have an approximate MERA with bond dimension $e^{c \log^{2(1+\delta)}(L)}$. We discuss extensively a number of subsidiary ideas and results necessary to make the main arguments, including field theory constructions. While our argument for the general area law rests on physically-motived assumptions (which we make explicit) and is therefore not rigorous, we may conclude that ``conventional" gapped phases obey the area law and that any gapped phase which violates the area law must be a dragon.

\end{abstract}

\vfill

\today

\end{titlepage}

\vfill\eject

\tableofcontents

\vfill\eject

\section{Introduction}

In this paper we make progress towards a proof of the area law for entanglement entropy in gapped phases of matter arising from local Hamiltonians. The area law conjecture states that if $\rho = |g \rangle \langle g|$ is a ground state of a local Hamiltonian with an energy gap to excitations, then given a subregion $A$ with state $\rho_A = \tr_{\bar{A}}(\rho)$ the entanglement entropy $S(A)$ of $A$ obeys $S(A) \equiv -\tr(\rho_A \log(\rho_A)) \leq |\partial A|$. Although the area law for gapped phases is widely believed to hold, at least for ``conventional" gapped phases, there are no rigorous proofs of the area law outside one dimension.  Hastings' seminal result \cite{2007JSMTE..08...24H} gave the first rigorous proof of an area law for local gapped Hamiltonians in one dimension. There have since been several alternative proofs and improvements of Hastings' result in one dimensional systems \cite{2012PhRvB..85s5145A, 2013arXiv1301.1162A, 2012arXiv1206.2947B,2010arXiv1011.3445A}. In more than one dimension there are various partial results including area laws for gapped free fermion systems, for certain special kinds of gapped frustration free systems, and numerous special cases which have been checked numerically \cite{2012PhRvB..85s5145A,
2013arXiv1301.1162A,2012arXiv1206.2947B,2010arXiv1011.3445A,
2010RvMP...82..277E,2012arXiv1206.6900M,2010PhRvL.105a0502B,2010NJPh...12b5002G,2008PhRvL.101r7202S,2006PhRvA..74e2326R,2010JMP....51b2101I,2009PhRvA..80e2104M,
2008PhRvL.100g0502W,2006PhRvL..97e0401B,2006PhRvL..97o0404E,2007PhRvB..76c5114H,2006PhRvL..96v0601V,2010NJPh...12i5007D,2013PhRvB..87o5120K,2014arXiv1405.0447W,2014arXiv1404.7616C,2013PhRvB..87c5108H,
RevModPhys.80.517,2009JMP....50i5213H}. It has also recently been shown that if one representative (meaning a particular Hamiltonian) within a phase obeys the area law, then all representatives within the phase obey the area law \cite{2013PhRvL.111q0501V}.

By contrast, the authors have long believed on physical grounds that ``conventional" gapped phases obey the area law and that the area law is robust within phases. Indeed, we believe that the area law is robust even within gapless phases like emergent $U(1)$ electrodynamics, but the existing rigorous techniques are unable to demonstrate this.
This circumstance raises the following questions:
Can we at least give a convincing, if not rigorous, physical argument for the area law in ``conventional" phases? And what about ``unconventional" phases where physical intuition provides a weaker guide? To make various physical intuitions into a real argument for the area law, three things are required. First, we must specify what is meant by ``conventional" phases (our answer,
for gapped states, is
the notion of ``topological quantum liquids"). Second, we must characterize the range of possible ``unconventional" phases. Third, we must show that all such phases obey the area law. In this paper we propose a classification scheme for gapped phases of matter
(which quantifies how conventional they are)
and give a detailed physical argument for the area law based on it. Our argument for a general area law is not rigorous and rests on our classification scheme. As part of the argument, we develop additional tools based on the idea of reconstructing global states from local data which are independent of the classification scheme but which rest on other assumptions. As anticipated in Hastings' original work, the techniques necessary to argue for the area law give additional insight into the structure of gapped phases.

Our approach is based on renormalization group (RG) ideas and takes a resource oriented perspective. We define the notion of an $s$ source RG fixed point in $d$ dimensions as a phase of matter where we need $s$ copies of the entangled ground state at linear size $L$ (the resource) along with initially unentangled degrees of freedom to produce the ground state at linear size $2L$ by acting with a quasi-local unitary transformation\footnote{In our formulation, the RG transformation is reversible. This assumption can be relaxed to give a more general construction, but we will not need it here. Relatedly, we are using the term `fixed point' in the metonymic sense that a fixed point of the RG labels a phase of matter. The systems we describe will often have finite correlation length.}. It follows from our definitions that all $s$ source RG fixed points with $s<2^{d-1}$ obey the area law. Much of the paper is concerned with demonstrating that various interesting models are $s$ source fixed points and with building tools that relate $s$ to spectral properties of the Hamiltonian. Ultimately, our approach is an attempt to make rigorous the simple intuition that violations of the area law are infrared phenomena, so to violate the area law a phase of matter should have many low-energy states.

It should also be emphasized that we are studying quantum phases of matter, not just isolated gapped Hamiltonians. In our analysis we make crucial use of the existence of families of Hamiltonians at varying length scales with uniform local properties which are all in the same phase; this leaves open the possibility of isolated cases outside our framework (a possibility we discuss further below). See Appendix~\ref{sec:phase} for a further discussion of what we mean by a phase of matter.

Besides the importance of understanding the entanglement structure of gapped phases of matter, e.g., for purposes of classical simulation, we have a seemingly different motivation for the constructions presented here. Holographic duality \cite{1993gr.qc....10026T,susskind,1999IJTP...38.1113M,1998PhLB..428..105G,1998AdTMP...2..253W} relates quantum many-body systems without gravity to quantum gravitational systems. It has long been known that entropy is related to geometry in gravitational systems, e.g., thermal entropy \cite{PhysRevD.7.2333} and entanglement entropy \cite{2006PhRvL..96r1602R}. \cite{Swingle:2009bg} proposed to construct the dual holographic geometry from entanglement in the quantum many-body system using a renormalization group construction like MERA \cite{mera} (see also \cite{2009arXiv0907.2939V}). Besides qualitatively matching many features of conventional holographic duality, it is now possible to directly derive the gravitational dynamics from the dynamics of entanglement plus the assumption that ``entanglement = geometry" \cite{2014JHEP...03..051F,2014arXiv1405.2933S}. The proposal of \cite{Swingle:2009bg} naturally produces the identification ``entanglement = geometry", but applying this to a particular model requires that a MERA representation (or something similar) exists. Our demonstration that such MERA representations exist for gapped field theories (including long-range entangled topological theories) thus strengthens the logic beginning from \cite{Swingle:2009bg} and ending at quantum gravity.

\subsection{Overview of results and axioms}

The overall structure of the argument for the area law is as follows. We first rule out very highly entangled states using a thermodynamic argument based on weak spectral assumptions. Then we discuss in detail two more-or-less independent approaches to the remaining range of gapped phases, the $s$ source RG fixed point approach and the reconstruction from local data approach.  With certain physical assumptions which can be proven in some cases and for which we offer general arguments, both approaches give an area law for phases with fewer than $e^{c L^{d-1}}$ ground states on various spaces. Finally, while neither approach seems able to give a general area law by itself, the combination of the two does permit us to argue for a general area law. 

In terms of the $s$ source framework, we argue that gapped phases with fewer than $e^{c L^{d-1}}$ torus ground states ($d$ dimensions, size $L$ torus, $c$ a constant) have $s< 2^{d-1}$ and obey the area law. We also show that, with a weak assumption about the thermal free energy, the area law may be violated at most logarithmically. This argument rules out phases with $s > 2^{d-1}$ and leaves one interesting case, $s=2^{d-1}$, which is dangerous to the area law. We treat the special case $s=2^{d-1}$ separately and argue such phases of matter either do not  exist or obey the area law.

Throughout this paper we will, with a few exceptions, consider gapped phases of matter that are stable to arbitrary weak Hamiltonian perturbations (sometimes this can be proven \cite{2010AnPhy.325.2120K,2011CMaPh.307..609B,2010JMP....51i3512B,2013CMaPh.322..277M}, but we take it as a physical assumption). Except for translation invariance, symmetry plays no role in our analysis, and translation invariance beyond rough homogeneity is not at all essential to the construction. We will also assume that when the phase of matter possesses degenerate ground states, those ground states are locally indistinguishable. Local indistinguishability is a consequence of stability, for if the degenerate ground states were locally distinguishable then the degeneracy could be split with a local field and the system would not be stable\footnote{For example, this restriction rules out a dilute array of decoupled spins for which there exist linear combinations of degenerate states with lots of entanglement. A more interesting case is a dilute array of non-abelian anyons. Unlike in the spin case, there are no operators localized at a single anyon that can split the degeneracy. Nevertheless, as we discuss in Appendix~\ref{sec:anyon} the anyon array also violates our assumptions.}. Quantitatively, we assume that the ground states are split by at most an exponentially small amount of order $e^{-cL^\alpha}$ for some constants $c$ and $\alpha$ (see the Ground State Degeneracy Lemma in \S\ref{subsec:basiclemma}).

Our fundamental assumption is that all stable gapped phases of matter are generalized $s$ source fixed points (defined below) for some $s$. We will discuss this assumption further, but for now let us simply note that we know of no gapped phase of matter that is not plausibly such a fixed point; we refer to a phase which is not a fixed point as a {\it dragon}.
(For work towards constructing a possible dragon, see \cite{dragon}\footnote{The system described in \cite{dragon} is similar to two infinite dimensional clusters coupled by a weak link. As such, it appears to violate our assumptions that the space must have a definite dimension and that the Hamiltonian arise from a Hamiltonian motif (Appendix~\ref{sec:phase}). Other examples of highly entangled states include \cite{2010NJPh...12b5002G,2010JMP....51b2101I}. These states also fall outside our assumptions since the gap vanishes with increasing system size, however, they are interesting in that they challenge standard field theory scalings of entanglement with spectral gap.}). In essence, we are assuming that all phases of matter are renormalization group fixed points.

Within our broader analysis an important role is played by what we call the ``weak area law" which asserts that all gapped states with a unique ground state on any closed manifold obey the area law for entanglement entropy.  We give several physical arguments for the weak area law. Using the weak area law plus our basic assumption that all gapped phases are generalized $s$ source fixed points, we show that all gapped phases with torus ground state degeneracy $G(L)$ scaling slower than $e^{c L^{d-1}}$ obey the area law. This leaves a small window of highly degenerate topological phases (with $s=2^{d-1}$) which, if they exist, may violate the area law. We give a special argument in this marginal case
(\S\ref{sec:arealaw}) to show that such phases in fact do not exist.  These arguments rely on ideas about reconstructing quantum states from local data and lead to additional arguments for the weak area law.

We further define the notion of topological quantum liquids (a subset of all possible topological phases) which are, roughly speaking, topological phases that are insensitive to the local details of the system and to the precise geometry. For example, the ground state degeneracy of a topological quantum liquid depends only on long distance data, so the ground state degeneracy on a $d$-torus is independent of torus size. The primary experimental realizations of topological quantum liquids are the fractional quantum Hall states \cite{PhysRevLett.48.1559,PhysRevLett.50.1395,PhysRevB.41.9377}. Any phase of matter which can be adiabatically deformed from linear size $L$ to linear size $2L$ is a topological quantum liquid. States with ground state degeneracy independent of system size and shape have also been singled out in \cite{2011AnPhy.326...15Y} and
more recently in \cite{2014arXiv1406.5090Z}.

We prove that topological quantum liquids obey the area law in $d>1$. We also show how to produce a MERA representation using modest resources for all topological quantum liquids in any dimension. For example, we show that chiral topological phases in $d=2$ have approximate MERA representations (see \cite{2008JSMTE..01L.001H,2010arXiv1004.2563G,2011PhRvL.106o6401B,2012PhRvB..86x5305Z,2013arXiv1307.7726D} for important prior work on this topic; see \cite{2014arXiv1404.4327H} for a discussion of some obstructions). The MERA representative has bond dimension $e^{c \log^{d(1+\delta)}(L)}$ in $d$ dimensions and achieves unit overlap with the ground state in the thermodynamic limit. The $\delta$ factor arises from truncating almost-exponentially decaying interactions; any $\delta>0$ will do, and we can even achieve a dependence like $e^{c (\log(L)\log(\log(L)))^d}$ as a limit $\delta \rightarrow 0$. Although such a MERA is not quite contractible in time polynomial in $L$, it is much more easily contracted than PEPS constructions \cite{2006PhRvB..73h5115H,2007PhRvB..76c5114H,2014arXiv1406.2973M} of similar bond dimension. Furthermore, if we don't require such fantastic accuracy in the thermodynamic limit, our results support the conjecture that universal properties can be captured with a constant bond dimension MERA. Our procedure for constructing a MERA is quite different from one which obtains a MERA by variational calculation, so it may lead to interesting new algorithms.

We conclude with discussion and conjectures about the extension of our results to gapless systems.  The ideas previewed in this final section will be discussed in greater detail in a forthcoming companion paper.

Given the length and complexity of the paper, here is a brief summary of results and a guide to notation. An attempt has been made to render the sections modular so that readers may skip around. The paper may be roughly divided into three parts. First, the basic $s$ source RG construction is introduced and developed (\S\ref{sec:rgintro}-\S\ref{sec:thermo}). Second, a number of concrete examples and some elaborations of the basic framework are discussed (\S\ref{sec:examples}-\S\ref{sec:gen-s-source}). Finally, the more advanced arguments for the area law, for MERA representations, and for all field theories having $s\leq 1$ are presented (\S\ref{sec:arealaw}-\S\ref{sec:fieldtheory}). The quickest way to proceed is to study the basic $s$-source definitions in \S\ref{sec:rgintro} and \S\ref{sec:s-source} and the examples in \S\ref{sec:examples}. After listing the main results and definitions, we briefly indicate the level of rigor of the various results. The main definitions and results include:
\begin{itemize}
\item Definition of $s$ source RG fixed points and demonstration that a large number of phases fall into this class. [\S\ref{sec:rgintro},\S\ref{sec:s-source},\S\ref{sec:examples}]
\item Definition of an inverse state $|\psi^{-1}\rangle$ for a state $|\psi\rangle$: a state $|\psi^{-1}\rangle $ such that $|\psi\rangle |\psi^{-1}\rangle $ is deformable to a product state by a quasi-local unitary. [\S\ref{sec:s-source}]
\item Definition of short-range entangled states: a state is short-range entangled if it has an inverse state. [\S\ref{sec:s-source}]
\item ``Wormhole array" argument that phases with a unique ground state on any closed manifold have an inverse state, implies the weak area law. [\S\ref{sec:s-source}]
\item Demonstration of at most logarithmic violation of the area law with a weak assumption about thermal free energy (generalizes Hastings' argument \cite{2007PhRvB..76c5114H}). [\S\ref{sec:thermo}]
\item Explicit demonstration that a chiral phase (Chern insulator) is an $s=1$ fixed point. [\S\ref{sec:examples}]
\item Definition of a topological quantum liquid (TQL): a phase which can be adiabatically locally deformed, proof that TQLs are $s\leq 1$ fixed points, proof of an area law for TQLs in $d>1$. [\S\ref{sec:tql}]
\item Definition of generalized $s$ source RG fixed points, conjecture that all stable gapped phases are such fixed points. [\S\ref{sec:gen-s-source}]
\item Argument for area law for generalized $s$ source fixed points with ground state degeneracy $G$ scaling slower than $e^{c L^{d-1}}$, assumes weak area law. [\S\ref{sec:gen-s-source}]
\item Reconstruction from local data argument for weak area law, argument for the entanglement entropy bound $S(A) \leq \mathcal{O}(|\partial A|) + \log(G(H_A))$ where $G(H_A)$ is the ground state degeneracy on a space with boundary. [\S\ref{sec:arealaw}]
\item Proof that entropy bound $S(A) \leq \mathcal{O}(|\partial A|) + \log(G(H_A))$ plus assumption that all phases are $s$ source fixed points implies area law. [\S\ref{sec:arealaw}]
\item Construction of approximate MERA representative with $e^{c \log^{d(1+\delta)}(L)}$ bond dimension ($\delta >0$) for TQLs in $d$ dimensions. [\S\ref{sec:mera}]
\item Conjecture that TQLs have an approximate MERA representative with polynomial bond dimension, argument for universal properties from bounded bond dimension. [\S\ref{sec:mera}]
\item Expanding universe construction for field theories, argument that all gapped field theories are $s\leq 1$ fixed points, explicit example with Dirac fermions (same universality class as Chern insulator), discussion of relation to dS/CFT. [\S\ref{sec:fieldtheory}]
\end{itemize}

Of these results, the area law for TQLs, the area law for $s< 2^{d-1}$ RG fixed points, the MERA construction for TQLs, the demonstration that Chern insulators are TQLs, the logarithmic bound on area law violations from thermodynamics, the Dirac fermion field theory construction, and the weak area law for frustration free Hamiltonians are (or can be made) rigorous. The weak area law in full generality, the general field theory constructions, the bound $S \leq \mathcal{O}(|\partial A|)+\log(G)$, and the suggestion that system-size-independent bound dimension in MERA suffices to capture universal properties are given strong physical arguments. The $s$ source RG framework (and the general area law result which relies on it) plausibly applies to all phases we are aware of, but we cannot rule out isolated cases outside the framework at the present time.

Notation: $A$ denotes a subregion, $L$ is the linear size of the whole system, $R$ is the linear size of $A$. $c$ and $k$ denote generic constants which don't depend on important parameters, $s$ specifies the number of copies or the matrix of RG dependencies. $d$ is the dimension of space, $\mathcal{D}$ is the local Hilbert space dimension. $\Delta$ and $m$ denote gaps, $J$ denotes the magnitude of terms in the Hamiltonian. $G(L)$ is the ground state degeneracy on a torus of linear size $L$; sometimes we use $G(R)$ to denote the ground state degeneracy on an open manifold of size $R$ and $G(H)$ to denote the ground state degeneracy of a Hamiltonian $H$. Ground states are often denoted $|g\rangle$ or $|g_i\rangle$. Couplings in the Hamiltonian are denoted $g_x$ or sometimes $\lambda$ and should not be confused with the labels of ground states.

\section{The RG-like transformation}
\label{sec:rgintro}

\begin{wrapfigure}{R}{.5\textwidth}
  \centering
    \includegraphics[height=5cm, width=0.48\textwidth]{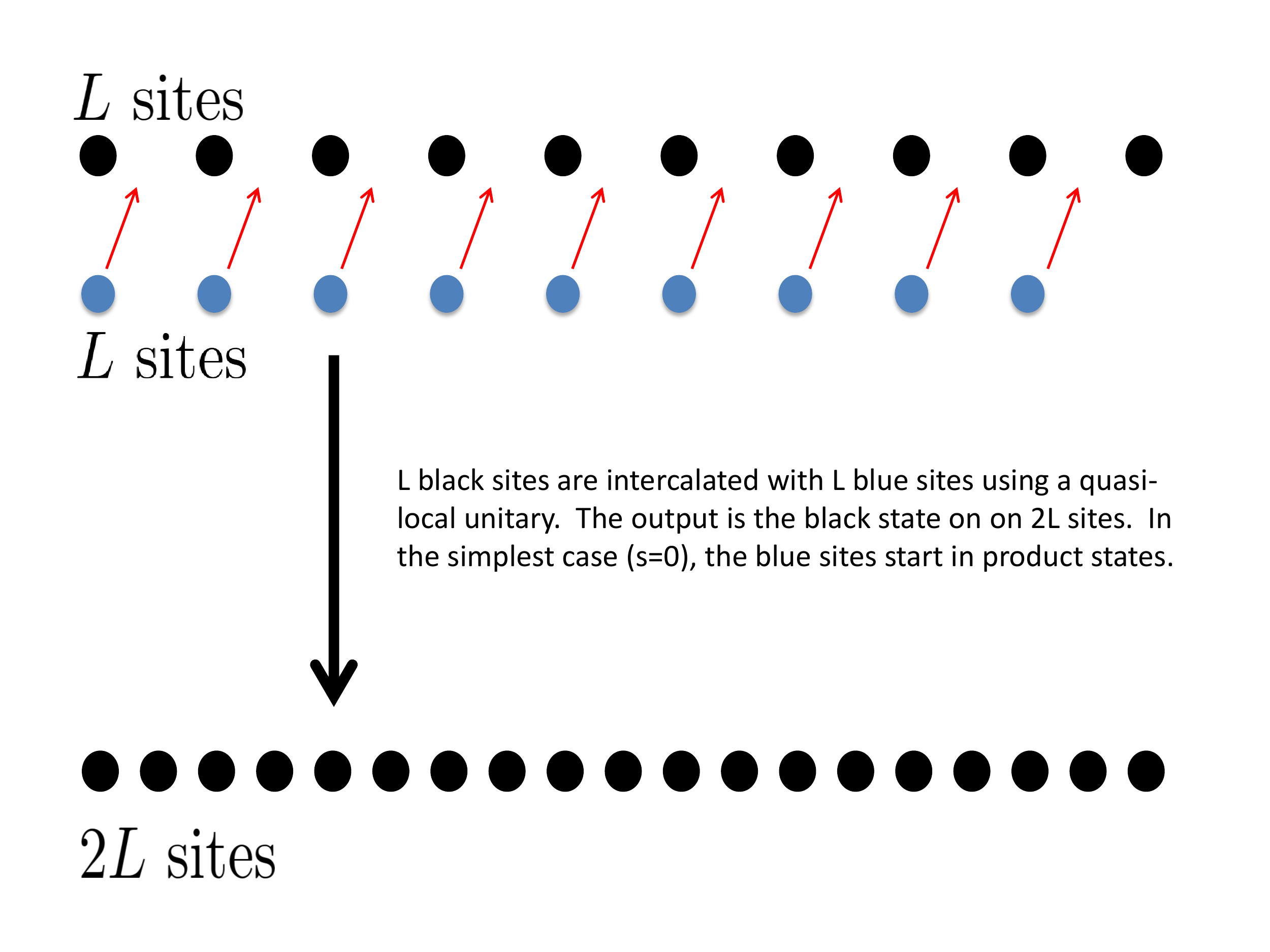}
  \caption{A $d=1$ version of the RG transformation.}
  \label{intercalate}
\end{wrapfigure}

We begin by defining an $s$ source RG fixed point in $d$ dimensions. The number $s$ specifies the number of non-trivial resources (``source states") needed to construct the state of a larger system in terms of states of smaller systems. Unentangled states always cost nothing and can be added or removed at will. Note that this first definition is a simplified version (single type theory) of the full theory (multi-type theory) where we restrict to source states that are identical. Below (\S\ref{sec:gen-s-source}) we define the notion of a generalized $s$ source RG fixed point (multi-type theory) which we conjecture is a sufficiently powerful notion to include all gapped phases. The single type theory is nevertheless quite useful as it illustrates the main ideas in a simpler setting and describes many cases of physical interest.

\begin{DEF}[$s$ source RG fixed point]
A $d$ dimensional $s$ source RG fixed point is a system where a ground state on $(2L)^d$ sites can be constructed from $s$ copies of the ground state on $L^d$ sites plus some unentangled degrees of freedom by acting with a quasi-local unitary as in Fig.~\ref{intercalate}. Unless otherwise noted, $s$ is assumed to be the smallest value for which the construction is possible.
\end{DEF}

A quasi-local unitary is a unitary $U$ generated by time evolution for a time of order $L^0$ by a Hamiltonian $K$ which is a sum of terms that are local up to tails decaying faster than any power. In detail, $K = \sum_x K_x$ and
\beq
K_x = \sum_{r} K_{x,r}
\eeq
where each term $K_{x,r}$ is supported on a disk of radius $r$ centered at $x$ and has norm $\|K_{x,r}\|$ decaying faster than any power of $r$.\footnote{The terms decay with distance $r$ as $e^{-r g(r)}$ where $g(r)$ is any function with the property that $\int_1^\infty dr \frac{g(r)}{r} < \infty$ \cite{Ingham01011934}. For example, $g(r) \sim r^{-\delta}$ or $g(r) \sim (\log(r))^{-2}$ are sufficient. This almost exponential decay is the origin of the $\delta$ factor in our MERA constructions.}

Recall that we restrict to stable phases, so the gap does not close under small Hamiltonian perturbations (i.e.~the $s$ source fixed point is a completely attractive RG fixed point). We will relax this assumption for future extensions to gapless states. In many cases we need only consider $ 0 \leq s \leq 2^{d-1}$ (see \S\ref{sec:thermo}). The case $s=0$ corresponds to ground states which can be produced at any size just from product states with a quasi-local unitary.

As a technical note, for concreteness we focus on coarse-graining schemes where linear dimensions are halved, e.g.~a decimation scheme where we map $2^d$ sites to one site.  Nothing in the formalism depends on this choice, so we may immediately extend our results to other kinds of decimation schemes. Indeed, some phases of matter in the recent literature behave best under coarse-graining transformations with different coarsening factors, and the formalism can easily accommodate this degree of freedom. We can even grow the system anisotropically, enlarging some dimensions while keeping others fixed, but we do not make use of this extra freedom in the present paper.

Let us also be clear about the notion of quasi-locality. What we are considering is a situation where the $s$ copies at scale $L$ are intercalated and then glued together by a quasi-local unitary as shown in Fig.~\ref{intercalate}. We are not gluing together regions at their boundaries. We are imagining that the $s$ copies exist in the same space and are merged together locally (like riffling a deck of cards) with respect to the usual Euclidean metric.
Note that the range of the quasilocal unitary which accomplishes this does not depend on the system size $L$.

Finally, we assume that there exists an $L_0$ such that the local Hilbert space is isomorphic at all scales $L \geq L_0$. If this were not so then we could always trivially realize a size $2L$ system with local dimension $\mathcal{D}$ as a size $L$ system with local dimension $\mathcal{D}^{2^d}$, so to get an interesting definition we must make a restriction on the local Hilbert space.  Furthermore, throughout we assume no symmetry besides translation invariance, and translation invariance primarily means that we consider Hamiltonians which are roughly homogenous in space. Clearly our approach can be refined by the inclusion of symmetry (leading to the physics of topological insulators \cite{2005PhRvL..95n6802K,2006Sci...314.1757B,2007Sci...318..766K,2007PhRvB..76d5302F,2008Natur.452..970H,2010RvMP...82.3045H,PhysRevB.87.155114,PhysRevX.3.011016,2012PhRvB..86l5119L}), but we leave this for future work.

Several detailed examples are presented below. As a preview, any trivial insulator is an $s=0$ RG fixed point while the toric code/$\IZ_2$ gauge theory is an $s=1$ RG fixed point. Haah's code \cite{2011PhRvA..83d2330H} is an example of an $s=2$ RG fixed point in $d=3$ \cite{2014PhRvB..89g5119H} (see also \cite{PhysRevB.88.125122}). The concept of an $s$ source RG fixed point has been latent for some time. In particular, the (gapless) case of fermions with a Fermi surface seems to realize an $s=2^{d-1}$ fixed point (although we do not prove this claim here). Indeed, it is the distinction between the RG for a conformal field theory (CFT) \cite{RevModPhys.47.773} and a Fermi surface \cite{RevModPhys.66.129,1992hep.th...10046P,PhysRevB.42.9967} which we are trying to capture with our notion of $s$ source fixed point. However, we will put these motivations aside for the present paper which is concerned almost exclusively with gapped phases. The notion of an $s$ source fixed point shares some similarities with branching MERA \cite{PhysRevLett.112.240502}, but we emphasize that our construction is different in various important ways. Chief among them, our formalism (making use of quasi-local unitaries) is sufficiently flexible to naturally describe a wide class of phases, while producing a MERA or branching MERA with its strict causality structure requires a blow-up of complexity\footnote{Our framework also preserves translation invariance, where as the MERA network breaks translation invariance. However, the MERA construction in \S\ref{sec:mera} shows that a nearly translation invariant MERA is achievable despite the bias in the network.}. Later we will discuss the precise relation to MERA (see \S\ref{sec:mera}).

The concept of a phase of matter is a primitive notion discussed in Appendix~\ref{sec:phase}. An important property of many phases is that they can be defined on any space\footnote{Extending the real time quantum theory to a Euclidean theory defined on an arbitrary Euclidean spacetime is a non-trivial further step.}. However, in some cases it is not clear at present how to make the definition, e.g.~with Haah's code. The phases we consider must have some part of this flexibility so that they may be defined on tori and disks of various sizes. Considerably greater flexibility is available for phases obtained from a Hamiltonian with two-body interactions after sufficient coarse-graining (e.g., by drilling a lattice of little holes). There is a large literature on related two-body constructions realizing interesting topological states, e.g., \cite{2009arXiv0901.1333K,2011PhRvL.107y0502O}. We conjecture that at least all phases with $s\leq 1$ have such two-body Hamiltonians. In any event, we will work for the most part in the simplified setting of tori and disks\footnote{Interesting phenomena occur when we deviate substantially from flatness or significantly complicate the topology, e.g., \cite{tc_neg_curved} which studied the toric code on a negatively curved space with extensive first Betti number. Fascinating as such examples are, we restrict to flat geometries and small perturbations thereof; one reason for this is that the area law is less well defined in a hyperbolic-like geometry where volumes and areas scale in the same way.}.

\section{The $s$ source framework}
\label{sec:s-source}

We now present a number of basic assumptions and results within the $s$ source RG framework that will be used extensively later. The first statement is our basic assumption, namely that all gapped phases of matter are generalized $s$ source fixed points. The second statement is the weak area law. We offer physical arguments for both these assumptions, we also later rigorously prove the weak area law for the restricted class of frustration free Hamiltonians. Then we characterize how the entropy of sub-systems depends on size via a recursive bound. We also characterize the growth of ground state degeneracy as a function of system size. Finally, we show that $s$ can be restricted to a certain reasonable range with an additional weak spectral assumption.

\begin{conj}[Fundamental Assumption]
All stable gapped phases are (generalized) $s$ source RG fixed points for some $s$.
\end{conj}
Evidence: As a warm-up note that all continuum topological field theories have $s=1$ or $s=0$. Indeed, we may place the field theory (mass gap $m$) into a slowly expanding universe with metric $ds^2 = - dt^2 + a^2(t) d \vec{x}^2$ with the scale factor obeying $\dot{a}/a \ll m$ (see \S\ref{sec:fieldtheory}).  The adiabatic time evolution from $a=1$ to $a=2$ generates an approximation to the desired quasi-local unitary transformation.  The short wavelength modes which expand with the universe are the analogs of the unentangled auxiliary degrees of freedom. Since we need only one copy of the state to do this evolution we have $s\leq 1$.  (More details of this protocol can be found in \S\ref{sec:fieldtheory}.)

More generally, as we show below, generalized $s$ source fixed points can accommodate a wide variety of scalings of entanglement entropy (up to volume law scaling) and can even support long-range correlations. In other words, the formalism is quite expressive in terms of the states it can accommodate.  Indeed, the authors know of no gapped phase which isn't plausibly in this class.

For a state to not be in this class, it must be the case that there is no path in the space of local Hamiltonians (of system-size-independent length) which connects the Hamiltonian on $(2L)^d$ sites to $2^d$ other decoupled Hamiltonians each on $L^d$ sites and which keeps the gap open. This must be true even if we permit the use of extra initially unentangled degrees of freedom which are returned to their unentangled state at the end of the adiabatic path. Note that stability implies that we have an open set in the space of Hamiltonians to work with, at least in the neighborhood of the fixed points and we need just one connection between these open sets. The preceding statements must also be true for all other choices of coarse-graining scheme. Given the considerable freedom this construction affords us, we believe it is a plausible fundamental assumption.

We also tend to the opinion to that a gapped Hamiltonian which is so radically disconnected from any other gapped Hamiltonian at smaller scales would be very unusual beast.  Our RG intuition probably provides very little guidance to the properties of this Hamiltonian. Nevertheless, it should be said that our frustration free results (if the Hamiltonian is in this class) still provide a measure of control independent of the assumption of being an $s$ source fixed point. For example, we can still show that to violate the area law the system would have to have many degenerate ground states on an open manifold. Thus our basic intuition that area law violations are related to the existence of many low energy states is still partially preserved.

As a final point in favor of the $s$ source framework, we observe that it produces conclusions in harmony with a variety of independent results. For example, assuming the entanglement entropy obeys an area law and the sub-leading terms have a certain structure, \cite{2013PhRvL.111h0503K} has shown that the number of locally indistinguishable ground states is bounded by certain combinations of entanglement entropies. The structure of sub-leading terms necessary to have $G(L)\sim e^{cL}$ is precisely what is predicted by the $s$ source framework.

In essence, our fundamental assumption claims that all phases of matter arising from local Hamiltonians are renormalization group fixed points. In any event, the very wide applicability of the $s$ source framework justifies its study even if phases outside the framework are eventually identified.

\subsection{Weak area law}
\label{subsec:wal}

\begin{physth}[Weak Area Law]
All gapped phases of matter with a unique ground state (on any closed geometry) obey the area law.
\end{physth}
Argument: We now present our first argument for the weak area law. In fact, we establish a stronger result: phases with a unique ground state on any closed geometry always have an ``inverse state" (defined momentarily). We give an independent argument for the weak area law in \S\ref{sec:arealaw}.

\textbf{Inverse state}

To begin, let us define the notion of an ``inverse state". Given a gapped ground state $|\psi\rangle$ defined on some local geometry, we say $|\phi\rangle$ (defined on the same geometry) is an inverse state for $|\psi\rangle$ if the tensor product $|\psi \rangle |\phi\rangle $ can be deformed into a product state with a quasi-local unitary. Note that $|\psi \rangle$ is also an inverse state for $|\phi\rangle$. As an example, if $|\psi_n\rangle$ is a quantum Hall state with $n$ filled Landau levels, then a state $|\phi_{-n}\rangle $ with $n$ filled Landau levels of the opposite magnetic field is an inverse state for $|\psi\rangle$. This is because we may cancel the chiral edge states between the two states, so while either state alone is non-trivial, the combination is a trivial insulator.\footnote{The related notion of an invertible
topological field theory has been used recently in \cite{Freed:2014eja}. Kitaev has independently developed a very similar notion \cite{Kitaev-unpublished}. Hastings has proven the existence of inverse states for free fermions \cite{2008JSMTE..01L.001H}.}

Now if $|\psi\rangle$ has an inverse state, then $|\psi\rangle $ obeys the area law.  Indeed, we have $|\psi \rangle |\phi\rangle = U^{-1} |0\rangle^{2 L^d}$ where $U$ is quasi-local and we have assumed without loss of generality that $\psi$ and $\phi$ are defined on $L^d$ sites.  Then the entropy bound for a quasi-local unitary implies that
\beq
S_\phi(R) + S_\psi(R) \leq R^{d-1},
\eeq
so both entropies obey the area law separately since they are both positive.  Our goal is thus to show that every phase of matter with a unique ground state on every closed geometry has an inverse state.

As an aside, the existence of an inverse state is a good criterion for calling a state short-range entangled (and is different from circuit definitions, e.g.~\cite{2010PhRvB..82o5138C}, which fail to classify integer quantum Hall states as short-range entangled). Since phases with an inverse have a unique ground state on any closed geometry, the ground state can be exactly reconstructed from local data \cite{2014arXiv1407.2658S,2010NatCo...1E.149C}, so the inverse-based definition of short-range entanglement seems closely related to Kitaev's definition of short-range entanglement \cite{Kitaev-unpublished}.

\textbf{Edge inverse}

\begin{wrapfigure}{R}{.5\textwidth}
  \centering
    \includegraphics[height=5cm, width=0.48\textwidth]{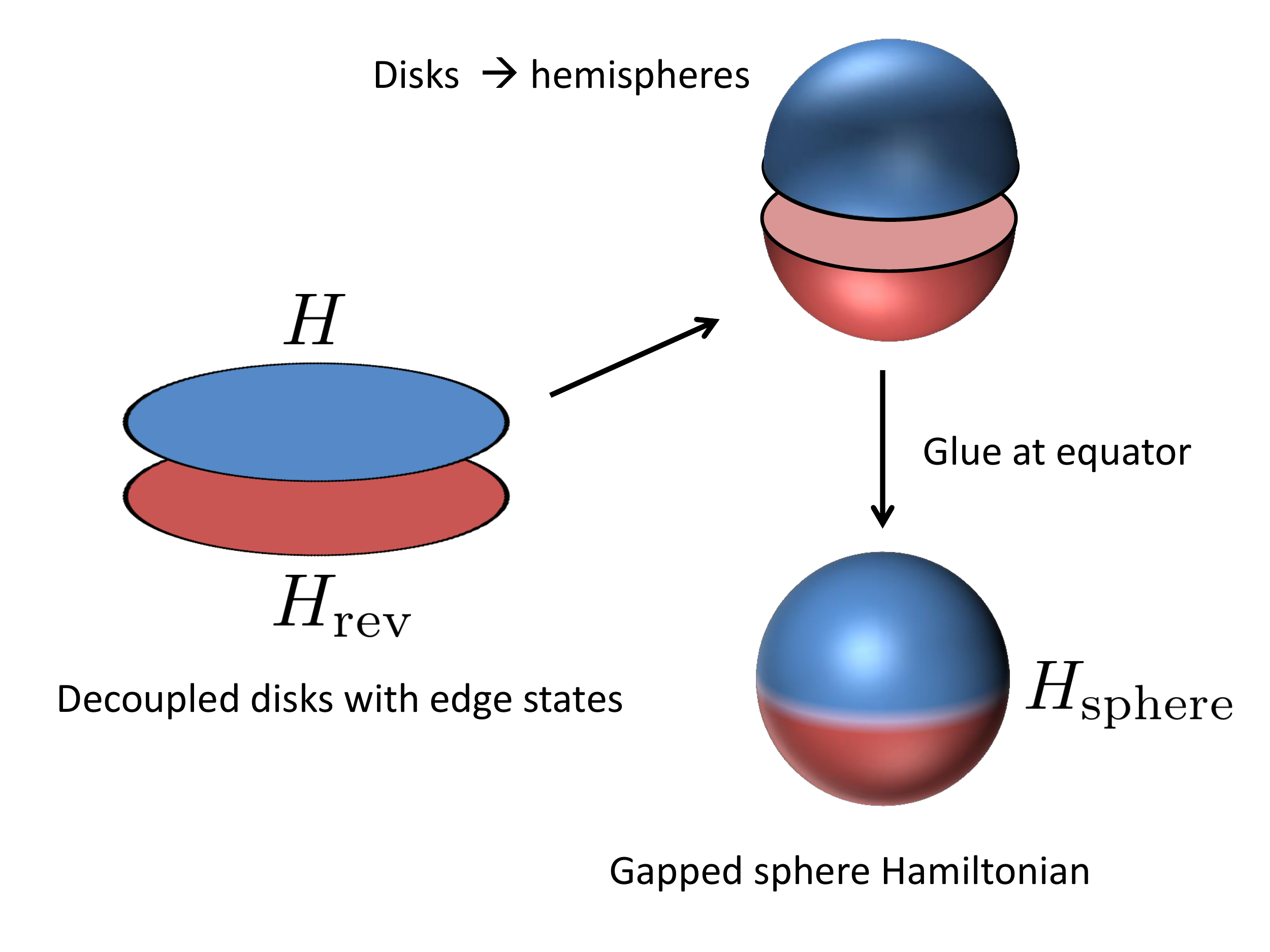}
  \caption{Coupling $H$ (blue disk) to its orientation reversed partner $H_{\text{rev}}$ (red disk) along their common boundary, we can produce the gapped sphere Hamiltonian.}
  \label{edgeinv}
\end{wrapfigure}

Intuitively, if a phase has a unique ground state then there is little interesting happening in the bulk of the phase. However, the system may display interesting physics if we place it on a manifold with boundary. In particular, we have the phenomenon of protected ``edge" or ``boundary" states which are boundary degrees of freedom that are necessarily gapless (or otherwise have some necessary low energy degeneracy). The canonical example here is chiral edge states in $d=2$ dimensions. An integer quantum Hall state has a unique ground state on any closed manifold, but on any open manifold the system necessarily possesses chiral edge modes which transport charge and heat.

These edge states will obstruct attempts to deform the system to a product state (making chiral states $s=1$ fixed points, for example). Fortunately, every phase has an ``edge inverse": another phase that can be coupled with the first phase just along the boundary to gap out the edge states.

To show the existence of an edge inverse, let $H$ be a Hamiltonian defined on a $d$-disk which may have protected edge states, e.g., the top blue disk ($d=2$) in Fig.~\ref{edgeinv}. Let $H_{\text{rev}}$ be the Hamiltonian defined on a $d$-disk which is obtained from $H$ by reversing the orientation, e.g., the bottom red disk ($d=2$) in Fig.~\ref{edgeinv}. For example, if $H$ were a quantum Hall system, the sign of the magnetic field would be reversed in $H_{\text{rev}}$. Now imagine deforming these two $d$-disks into the northern and southern hemispheres of a $d$-sphere as in Fig.~\ref{edgeinv}. Then couple the boundary of $H$ to the boundary of $H_{\text{rev}}$ while keeping them decoupled in the bulk. The resulting state, for suitable couplings and perhaps after passing through an edge phase transition, is the ground state of the original system but defined on a $d$-sphere with Hamiltonian $H_{\text{sphere}}$. Since this is a closed manifold, the Hamiltonian $H_{\text{sphere}}$ possesses an energy gap, so every protected edge state may be gapped out by pairing it with its reverse. Furthermore, if the phase in question has a unique ground state on any closed manifold, then the edge inverse, which can be defined analogously for arbitrary open geometries, always leads to a unique gapped bulk state.

\textbf{Wormhole array}

\begin{wrapfigure}{R}{.5\textwidth}
  \centering
    \includegraphics[height=5cm, width=0.48\textwidth]{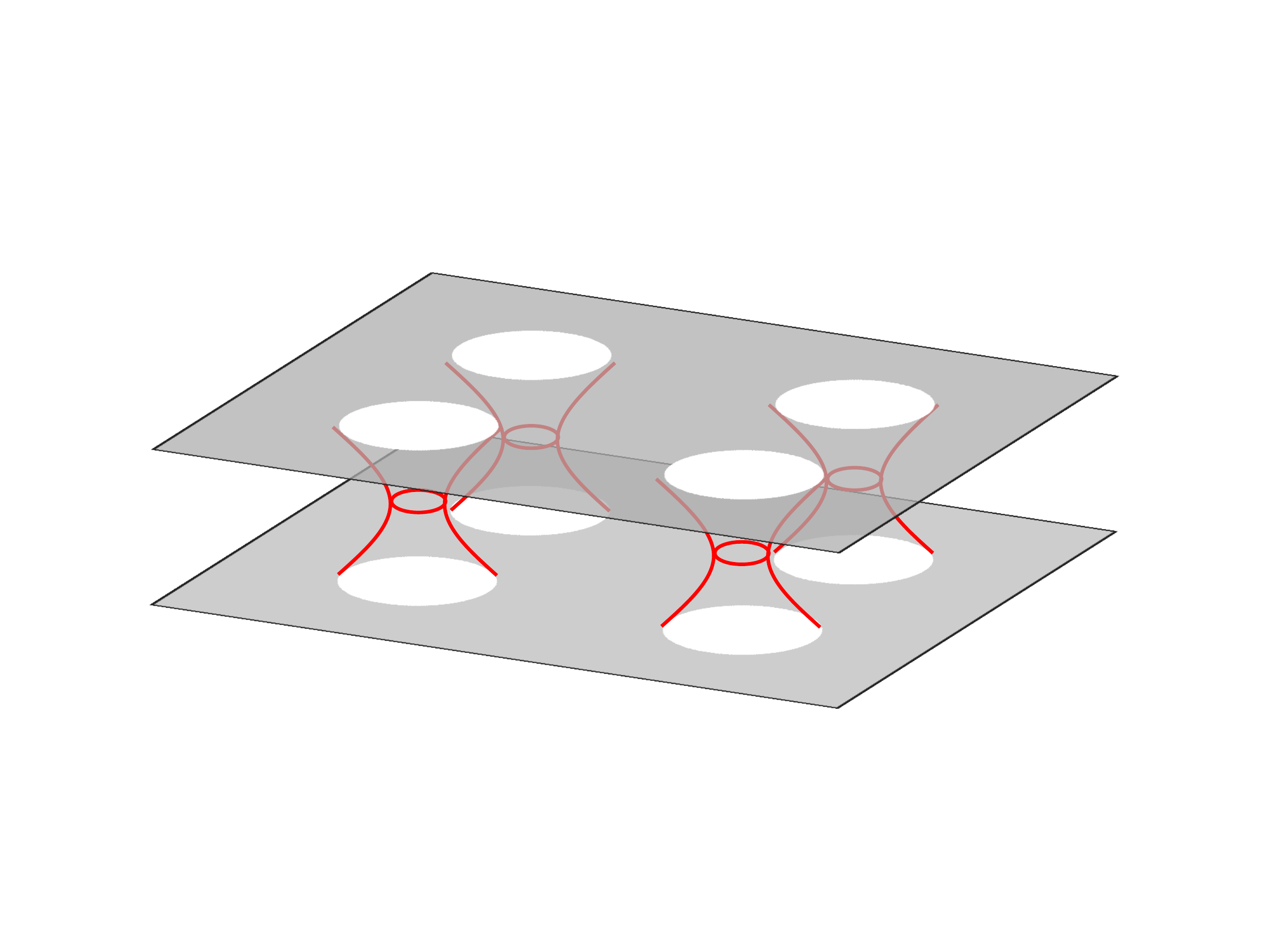}
  \caption{An array of wormholes in $d=2$.}
  \label{wharray}
\end{wrapfigure}

At this point, it is important to note that edge gappability by itself does not imply that an inverse state exists. Edge states can always be gapped by coupling to $H_{\text{rev}}$, but the bulk remains non-trivial if it has non-trivial ground state degeneracy.  Indeed, if the ground states are locally indistinguishable, then no quasi-local unitary can connect the ground states to product states because product states are locally distinguishable. If such a quasi-local unitary did exist, we could locally distinguish the supposedly locally indistinguishable ground states by choosing a local operator that distinguishes the corresponding product states and conjugating it with the quasi-local unitary.

However, we now argue that edge gappability plus trivial ground state degeneracy on any closed manifold implies that an inverse state exists. To begin, consider such a system with Hamiltonian $H$ on an open manifold consisting of a $d$-torus of linear size $L$ with a periodic array of holes of linear size $L_h$ and separation $L_s$.  The system may have gapless edge states around these holes, but we know that such edge states can be gapped by coupling to $H_{\text{rev}}$. Hence we introduce an identical torus with holes supporting $H_{\text{rev}}$ and couple the two systems along the boundary of the holes.  The resulting coupled system is equivalent to the original system but defined on a closed ``wormhole array" geometry which is illustrated in $d=2$ in Fig.~\ref{wharray}.  As shown there, we have two layers, corresponding to $H$ and $H_{\text{rev}}$, and the layers are coupled with ``wormholes" connecting the boundaries of the corresponding holes. Since this wormhole array is a closed geometry, the system,
by assumption, possesses a unique ground state on it.
A similar construction was used in \cite{Kitaev:2005dm} to relate topological
groundstate degeneracy to topological entanglement entropy.

\textbf{Construction of adiabatic path}

\begin{wrapfigure}{R}{.5\textwidth}
  \centering
    \includegraphics[height=5cm, width=0.48\textwidth]{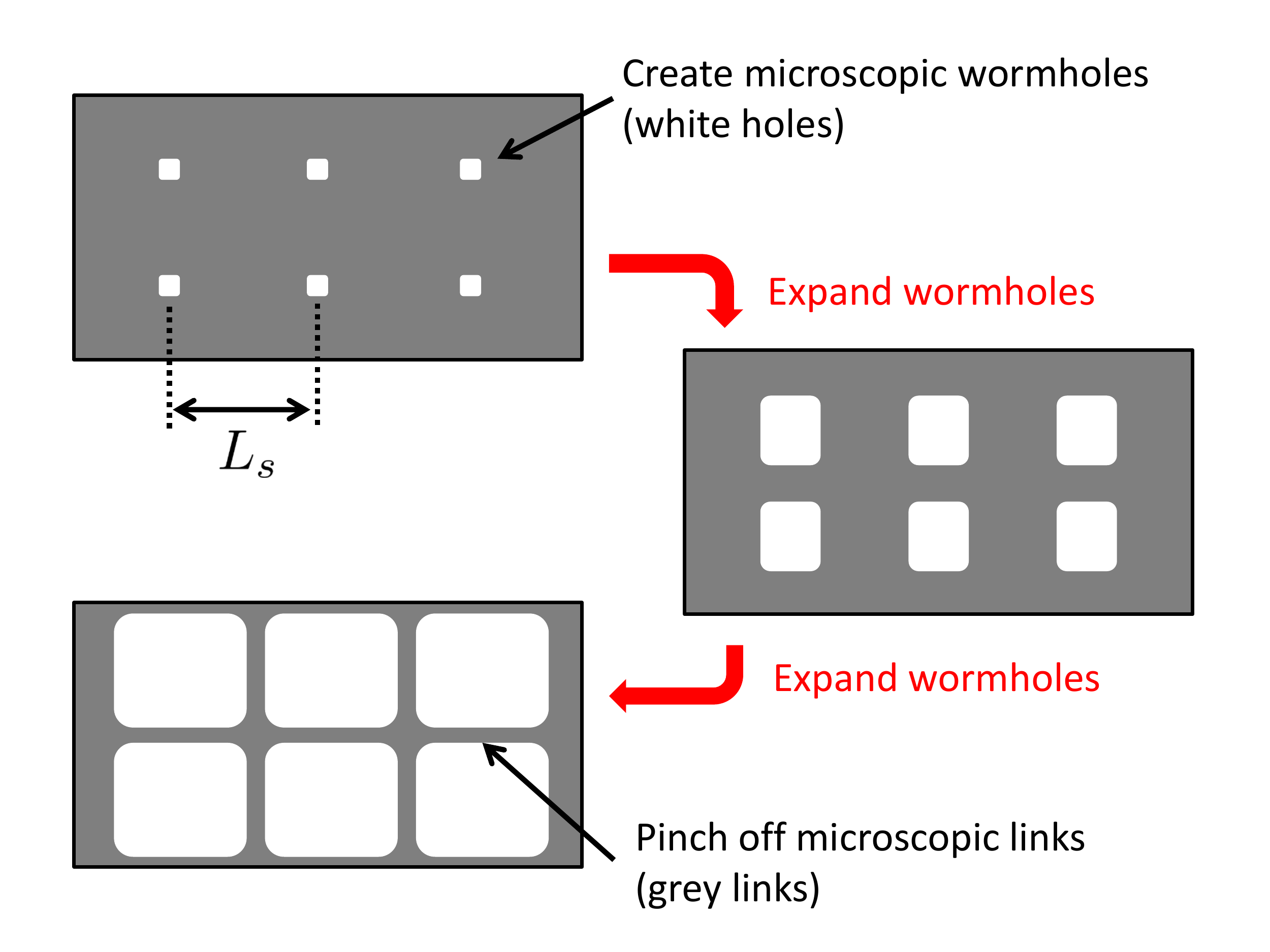}
  \caption{The transformation to a trivial state using an expanding wormhole array.  The white spaces denote product states or just empty space.  We have suppressed the wormholes and are effectively viewing the whole system as a composite of $H$ and $H_{\text{rec}}$ on a system with boundary.}
  \label{whseq}
\end{wrapfigure}

To complete the argument we make two physical assumptions.  Assumption 1: [Deformability] Because the system possesses a unique ground state on the wormhole array for any set of parameters $L$, $L_h$, and $L_s$, we assume that it is possible to deform the size and shape of the wormhole array without closing the gap.  Assumption 2: [Micro-insensitivity] We assume that we may make local microscopic deformations, e.g.~creating and pinching off microscopic wormholes without closing the gap.  Both assumptions essentially say that because the initial and final Hamiltonians are gapped, because the changes are local, and because the state is completely featureless, i.e.~no bulk physics, no edge physics, and hence nothing to require a phase transition, it should be possible to find a gapped path in Hamiltonian space connecting the initial and final points.  In other words, surely we can drill a dilute array of small holes in the system without closing the gap.

For example, to drill a single hole, consider the Hamiltonian $H(\eta) = (1-\eta) H_{\text{no hole}} + \eta H_{\text{hole}}$. Since this is a local perturbation, finite size effects may be sufficient to keep the gap open. However, suppose the gap does close along this path, say at $\eta_0$. Then we should be able to add to the Hamiltonian a local perturbation $V(\eta)$ which only turns on near $\eta_0$ and which keeps the gap open. Suppose the two states that are about to cross are $|0\rangle $ and $|1\rangle$. Zooming in on these two states, the Hamiltonian can be put in the form $H(\eta) \sim (\eta -\eta_0)(|0\rangle \langle 0| - |1 \rangle \langle 1|) = (\eta-\eta_0) Z$. The gap may be kept open by adding an $X$ perturbation, e.g., $V(\eta) = v(\eta) X$ with $v(\eta)$ a coupling localized in $\eta$ near $\eta_0$. We must only show that $X$ is a local operator, but this follows because $|0\rangle$ and $|1\rangle$ are indistinguishable far from the hole (they are gapped ground states of the same stable unique ground state Hamiltonian modulo a localized perturbation). Hence the operator which sends $|0\rangle$ to $|1\rangle$ is local and we can drill a hole in the system without closing the gap. Using the locality of the process plus the stability of the system to weak perturbations (due to effects from distant holes), we should also be able to drill a dilute array of holes without closing the gap. A similar argument applies to the process of expanding the holes, e.g., done a site at a time.

Using [Deformability] and [Micro-insenstivity], an adiabatic path to a product state may be found as illustrated for $d=2$ in Fig.~\ref{whseq}. Begin with two decoupled layers, one containing $H$ and one containing $H_{\text{rev}}$, which are shown as a single system in Fig.~\ref{whseq}. Then introduce an array of microscopic wormholes coupling the two layers. This can be done without closing the gap by [Micro-insensitivity]. Next expand the wormholes to larger and larger sizes. This can be done without closing the gap because we know $H$ coupled to $H_{\text{rev}}$ has a unique gapped ground state and by [Deformability]. Finally, when the wormholes have expanded to consume almost the entire system, pinch off the remaining thin tubes connecting different bulk regions. This can be done without closing the gap by [Micro-insensitivity]. At the end of this process we have reduced the system to product states.
Our assumptions imply that a system-size-independent gap may be maintained throughout this process
and that therefore 
the duration of the required adiabatic time evolution
(as well as the depth of the resulting circuit approximation) is 
also independent of system size.

In $d=1$ the introduction of wormholes simply disconnects the space into many small pieces, so we immediately obtain a product state. In $d>2$ a slightly more complicated recursive protocol is required. To begin, take two ``layers" consisting of $H$ and $H_{\text{rev}}$ and introduce an array of microscopic wormholes coupling them as before. Let the wormhole spacing be $L_s$. Expand these wormholes until their size is close to $L_s$ (the generalization of the process in Fig.~\ref{whseq} to higher dimensions). The expanding wormholes eat most of the $d$-dimensional bulk of the system, but leave a set of $d-1$-dimensional faces which are still entangled (analogous to the thin grey tubes at the end of Fig.~\ref{whseq}). Now repeat the procedure by introducing wormholes in the $d-1$-dimensional faces and expanding the wormholes to consume the faces. This leaves $d-2$-dimensional objects which are then eaten with still more wormholes and so on. The process terminates when we reach a one-dimensional network at which point introducing wormholes simply disconnects the remaining degrees of freedom into product states.

Thus the ground state of $H+H_{\text{rev}}$ is an $s=0$ fixed point in any dimension $d$, and the ground state of $H_{\text{rev}}$ is an inverse state for the ground state of $H$. This more general result, the existence of an inverse state, implies the weak area law. Note the crucial role played by the wormhole array and the assumption that the system has a unique ground state on it.

\subsection{Basic $s$ source results}
\label{subsec:basiclemma}

\begin{lem}[Entropy Lemma]
The entanglement entropy $S(R)$ of a region of size $R$ in any $s$ source RG fixed point obeys $S(2R) \leq s S(R) + k R^{d-1}$ where $k$ depends on the details of the quasi-local unitary.
\end{lem}

Proof: The entropy of a region of size $2R$ can be no more than the sum of the entropies of the $s$ regions of size $R$ used to make it plus a term coming from the quasi-local unitary. Such a quasi-local unitary can generate at most area law entanglement \cite{2013PhRvL.111q0501V}.
(This result is illustrated in Appendix~\ref{sec:TEE}.)
Although we have phrased this as a bound, it should describe the asymptotic behavior provided all $s$ copies are actually being used at every RG step and the quasi-local unitary is adding some entropy. Note that we have implicitly assumed that the entropy $S(R)$ is independent of $L$ provided $L \gg R$; this is one example of what we call insensitivity to boundary conditions in Appendix~\ref{sec:phase}. This assumption is not essential to the bound, but it is part of what we mean by a phase of matter and can be proven in some cases.

This bound is similar to the entropy accounting in branching MERA \cite{PhysRevLett.112.240502}, but our bound is more general because we allow quasi-local unitaries instead of strictly local circuits.  The extension to quasi-locality, while intuitively plausible, is not immediate and requires the technology in \cite{2013PhRvL.111q0501V}. Furthermore, for strictly local circuits one has much more control, e.g.~over even the Schmidt rank, but such control is currently lacking for quasi-local unitaries.

\begin{lem}[Ground State Degeneracy Lemma]
The ground state degeneracy $G(L)$ of a $s$ source RG fixed on a $d$-torus of linear size $L$ obeys the recursion relation $G(2L) = G(L)^s$.
\end{lem}

Proof: Recall that we are assuming all ground states are locally indistinguishable.  Choose one ground state from each of the $s$ copies at scale $L$.  By assumption we can construct a ground state at scale $2L$ using a quasi-local unitary.  However, because the unitary is quasi-local and because the ground states are locally indistinguishable, we can actually produce a different orthogonal ground state at size $2L$ for every choice of ground state from each of the $s$ sources at size $L$ using the same quasi-local unitary.  Indeed, the action of all local Hamiltonian terms, modulo the slight spreading due to the quasi-local unitary, remains local throughout and so has the same effect on all ground states.  In other words, if we get a ground state from one choice, we get a ground state for another choice, because all Hamiltonian terms act the same on the locally indistinguishable states.

To be precise, we take local indistinguishability to mean that we have a set of ground states $\{ |g_i \rangle\}_{i=1,...,G}$ such that we have
\beq \label{eq:local-indistinguish1}
|\langle g_i | O | g_j \rangle | < \epsilon \,\,\, (i \neq j),
\eeq
and
\beq\label{eq:local-indistinguish2}
|\langle g_i |O | g_i \rangle  - \langle g_j |O | g_j \rangle | < 2 \epsilon,
\eeq
for any normalized local operator $O$ and with $\epsilon \sim e^{- c L^{\alpha}}$.  To distinguish ground states we need to act with some operator supported on $L^{\alpha}$ sites (called the code distance), thus any exact ground state degeneracy is broken at order $L^\alpha$ in perturbation theory which is the origin of the above estimate for $\epsilon$.

Then let $|\psi_I(2L) \rangle = U |i_1\rangle  ... |i_s\rangle$ denote the state obtained at scale $2L$ from ground states labelled $I= \{i_1, ...,i_s\}$ (plus product states) at scale $L$.  By definition we have $\sum_x \langle \psi_I(2L)| H_x(2L) | \psi_I(2L)\rangle = E_g(2L)$ where $\{ H_x(2L)\}$ are the local Hamiltonian terms at size $2L$ and $E_g(2L)$ is the ground state energy at size $2L$.  Since $U$ is quasi-local and $H_x(2L)$ is local, the conjugated operators $U^\dagger H_x(2L) U$ are also quasi-local.  Hence by local indistinguishability we have $\langle i_1|  ... \langle i_s | U^\dagger H_x(2L) U |j_1\rangle  ... |j_s\rangle = h_x(2L) \delta_{i_1 j_1} ... \delta_{i_s j_s}$ with $h_x(2L)$ a c-number up to exponentially small corrections.  Hence $\sum_x \langle \psi_I(2L)| H_x(2L) | \psi_J(2L)\rangle = E_g(2L) \delta_{IJ}$ up to exponentially small corrections and we have $G(L)^s$ ground states at scale $2L$.

\begin{lem}[Restriction Lemma]
Under weak spectral assumptions, we may restrict to $s \leq 2^{d-1}$.
\end{lem}

Proof: As discussed just below, a weak spectral assumption on the low temperature thermal free energy implies that gapped phases obey the area law up to logarithmic corrections.  Assuming all $s = 2^{d-1+\alpha}$ copies of the state at size $L$ are needed to produce the state at size $2L$ (otherwise, choose a smaller $s$) and that the quasi-local unitary is adding entropy, the bound in the entropy lemma will be asymptotically saturated. Then the entanglement entropy scales as $S(R) \sim R^{d-1+\alpha}$, but this violates the area law worse than logarithmically when $\alpha > 0$. When $\alpha =0$ a logarithmic violation is consistent with the entropy lemma.  Hence we must have $\alpha \leq 0$ as claimed.

If the quasi-local unitary is not adding any entropy, then we have a decoupled system which can be understood within the layer construction (see \S\ref{sec:examples} just below). Apply the argument of the previous paragraph to the non-trivial components making up the layers to reach the same conclusion.

Alternatively, suppose the quasi-local unitary adds no entropy. In this case the entropy obeys $S(2R) = s S(R)$ and $S(R) \sim s^{\log(R)} S(1)$. Assuming there is some entanglement to begin with, this formula gives an entanglement entropy growing faster than $R^{d-1}\log(R)$ for $s>2^{d-1}$. In fact, even considering the perverse possibility of the quasi-local unitary removing entropy, it can only remove an area's worth of entropy, so there is a lower bound $S(2R) \geq s S(R) - k R^{d-1}$. For $s>2^{d-1}$ it may be verified that the only consistent possibilities are $S=0$ or $S$ growing faster than $R^{d-1}\log(R)$. Hence as claimed $s>2^{d-1}$ implies worse than logarithmic violation of the area law.

\section{$S(R) \leq R^{d-1} \log(R)$ from spectral assumptions}
\label{sec:thermo}

Following \cite{2008PhRvL.100g0502W} and \cite{2007PhRvB..76c5114H} (see also \cite{2009PhRvA..80e2104M} for an argument for at most logarithmic violations of the area law with somewhat different spectral assumptions), we can show that with a weak spectral assumption the area law can be violated at most logarithmically. We first review Hastings' original argument bounding ground state entanglement by thermal mutual information at low temperature and then generalize the argument to a wider class of physically relevant systems, e.g., perturbed conformal field theories which flow to massive infrared fixed points and which violate Hastings' density of states assumption.  We also give an explicit argument that the thermal mutual information bounds the ground state entanglement even when we have many locally indistinguishable ground states.

To motivate the assumption, consider a trivial paramagnet on $L^d$ sites.  The Hamiltonian is just a local magnetic field which favors the spins to align with the field, and the gap is $\Delta$.  The number of excited states at energy $E$ (between $E$ and $E+\Delta$, say) scales like
\beq \label{Dbound}
D(E) \sim D_0 \frac{\left(L^d\right)^{E/\Delta}}{(E/\Delta)!}
\eeq
where $D_0$ is some constant.  In other words, we can flip $E/\Delta$ spins, and these flipped spins can be on any of the sites, but spin flips are indistinguishable so we must divide by the factorial.  In fact, the above formula overestimates $D(E)$ because once a spin is flipped, we cannot flip it again, so the correct formula is actually
\beq
D(E) \sim D_0 \frac{1}{(E/\Delta)!} \underbrace{L^d (L^d-1)(L^d-2) ...(L^d-(E/\Delta)+1)}_{(E/\Delta)\, \text{factors}}.
\eeq
This formula is bounded by the form in Eq.~\ref{Dbound}, so below we assume that Eq.~\ref{Dbound} bounds the true spectral density (with some system-size-independent constant $\Delta$).

Now let $P$ denote the ground state projector and let $\rho(T) = e^{-H/T}/Z$ be the thermal state of the system.  Clearly we have
\beq
\lim_{T\rightarrow 0} \rho(T) = \frac{P}{G}
\eeq
where $G = \tr(P)$ is the ground state degeneracy.  We would like to now approximate the ground state projector by taking a small but non-zero temperature.

We can easily compute the trace distance between $\rho(T)$ and $P/G$ to be
\beq
\left|\frac{P}{G} - \rho(T) \right|_1 = \frac{1}{G} - \frac{1}{Z(T)} + \frac{Z(T)-G}{Z(T)}.
\eeq
We have set the ground state energy equal to zero and then used the fact that $Z \geq G$.  We want the difference in trace norm to go to zero like $L^{-q}$ to achieve high overlap in the thermodynamic limit.

To achieve this, we may set $T = \frac{\Delta}{\kappa \log(L)}$.  Then we write the partition function as
\beq
Z = G + \sum_{E >0} D(E) e^{-E/T} \leq G + \sum_E D_0 \frac{\left(L^d\right)^{E/\Delta}}{(E/\Delta)!} e^{-\kappa E \log(L)/\Delta}.
\eeq
Introducing the variable $x = E/\Delta$ we have
\beq
Z \leq G + D_0 \sum_{x=1}^{\infty} \frac{(L^{d-\kappa})^x}{x!} = G + D_0 \left(e^{L^{d-\kappa}} - 1\right).
\eeq
If $\kappa > d$ then the term in the exponent is going to zero at large $L$ and we have
\beq
Z \leq G + D_0 L^{d-\kappa},
\eeq
and setting $\kappa = d+q$ provides the desired accuracy in trace norm.

We now use the property that the mutual information between any region $A$ and its complement $\bar{A}$ in such a thermal state obeys \cite{2008PhRvL.100g0502W}
\beq
I(A,\bar{A}) \leq J |\partial A|/T
\eeq
where $J$ is the norm of the local Hamilonian terms.  This bound is proven by comparing the free energy, defined as $F(\rho) = \tr(\rho H) - T S(\rho)$, of $\rho_A \rho_{\bar{A}}$ to the free energy of $\rho_{A\bar{A}}$ and using $F(\rho_A \rho_{\bar{A}}) \geq F(\rho_{A\bar{A}})$ (the thermal state $\rho_{A\bar{A}}$ minimizes the free energy).  Using our expression for $T$ we find that the mutual information is bounded by
\beq
I \leq \frac{\kappa J}{\Delta} |\partial A| \log(L).
\eeq
The mutual information in the thermal state is also close to the mutual information in the equal weight mixture of ground states as follows from Fannes' inequality \cite{1751-8121-40-28-S18,fannes} provided $q > d$.  Hence the mutual information of any region $A$ of linear size $R$ in the equal weight mixture of ground states is bounded by $\sim R^{d-1}\log(L)$.

The above arguments also go through if the ground states are only approximately locally indistinguishable and split by an exponentially small amount.

\subsection{Entanglement entropy from mutual information}

To compute the actual entropy of the region $A$ in the equal weight mixture we need a little more work.  First, suppose that $A$ is small enough so that the ground states are still locally indistinguishable with respect to observables supported on $A$.  
Recall we assume locally indistinguishable ground states up to exponential correctons, 
in the sense of Eqs.~\eqref{eq:local-indistinguish1}, \eqref{eq:local-indistinguish2}.
With these definitions, any finite region of a sufficiently large system satisfies the criterion of local indistinguishability.  Now the state of region $A$ is
\beq
\rho_A = \tr_{\bar{A}}\left(\frac{P}{G}\right) = \frac{1}{G} \sum_{a=1}^G \tr_{\bar{A}}(|\psi_a\rangle \langle \psi_a |)
\eeq
where $|\psi_a\rangle $ are the ground states.  By the assumption of local indistinguishability we have $\tr_{\bar{A}}(|\psi_a\rangle \langle \psi_a |) \approx \tr_{\bar{A}}(|\psi_b\rangle \langle \psi_b |)$ for all $a$ and $b$.  Thus the sum over $a$ is a sum over identical terms, so the sum cancels the overall factor of $G$ and we find that the state of $A$ in the equal weight mixture of ground states is approximately the state of $A$ in any particular ground state.

What about the state of $\bar{A}$?  We have
\beq
\rho_{\bar{A}} = \frac{1}{G} \sum_{a=1}^G \tr_{A}(|\psi_a\rangle \langle \psi_a |),
\eeq
but now $\bar{A}$ is too large to guarantee local indistinguishability.  However, this is now useful because the states $\rho_{\bar{A},a} = \tr_{A}(|\psi_a\rangle \langle \psi_a |)$ must be orthogonal.  Let $\{P_a,1-P_a\}$ be the projective measurement which distinguishes $\rho_{\bar{A},a}$ from all other states $\rho_{\bar{A},b}$ so that we have $\tr(P_a \rho_{\bar{A},b}) = \delta_{a,b}$.  It follows from positivity of $P_a \rho_{\bar{A},b} P_a$ that $P_a \rho_{\bar{A},b} P_a = \rho_{\bar{A},a} \delta_{a,b}$ and that $(1-P_a)\rho_{\bar{A},b} (1-P_a) = \rho_{\bar{A},b}(1-\delta_{a,b})$.  Hence we have $\rho_{\bar{A},a} \rho_{\bar{A},b} = P_a \rho_{\bar{A},a} P_a (1-P_a) \rho_{\bar{A},b} (1-P_a) = 0$ as desired.  With the assumption of orthogonality plus the above results for the entropy $S(\rho_A)$ we have
\beq
S(\rho_{\bar{A}}) = \log(G) + S(\rho_A).
\eeq

Hence the mutual information $I(A,\bar{A}) = S(\rho_A) + S(\rho_{\bar{A}}) - S(\rho_{A\bar{A}})$ is given by $I(A,\bar{A}) = 2 S(A)$, so the entanglement entropy of region $A$ in any ground state is approximately half the mutual information obtained above.  Finally, taking $R \sim L^\delta$ for some positive $\delta$ (e.g.~$0 < \delta < \alpha$), we have $S(R) = R^{d-1}g(R)$ with $g(R) \leq k \log(R)$ for some $k$.  If this were not so, we could overwhelm the bound $S(R) \leq R^{d-1}\log(L)$ by taking $R\sim L^\delta$ and using the fact that for sufficiently large $L$ we have $g(L^\delta) >  k \delta \log(L)$ for any $k$.

\subsection{Generalized argument for massive deformations of CFTs}

Unfortunately, not all systems obey the spectral assumption discussed above. For example, consider a massive relativistic field $\phi$. Even with the mass gap $m$, the density of states at high energies scales like
\beq
D(E \gg m) \sim \exp\left(c_T (EL)^{d/(d+1)} \right),
\eeq
a result which is fixed by scale invariance and thermodynamics at high temperature ($c_T$ is a constant). Since we will consider field theories explicitly in \S\ref{sec:fieldtheory}, it is important to understand this case.

Of course, one may object that if a field theory is properly regulated, then perhaps the scaling of $D(E)$ in the previous section can be recovered from the physics of the regulator.  Perhaps this is so in some cases, but it is a physically irrelevant objection, because violations of the area law should have nothing to do microscopic details. In fact, we can show directly in the continuum that the same argument of the previous section, even with the CFT scaling of $D(E)$ at high energies, gives at most logarithmic violation of the area law for gapped field theories.  The argument is identical except that we assume the thermal free energy scales like $F = - T \log(Z) \sim L^d e^{-\Delta/T}$ at low temperature.  This scaling is satisfied by all gapped field theories, for example, despite the fact that these field theories violate the above spectral assumption above at high energies.  We now give a free field example to demonstrate the argument.

To carry out the argument in the previous section we need to know the density of states $D(E)$ of a massive free boson or fermion field in $d$ dimensions.  Since we will be especially interested in the limit of low temperatures, where we have a dilute gas of particles, both types of particles are effectively classical and their statistics become irrelevant.  We focus on the boson case for simplicity.

The density of states is by definition
\beq
D(E) = \sum_{\{n_k\}} \delta(E - E(\{n_k\}))
\eeq
where $E(\{n_k\}) = \sum_k \epsilon_k n_k$, $\epsilon_k = \sqrt{k^2+m^2}$ is the relativistic dispersion relation, and $n_k$ is the number of particles in mode $k$.  We can easily develop an expression for this quantity, but in fact what we really need is the partition function.  Since the total density of states can be written as a many-fold convolution (over all $k$) of each mode's density of states and since the partition function is essentially the Laplace transform of the density of states, we have a simple formula for the partition function as a product of the partition functions of the individual modes.

That is, we have
\beq
Z(T) = \int dE D(E) e^{-E/T} = \prod_k \sum_{n_k} e^{- \epsilon_k n_k/T} \underset{T\rightarrow 0}{\approx} \prod_k (1 + e^{-\epsilon_k/T}).
\eeq
Taking the logarithm of both sides we obtain
\beq
\log(Z(T)) = \sum_k \log(1 + e^{-\epsilon_k/T}) \approx L^d \int \frac{d^d k}{(2\pi)^d} e^{-\epsilon_k/T}.
\eeq
When $T \ll m$, the integral over $k$ may be well approximated as
\beq
\int \frac{d^d k}{(2\pi)^d} e^{-\epsilon_k/T} \approx c_d e^{-m/T} (mT)^{d/2}
\eeq
with $c_d$ a dimension dependent constant.

If we now wish to have the total partition function close to one we must have
\beq
L^d c_d e^{-m/T} (mT)^{d/2} \sim \epsilon \ll 1.
\eeq
This is easily satisfied if we take
\beq
T \sim \frac{m}{\log(L^d/\epsilon)},
\eeq
and even if we demand $\epsilon \sim 1/L^q$, we can achieve this with only a logarithmically small $T$.

More generally, it should be clear that what we require to demonstrate an area law up to at most logarithmic violations is a low temperature free energy $F$ of the form
\beq \label{eq:Fgap}
F = - T \log(G) + F_{\text{excited}}
\eeq
where $F_{\text{excited}}$ is extensive and decays as $e^{-\Delta/T}$.  This suffices to bound the mutual information. With the same assumptions about the scale of the failure of local indistinguishability, the entanglement entropy can be bounded as well. This formula will also be useful later when we argue for the area law in \S\ref{sec:arealaw}.

\section{Examples}
\label{sec:examples}

In this section we give numerous examples to flesh out the formalism.  We also discuss in more detail how to construct the quasi-local unitary which maps size $L$ to size $2L$.

\subsection{Trivial insulators, any $d$}

Any trivial insulator with an energy gap $\Delta$ that is independent of system size is an $s=0$ RG fixed point (even if it is protected by a symmetry) because we can construct a quasi-local unitary transformation (which may not commute with the symmetry) which produces the ground state from a product state\footnote{It is worth mentioning that the ground state of quantum chromodynamics (QCD) in the context of Hamiltonian lattice gauge theory, say, is likely an $s=0$ fixed point.}.  This is because by the definition of a trivial insulator there is a path $H(\eta)$ in the space of Hamiltonians such that $H(0)$ has a product ground state and $H(1)$ is the Hamiltonian of the trivial insulator. With this path in Hamiltonian space we can construct the required quasi-local unitary.

Define the quasi-adiabatic generator
\beq\label{eq:quasi-adiabatic-generator}
-i K(\eta) = \int_{-\infty}^\infty dt F(t) e^{i H(\eta) t} \partial_\eta H(\eta) e^{-i H(\eta) t}
\eeq
with $F$ a filter function \cite{2004PhRvB..69j4431H,2005PhRvB..72d5141H}. $F(t)$ is a fast-decaying, odd function of $t$ with the following properties.  First, its Fourier transform $\tilde{F}$ satisfies $\tilde{F}(\omega) = -\frac{1}{\omega}$ for $|\omega| \geq \Delta $ and second, $\tilde{F}(\omega=0)=0$.  $K(\eta)$ is designed to do the following job: when acting on the ground state $|\psi(\eta)\rangle$ of $H(\eta)$ it outputs $i \partial_\eta |\psi(\eta)\rangle$ as defined by first order perturbation theory.  The assumption of a finite gap is necessary to keep $K(\eta)$ quasi-local.  Once we know that $K(\eta)$ is quasi-local, then we know that it generates a quasi-local unitary that maps the product state to the trivial insulator ground state.  The quasi-locality of $K(\eta)$ also implies that all trivial insulators obey an area law \cite{2013PhRvL.111q0501V}.

To get a sense of what $K(\eta)$ is doing, consider a family of gapped Hamiltonians defined on $L$ spins of the form
\beq
\label{eq:rotating-paramagnet}
H(\theta) = \sum_{x=1}^L \frac{\Delta}{2} \left(Z_x \cos(\theta) + X_x \sin(\theta) \right).
\eeq
Observe that $H(\theta)$ is gapped with gap $\Delta$ for all $\theta$.  We could then appeal to the adiabatic theorem to argue that if we vary $\theta$ slowly the time evolved state will approximately follow the instantaneous ground state.  However, even with a finite gap there will typically be some small probability $p$ of error, e.g., a transition into a local excited state.  Since the probability of error is independent between sites, it follows that the total probability to remain in the ground state of the whole system is $(1-p)^L$.  Even if $p$ is quite small, for a sufficiently large $L$ there will always be an excitation somewhere in the system and we will no longer be in the global ground state.

There are three responses to this fact. The first response is to say that we just don't care if the system has a (roughly) exponentially small density of excitations, since this is not expected to modify physical properties except perhaps at exponentially long times, etc. And such a nearly exponential scaling is achievable since the probability $p$ of error can typically be made nearly exponentially small in the gap times the timescale, $\tau$, of the adiabatic evolution: $p \sim e^{-(\Delta \tau)^{1-\delta}}$.  The second response is to say that if we really want close to zero excitations, we need only make $p \sim 1/L^q$ for some sufficiently large $q$.  Assuming $p \sim e^{-(\Delta \tau)^{1-\delta}}$ we may take $\tau \sim \log^{1+\delta}(L)$, an evolution time growing very modestly with system size.  The third response reminds us that the above concern is silly: there is another Hamiltonian $\breve{H}(\theta)$ which is a sum of single spin operators which generates a time evolution that exactly maps the ground state of $H(0)$ to the ground state of $H(\theta)$.  Identifying $\theta = \eta$, the quasi-adiabatic generator $K(\theta)$ is nothing but an explicit construction of a Hamiltonian like $\breve{H}(\theta)$ which generates a time evolution that exactly maps ground state to ground state. (For the example \eqref{eq:rotating-paramagnet},
the quasi-adiabatic generator
\eqref{eq:quasi-adiabatic-generator}
evaluates to $ K(\eta) = \sum_x Y_x $, which clearly generates a rotation from $Z$ to $X$ without incident.)

Returning to the general case of a gapped local Hamiltonian, we still have the first two responses above and they may be sufficient for many purposes.  However, it is now less obvious that the third response remains valid, that there exists a local Hamiltonian like $\breve{H}(\eta)$ which generates a time evolution that maps ground states to ground states.  Remarkably, the quasi-adiabatic generator $K(\eta)$ can still be defined and, at the cost of a mild weakening of strict locality to quasi-locality, generates a time evolution which exactly maps ground states to ground states.  Note that this doesn't preclude the existence of a strictly local Hamiltonian which does the same job, but a quasi-local generator like $K(\eta)$ is sufficient for almost all purposes.

\subsection{Chiral insulators, $d=2$}

Examples here include $p+ip$ superconductors, integer quantum Hall states \cite{PhysRevLett.45.494,1981PhRvB..23.5632L}, the $E_8$ state of bosons \cite{2006AnPhy.321....2K},
and various kinds of chiral topological states such as fractional quantum Hall states \cite{PhysRevLett.48.1559,PhysRevLett.50.1395}. The distinguishing feature of this class is that any system in it, when placed on a manifold with boundary, supports chiral edge modes which cannot be gapped \cite{PhysRevB.25.2185}.

On general grounds, we can argue that such insulators are $s=1$ fixed points.  For example, in the context of a coupled island construction it is possible to remove some faction of the islands and place them into product states provided the rest of the islands remain coherent. We cannot do this simultaneously for all islands, because the system is not an $s=0$ fixed point, but it is possible remove a finite fraction of the islands.  Later we will give a very general field theory argument for $s=1$ which applies to various field theoretic representations of these phases.

In the remainder of this sub-section we would like to give another construction for chiral insulators using band engineering in a free fermion limit.  For concreteness, we consider the case of integer quantum Hall states in the guise of Chern insulators.  A simple lattice model for a Chern insulator \cite{2008PhRvB..78s5424Q} is obtained by taking a square lattice with two orbitals $c_{ra}$ per site $r=(x,y)$ and Hamiltonian $H = \sum_k c_k^\dagger h_k c_k$ ($c_{ka} = L^{-1} \sum_r e^{ikr} c_{ra}$) where
\beq
h_k = t_{AB} \( \sin(k_x) X + \sin(k_y) Y\) + (m + t_{AB} \(  \cos{k_x} +\cos{k_y} \) ) Z .
\eeq
If $0 < m < 2 t_{AB}$
then
at half-filling this system is a gapped Chern insulator with Chern number $C=1$.

Introduce a $Q=(\pi,\pi)$ perturbation which doubles the unit cell.  The perturbation has the form $\Delta_1 H = \sum_r V (-1)^{x+y} c_r^\dagger c_r$.  The resulting $k$ space Hamiltonian is thus
\beq
\label{eq:split-chern}
\tilde{h}_k = \left(
  \begin{array}{cc}
    h_k & V \\
    V & h_{k+Q} \\
  \end{array}
\right).
\eeq
We then obtain the band structure for all $V$ and find that up to $V=1$ the bulk gap remains open. At $V=.8$ the two filled bands below the gap are themselves separated by a gap with the band nearer the chemical potential carrying $C=1$ and with the other band carrying $C=0$. Having achieved a non-trivial separation of the original Chern band into a non-trivial band and a trivial band, we may now add additional perturbations to manipulate the trivial band. In fact, we may deform the trivial $C=0$ band into a perfectly flat perfectly localized band which forms an independent trivial insulator supported on $L^2/2$ sites.

The specific perturbations which accomplish this are
\beq \Delta_2 H = \half \sum_{r \in A} t_{AA} \(  c_r^\dagger ( Z - i X ) c_{r+x-y}
+ c^\dagger_r (Z - i Y) c_{r+ x+y}  + h.c. \)
+ \sum_{r \in B} u c_r^\dagger Z c_r
\eeq
where $A/B$ refer to the two sites of the enlarged (by the $V$ term) unit cell.
In momentum space, this is
$ \Delta_2 H =  \sum_{k}  c_k^\dagger \Delta_2 \tilde h_k c_k  $
with
$$ \Delta_2 \tilde h_k =  \begin{pmatrix}
t_{AA} \( 2 \cos(k_x) \cos(k_y) Z + \sin(k_x - k_y) X + \sin(k_x + k_y ) Y \) & 0 \\
0 & u  Z
\end{pmatrix}
$$
The $t_{AA}$ term is a hopping term within the A-sublattice
of the same form as the original hopping;
to preserve the gap, we must turn $t_{AA}$ on as we turn off the hopping between sublattices $t_{AB}$.
The $u$ term freezes the spins of the B-site particles.
A specific protocol for varying parameters between
\eqref{eq:split-chern}
and a Hamiltonian where the B-sites are decoupled and host a completely trivial insulator,
without closing any gaps,
is given in Table \ref{table:protocol}.

\begin{table}[h]
\begin{center}
\begin{tabular}{|c|c|c|c|c|}
\hline
$t_{AB}$ & $-V$ & $u$ & $-t_{AA}$ & description of step \cr
 \hline
 \hline
 \hline 1 & 0 & 0 & 0 & Original bandstructure.
\\ \hline 1 &  .8 &  0&  0  &  Turn up $(\pi, \pi)$ potential  $V$, double unit cell.
\\ \hline 1 & .8  &  1 & 0  &  Turn on B-site field $u$.
\\ \hline 1 & .8  &  1 &  .4  &   Turn on AA hopping. The sign is important.
\\ \hline  1 &  1.3 &  1&  .4  & Turn up $(\pi,\pi)$ potential.
\\ \hline   .5 &  1.3 &  1&  .4 & Turn down AB hopping.
\\ \hline   .5 &  1.3 &  1&  .6 & Turn  up AA hopping.
\\ \hline   .3 &  1.3 &  1&  .6 & Turn down AB hopping.
\\ \hline   .3 &  1 &  1&  .6 & Turn down $(\pi, \pi)$ potential.
\\ \hline   .3 &  1 &  1&  .8 &  Turn  up AA hopping.
\\ \hline   0 &  1 &  1&  .8 &  Turn {\it off} AB hopping.
\\ \hline   0 &  1 &  1&  1 &  Turn  up AB hopping.
\\ \hline   0 &  1 &  5&  1 &  Crank up B-site field with impunity.
\\ \hline 0 & 0 & 5 & 1  & Turn off $V$ $\to$
$\begin{matrix}\text{B-site bands } \\
\text{become flat,} \\ \text{decouple.}\end{matrix}$
$\begin{matrix}\text{A-sites have } \\
\text{original bandstructure,}\\
\text{rotated by $ \pi/4$.}\end{matrix}$
\cr\hline
\end{tabular}
\caption{\label{table:protocol}  The details of the Chern-band-folding protocol.
Parameters are in units of $m$.
We checked that the gap stays open along a
linear interpolation between each of these these checkpoints,
and (therefore) that the Chern numbers of the four bands
remain $(0, -1, 1, 0) $ from top to bottom.
The tricky part is gradually turning off $t_{AB}$ while
turning on $t_{AA}$.
A movie of the resulting band-folding is available
upon request.}
\end{center}
\label{default}
\end{table}%

This construction can be performed two times to go from $L^2$ to $L^2/2$ to $L^2/4$ sites supporting the Chern insulator. Furthermore, since all manipulations preserved the bulk gap, the quasi-local generator $K(\eta)$ defined above generates a quasi-local unitary that implements the coarse-graining. Hence such Chern insulators are $s=1$ fixed points. This also implies that they obey an area law.

Given a quasi-local evolution generated by $K$ implementing an $s$ source RG transformation, the entropy $S(2R)$ of a region of linear size $2R$ in the new $2L$ linear size system obeys
\beq \label{eq:entropy_bound}
S(2R) \leq s S(R) + k R^{d-1}
\eeq
where $S(R)$ is the entropy of the same region type at size $R$ in the linear size $L$ system and $k$ is a number dependent on the details of $K$.  With $s=1$ the bound \eqref{eq:entropy_bound} is easily iterated to obtain
\beq
S(2^{\log(R)}) \leq \sum_{m=1}^{\log(R)} k \left(\frac{2^{\log(R)}}{2^{\log(R)-m}}\right)^{d-1} \leq k' (2^{\log(R)})^{d-1} = k' R^{d-1}.
\eeq
Hence the entropy is consistent with the area law.

One can also extend the argument to phases with chiral edge states and anyon excitations. In this context it is useful to note that discrete gauge theories in $d=2$ have exact MERA representations and hence are $s=1$ fixed points \cite{2008PhRvL.100g0404A,2009PhRvB..79h5118G,2009PhRvB..79s5123K}, so there is no obstruction to bringing anyons into the picture. Using a similar gauge theory picture, we can exhibit wavefunctions for fractional quantum Hall states by projecting copies of free fermion chiral states onto a gauge invariant subspace \cite{PhysRevB.60.8827}.  Adiabatically deforming the state of the partons from size $L$ to size $2L$ produces a short-ranged quantum Hall state which adiabatically deforms as well, and it is quite plausible that such a state is the ground state of a local Hamiltonian. The analysis of discrete gauge theories can also be extended to higher dimensions to exhibit exact MERAs for a variety of $p$-form gauge theories.

Before ending this subsection, we give one example with chiral edge states and topological order analogous to the model in \cite{2006AnPhy.321....2K}. Consider spinless fermions $f_r$ hopping on some two dimensional lattice with mean-field-like pairing Hamiltonian
\beq
H_f = \sum_{rr'} w_{rr'} f_r^\dagger f_{r'} + \sum_{rr'} \Delta_{rr'} f_r f_{r'}.
\eeq
The couplings in $H_f$ are chosen so that the ground state of $H_f$ is a $\nu=1$ $p+ip$ superconductor. Now introduce spins $\sigma^z_{rr'}$ living on the links, and for every term in the mean-field fermion Hamiltonian choose a path $\gamma_{rr'}$ connecting $r$ and $r'$. Defining $W[\gamma_{rr'}] = \prod_{\ell \in \gamma_{rr'}} \sigma^z_\ell$, we form the Hamiltonian
\bea
&& H_{f+\IZ_2} = \sum_{rr'} w_{rr'} f_r^\dagger W[\gamma_{rr'}] f_{r'} + \sum_{rr'} \Delta_{rr'} f_r W[\gamma_{rr'}] f_{r'} \cr \nonumber \\
&& - K \sum_p \prod_{\ell\in p} \sigma^z_\ell - U \sum_r (-1)^{f^\dagger_r f_r} \prod_{\ell|r\in \ell} \sigma^x_\ell.
\eea
The Hamiltonian $H_{f+\IZ_2}$ describes the $f$ fermions coupled to a $\IZ_2$ gauge theory in the tensionless limit.

A $\pi$-flux defect where $\prod_{\ell\in p} \sigma^z_\ell = -1$ supports a single Majorana zero mode and this system has Ising topological order \cite{1991NuPhB.360..362M}. Furthermore, because the Hamiltonian $H_{f+\IZ_2}$ is solvable we can exhibit a quasi-local unitary mapping size $L$ to size $2L$. We use a combination of the free fermion unitary which implements the mapping for the $f$s and the $\IZ_2$ circuit which implements the mapping for the spin degrees of freedom to produce a mapping for the total system. Thus Ising topological order is (as expected) an $s=1$ fixed point.

\subsection{Layer construction}

A class of examples of $s$ source RG fixed points for $s > 1$ is provided by the layer construction. Consider an $s_0$ source RG fixed point in $d_0$ dimensions, the ``layer", and stack $L^{d-d_0}$ copies of these layers, which are size $L^{d_0}$ objects, to form a torus of $L^d$ sites. We may also add local perturbations and couplings between the layers provided the individual layers remain incoherent with each other. This layered system is an $s = s_0 2^{d-d_0}$ RG fixed point in $d$ dimensions.  By the restriction lemma $s_0 \leq 2^{d_0-1}$, so the layered system also obeys the restriction lemma.

As a concrete example, consider a $d=3$ system composed of $L$ layers of integer quantum Hall states. Such a system, when cut open along a boundary piercing through the layers, supports $L$ chiral edge states that cannot be gapped. Furthermore, since no individual integer quantum Hall state can be produced from a product state using a quasi-local unitary, it follows that we need $s=2$ copies of the $L$ layer system to make a $2L$ layer system using a quasi-local unitary.

Note that in the context of the layer construction, some cancellation may arise. For example, it may be the case that multiple layers of a lower dimensional state can be deformed into a product state even if a single layer cannot (as with the notion of inverse states in \S\ref{subsec:wal}). In such a case, the effective value of $s$ will be reduced. In other words, we do not require the full $s$ copies since multiple layers can be produced from product states.

There is one interesting line of thought suggested by the layer construction. Observe that as $s$ increases we come closer to violating the area law. However, in the layer construction having large $s$ requires stacking low dimensional objects.  This intuition is precisely the same as for Fermi liquid entanglement \cite{PhysRevLett.105.050502}. Following for a moment this gapless line of thought, ordinary CFTs, being like $s=1$ fixed points, are hopelessly far from defeating the area law. We can do better by bundling lower dimensional gapless systems, and when we bundle gapless one dimensional systems we finally manage to violate the area law.

To obtain a gapped state, we want to stack lower dimensional topological objects.  If we could stack one dimensional topological objects with $s=1$, then we would obtain a $d$ dimensional topological system which violated the area law. However, such one dimensional $s=1$ gapped states do not exist (see \S\ref{sec:arealaw}). Of course, the layered construction is an amusing toy, but it is too trivial to cover the interesting examples (like Haah's code).  We speculate that some local generalization of the layer construction, a ``bundle" of layers, similar to the idea of a Fermi surface's worth of $1+1$ CFTs, would provide a more robust framework in which to understand the area law.

\section{Topological quantum liquids}
\label{sec:tql}

In this section we make good on our promise to define ``conventional" gapped phases.  We call our proposal ``topological quantum liquids" since they have the have the ability to ``flow" and take the local ``shape" of the system. We prove that all topological quantum liquids obey the area law and have $s \leq 1$.  We conjecture that all systems with $s\leq 1$ are topological quantum liquids.  Our definition of a topological quantum liquid is strong, so proving that $s\leq 1$ implies liquidity requires some work establishing local deformability from the global ability to map $L$ to $2L$.

A topological quantum liquid is, informally, a gapped (topological) quantum phase of matter which is insensitive to the local details of the system (liquid).  Continuum field theories with a mass gap, by their very definition, are topological quantum liquids.  This is because in order for a continuum limit to exist, the microscopic details of the space must be irrelevant.  A reason for singling out topological quantum liquids is that they represent, almost exclusively, the type of gapped states encountered in Nature so far.  Indeed, all experimental realizations of gapped phases are, to the best of the authors' knowledge, topological quantum liquids, or layers thereof.  These realizations include most prominently all integer and fractional quantum Hall states.

An example of a gapped state which is not a topological quantum liquid is Haah's code. This interesting Hamiltonian has the property that the ground state space manifold depends sensitively on the precise number of sites in the lattice.  This is not to say that Haah's code is uninteresting, only that it is not liquid-like.  Indeed, it displays features much more reminiscent of a glass. Following \cite{2012PMag...92..304C}, we might call such a phase a \textit{topological quantum glass}. We do want to imply that the dichotomy between topological quantum liquids and topological quantum glasses is exhaustive. At the very least, the layer construction demonstrates that we may have layers of topological quantum liquids which do not form a higher dimensional topological quantum liquid and have crystalline as opposed to glassy features.

The intuitive properties of topological quantum liquids include a ground state manifold that depends only on global features of the system as well as the ability to relax into thermal equilibrium on a reasonable timescale.  We may formalize these criteria by saying that a topological quantum liquid has the property that the shape of the underlying geometry may be changed without closing the gap.  We define a topological quantum liquid as a gapped phase of matter with the property that any ground state on a manifold $M$ may be deformed into a ground state on a manifold $M'$ without closing the gap provided there is a homeomorphism from $M$ to $M'$. As a technical point, $M$ and $M'$ should also support Riemannian metrics such that the deformation from $M$ to $M'$ is slowly varying compared to the correlation length $\xi$.

In the discrete setting, we demand that for any two graphs $\hat{M}$ and $\hat{M}'$ which differ only locally, there exists a gapped Hamiltonian path mapping ground states of the TQL on $\hat{M}$ to ground states of the same TQL on $\hat{M}'$. This gapped Hamiltonian path may be defined on a third graph $\hat{M}''$ having the property that both $\hat{M}$ and $\hat{M}'$ may be obtained from $\hat{M}''$ by locally deleting or identifying edges and vertices. Equivalently, we may imagine that both $\hat{M}$ and $\hat{M}'$ form locally equivalent triangulations of some manifold $M$.

To give a few examples, in the continuous context the two manifolds $M$ and $M'$ could have different sizes.  In the discrete setting, $\hat{M}$ could be a torus with $L^d$ sites while $\hat{M}'$ could be a torus with $(L+1)L^{d-1}$ sites, i.e.~having one extra layer of sites.  Topological quantum liquids have the property that any ground state on one such manifold or graph can be deformed into a ground state on the other manifold or graph using a quasi-local unitary without closing the gap.

\begin{thm}[TQL structure theorem]
All topological quantum liquids have $s \leq 1$.
\end{thm}
Proof: (trivial) Let $\hat{M}$ be an isotropic $d$-torus of length $L$ and let $\hat{M}'$ be an isotropic $d$-torus of length $2L$.  Then $\hat{M}$ and $\hat{M}'$ are locally equivalent.  Indeed, we may take $\hat{M}'' = \hat{M}'$ so that $\hat{M}$ is obtained by identifying every $2^d$ sites into one site.  By assumption there exists a gapped path connecting ground states on $\hat{M}$ to ground states on $\hat{M}'$.  This is precisely the definition of an $s=1$ fixed point.  $s=0$ fixed points are also allowed.

As a trivial consequence of the structure theorem, all topological quantum liquids obey the area law and have system-size-independent ground state degeneracy.  There is one subtlety, however.  System size independent ground state degeneracy is not, by itself, enough to guarantee $s \leq 1$.  Indeed, layers of $s=1$, $G=1$ states are not topological quantum liquids (for example, we cannot in general add layers with a quasi-local unitary), but they continue to have $G=1$.  However, it does seem that system-size-independent $G$ plus some measure of isotropy is often sufficient to give a topological quantum liquid.

This section may be summarized with the following brief statement.  If size $L$ and size $2L$ are in the same phase (meaning connected by a quasi-local unitary) and if $d>1$, then the phase obeys the area law.

\subsection{Local stability for $s=1$ fixed points}

Here we sketch an argument that gapped $s=1$ fixed points are stable to local deformations of the space. This is of course plausible since the ground state degeneracy is independent of system size. It is obvious for $s=0$ fixed points. Local gapped quantum field theories also boast this kind of local stability; this follows because local changes in the geometry couple to a local operator, the stress tensor, which is short-range correlated (and in fact identically zero in the topological limit). The motivation for this sketch is simple: since the definition of a topological quantum liquid is naively quite strong, it is helpful to show that local deformability follows from simpler assumptions.

What we are after is a spatially varying Hamiltonian $H_{\text{interpolate}}$ which, given a local region $A$, interpolates between $H_L$ far away from $A$ and $H_{2L}$ deep inside $A$. Let $H(\eta)$ be a gapped path between $H_L$ at $\eta=0$ and $H_{2L}$ at $\eta=1$. Decompose $H(\eta)$ into local terms as
\beq
H(\eta)= \sum_x H_x(\eta),
\eeq
and then construct $H_{\text{interpolate}}$ as
\beq
H_{\text{interpolate}} =\sum_x H_x (\eta_x),
\eeq
where $\eta_x$ is a slowly varying function that asymptotes to zero far from $A$ and to one deep inside $A$. Since both size $L$ and size $2L$ are in the same phase, since the phase is stable, and since the perturbation is slowly varying, it must be that $H_{\text{interpolate}}$ is also gapped. We will give a more detailed argument for this conclusion in \S\ref{sec:arealaw}.

Combined with the ability to rearrange local regions without closing the gap (see the discussion of [Micro-insensitivity] in \S\ref{subsec:wal}), the ability to insert and remove degrees of freedom strongly suggests that the phase possesses local deformability. Indeed, given any two homogeneous Hamiltonians $H_1$ and $H_2$ connected by an adiabatic path $H(\eta)$, we should be able to construct a gapped interpolating Hamiltonian which asymptotes to $H_1$ or $H_2$ in different regions. Thus adiabatic deformability in time implies adiabatic deformability in space (but not vice versa:
there are states with topological order but no gapless edge modes)\cite{Kitaev-2011}, and the phase appears to be locally stable. Hence our original definition of a topological quantum liquid is apparently essentially equivalent to having $s\leq 1$.

\section{Generalization of the $s$ source framework}
\label{sec:gen-s-source}

In this section we generalize the $s$ source framework to effectively allow fractional $s$. The data defining a generalized $s$ source fixed point are as follows. We have a label set $\Lambda$ whose elements label distinct gapped phases which transform into each other under the RG. We also have an RG rule which specifies that a type $i$ phase can be obtained from $s_{i1}$ copies of a type $1$ phase plus $s_{i2}$ copies of a type $2$ phase plus and so on. As a technical assumption, we assume that the total number of types of phases over all scales involved in producing a given phase at a given scale is bounded by a system-size-independent constant. We also assume that the quasi-local unitary at each RG step always adds some entropy (so the entropy recursion relation determines the asymptotic entropy instead of simply bounding it). We believe these assumption can be relaxed, but they make the arguments much simpler, so we leave their relaxation for future work.

\begin{DEF}[Generalized $s$ source RG fixed point]
A $d$ dimensional generalized $s$ source RG fixed point is a phase, denoted $i$, with the property that a ground state on $(2L)^d$ sites can be constructed from a set of ground states on $L^d$ sites using a quasi-local unitary $U$.  We write $|\psi_i(2L)\rangle = U \left(\prod_j |\psi_j(L)\rangle^{s_{ij}}\right)$.  Unless otherwise noted, we assume that $s$ represents the smallest set of states for which the construction is possible.
\end{DEF}

What follows are generalized entropy and ground state degeneracy lemmas.

\begin{lem}[Generalized Entropy Lemma]
The entanglement entropy $S_i$ of a type $i$ phase obeys $S_i(2R) \leq \sum_j s_{ij} S_j(R) + k R^{d-1}$ where $k$ depends on the details of the quasi-local unitary.
\end{lem}

\begin{lem}[Generalized Ground State Degeneracy Lemma]
The ground state degeneracy $G_i(L)$ of a type $i$ phase on a $d$-torus of linear size $L$ obeys the recursion relation $G_i(2L) = \prod_j G_j(L)^{s_{ij}}$.
\end{lem}

\begin{lem}[Generalized Restriction Lemma]
For each type $i$, we must have $\sum_j s_{ij} \leq 2^{d-1}$.  In particular, only a finite number of the $s_{ij}$ can be non-zero even if the index set is infinite.
\end{lem}
Proof: The argument is identical to the case of the single type theory.  Roughly speaking, if we are really using more than $2^{d-1}$ copies, then the entropy must violate the area law worse than logarithmically.  This contradicts the bound from thermodynamics in \S\ref{sec:thermo}.

\subsection{Entropy scaling in generalized $s$ source fixed points}

Let us characterize the set of phases that can violate the area law.  Let $Y_0$ be the set of phases obeying the area law and let $Y_{\log}$ be the set of phases violating the area law logarithmically.  We can imagine other types of violation, weaker than logarithmic, which could also arise in the generalized $s$ source framework.  Let $Y_f$ denote the set of phases which violate the area law like $S(R) \sim R^{d-1} f(R)$.  We must have $k_1 \leq f \leq k_2 \log(R)$.

Assume that the recursion relation in the entropy lemma is saturated (this being the worst case for the growth of $S(R)$).  Then the entropy at size $R=2^{\log(R)}$ scales like
\beq
S(2^{\log(R)}) \sim \sum_{\ell=0}^{\log(R)} \left(\frac{R}{2^\ell}\right)^{d-1} s^\ell k
\eeq
where $s$ is a matrix and $k$ is some vector of entropies added by the local unitary.  The fastest growth of entropy occurs if
\beq
s^\ell = \lambda^\ell \hat{s}_\lambda + ...
\eeq
or if this is obeyed after taking some number of RG steps as one.  Then the entropy is given by
\beq
S(R) \sim R^{d-1} \hat{s}_\lambda k \sum_\ell \left(\frac{\lambda}{2^{d-1}}\right)^\ell \sim \begin{cases} R^{d-1}, & \lambda < 2^{d-1} \\ R^{d-1}\log(R), & \lambda = 2^{d-1} \\
R^{d-1+\alpha},  & \lambda = 2^{d-1+\alpha} \end{cases}.
\eeq
So long as $\lambda < 2^{d-1}$ the area law is obeyed.

Next we analyze the ground state degeneracy.  Taking logarithms of the terms in the ground state degeneracy lemma gives us
\beq
\log(G_i(2L)) = \sum_j s_{ij} \log(G_j(L)).
\eeq
Thus the logarithm of the ground state degeneracy obeys a very similar recursion relation to the entanglement entropy.  With the same assumptions on $s$, we find
\beq
\log(G_i(L)) \sim s^{\log(L)} \log(G(2))
\eeq
where $\log{G(2)}$ denotes the ground state degeneracy of all types on some fixed small system size in the ideal limit of no ground state mixing.

Two cases are relevant.  If $\hat{s}_\lambda \log(G(2)) \neq 0$, that is if some state with non-trivial ground state degeneracy is participating in the asymptotics controlled by $\hat{s}_\lambda$ and $\lambda$, then the ground state degeneracy grows like $\lambda^{\log(R)}$.  Thus if $\lambda = 2^{d-1}$, so that the area law is violated, then the number of ground states must also grow like $\log{G} \sim L^{d-1}$ as claimed.

If no phases with non-trivial ground state degeneracy participates in the asymptotics, then at large scales all source terms have no ground state degeneracy and hence obey the area law (by the weak area law).  The generalized entropy lemma with all sources obeying the area law is then only consistent with an area law for the state at larger scales.

\subsection{Example: Haah's code}
\label{sec:haah}

Within the layered construction we can construct various examples which make use of the generalized $s$ source framework.  As a non-trivial example, Haah has shown that his code is a generalized $s$ source RG fixed with $\Lambda = \{1,2\}$ and source rules $s_{11} = 1$, $s_{12} = 1$, $s_{21} = 0$, $s_{22} = 2$ \cite{2011PhRvA..83d2330H,2014PhRvB..89g5119H}.

A simple calculation then gives
\beq
s^\ell = \left(
           \begin{array}{cc}
             1 & 2^\ell-1 \\
             0 & 2^\ell \\
           \end{array}
         \right).
\eeq
$s^\ell$ grows at large $\ell$ like $\lambda^\ell \hat{s}_\lambda$ with $\lambda=2$ and
\beq
\hat{s}_\lambda = \left(
           \begin{array}{cc}
             0 & 1 \\
             0 & 1 \\
           \end{array}
         \right).
\eeq
Hence both phases have roughly $\log(G) \sim 2^\ell = L$ on a size $L = 2^\ell$ system.

Since Haah's code is a stabilizer code with locally indistinguishable ground states, the entanglement entropy in any ground state can be computed exactly, see e.g., the discussion in \cite{2005PhRvA..71b2315H} for the toric code. The general formula for the entropy is $S(A) = \text{qubits in A} - \text{stabilizers in A}$. Haah's code is defined on a cubic lattice with two qubits per site and two stabilizers per cube \cite{2011PhRvA..83d2330H}. In a cube of $R^3$ sites there are $2 R^3$ qubits and $2 (R-1)^3$ stabilizers, so the entanglement entropy of the cube is $S(R) = 2 R^3 - 2(R-1)^3 = 6 R^2 - 6 R + 2$. This formula obeys the area law but has a peculiar subleading term proportional to $R$.

By contrast, the entanglement entropy of $\IZ_2$ gauge theory in $d=3$, which is also a stabilizer code, has no such term. $\IZ_2$ gauge theory in $d=3$ dimensions can be defined on a cubic lattice with one qubit per link and stabilizers for each vertex (Gauss' law) and face of the lattice (flux constraint). Given a cube with $R$ links on a side, the number of qubits is $3 R(R+1)^2$, the number of vertex stabilizers is $(R-1)^3$, and the number of plaquette stabilizers is $3 R^3 + 3 R^2 - R^3$ (the last subtraction accounts for the fact that only $5$ of the $6$ plaquette stabilizers for each elementary cube are independent). Hence the entanglement entropy is $S(R) = 6 R^2 + 1$. To understand this formula, note that the number of surface sites is $6 R^2 + 2$, so the entropy is the number of surface sites minus one. In gauge theory language, each surface site gives one bit of freedom (electric flux or no electric flux entering the site from outside) and there is one overall constraint of total $\IZ_2$ charge neutrality.

\section{Towards a general area law}
\label{sec:arealaw}

In this section we discuss the structure of states that could, within our RG framework, violate the area law, and we give a physical argument that such states do not exist. The tools developed here, based on reconstructing states from local data, also provide independent arguments for the weak area law and for the stronger claim that phases with ground state degeneracy scaling slower than $\log(G)\sim L^{d-1}$ obey the area law. Thus we provide an independent check on the results from the $s$ source RG framework.

For convenience in this section we will typically assume that the phases in question have no protected edge states. This assumption entails no loss of generality as regards the area law.  Intuitively, this is because edge states can only contribute area law entropy. Alternatively, the general existence of an edge inverse (\S\ref{subsec:wal}) implies that for every phase which has protected edge states and violates the area law, there is another phase which violates the area law and has no protected states. Hence ruling out area law violations in all phases with no protected edge states rules out area law violations in phases with protected edge states. Except where explicitly stated otherwise, we assume the ``accidental" edge states which arise in the constructions below can be removed with local perturbations.

Our goal is to establish a bound of the form $S(\rho_A) \leq \mathcal{O}(|\partial A|) + \log(G)$ for an appropriate ground state degeneracy $G$\footnote{Assuming the bound is saturated, as is plausible, states with unusual ground state degeneracy will have unusual terms in their entanglement entropy, e.g., a linear in $L$ term in $d=3$ for $G \sim e^{cL}$.}. We do this by constructing a local Gibbs state (exponential of a local ``effective Hamiltonian") which is locally consistent with the state $\rho_A$ and which upper bounds the entropy of $\rho_A$. The quickest route through the argument is to jump to the main argument in \S\ref{subsec:mainarealaw} referring back to the preliminaries in \S\ref{subsec:prelim} as needed.

Although the arguments in this section are powerful by themselves, we still need the $s$ source framework to argue for the general area law. Furthermore, the $s$-source RG also provides a powerful method to argue for the existence of frustration free Hamiltonians (see the MERA discussion in \S\ref{sec:mera}). Frustration free Hamiltonians are an important special case in our analysis, and we expect such Hamiltonians to exist on general RG grounds provided we are deep within the phase where spatial correlations are minimized. We also use ideas from \S\ref{sec:thermo}, in particular the free energy estimate \eqref{eq:Fgap}, in the arguments below.

\subsection{Preliminaries}
\label{subsec:prelim}

Several tools are needed to proceed with the arguments. First we discuss reconstruction of global states from local data. Then we discuss the idea of a local gap and the stability of spatially varying local Hamiltonians. Finally, we describe the idea of a diverging local gap.

\textbf{Local reconstruction}

It is useful to consider trying to reconstruct the ground states from local data (for important early work in this direction see \cite{petz1986,2004CMaPh.246..359H,2010NatCo...1E.149C,2011PhRvL.106h0403P}). This reconstruction is more feasible than one might at first imagine. For example, given access to the states of all local $d$-disks of sufficiently large (but still microscopic) size, \cite{2010NatCo...1E.149C,2014arXiv1407.2658S} have shown that the maximum entropy global state approximately consistent with this local data is close to the ground state projector. In other words, one can reconstruct global ground states from local data (even in topological phases). Here we consider a variant of this situation: the problem of reconstructing the state of a subsystem $A$ of size $R$ from local data.

Suppose we have a set of local operators $\{O_i\}$ supported in a region $A$. We want to find the maximum entropy state which gives expectation values for the $O_i$ that agree with expectations taken in the true state, $\rho_A$, of $A$. In other words, among all possible states $\sigma_A$ such that $\tr(\sigma_A O_i) = \tr(\rho_A O_i)$ for all $i$, we want the state that maximizes the entropy $S(\sigma_A)$. This problem has a known solution. Construct the variational function $f(\sigma,\{\lambda_i\})$ given by
\beq
f(\sigma,\{\lambda_i \}) = S(\sigma) + \sum_i \lambda_i (\tr(\sigma O_i) - \tr(\rho_A O_i)) + \lambda (\tr(\sigma)-1).
\eeq
Then maximize $f$ with respect to $\sigma$ and the $\lambda$s. The resulting maximum entropy state has the form
\beq
\sigma^\star = \frac{\exp\(-\sum_i \lambda_i^\star O_i\)}{Z},
\eeq
and hence is a local Gibbs state.

Suppose the operators $O_i$ form a complete set of observables for a set of small regions $\{A_j\}$ such that $\cup_j A_j = A$. Then we say $\sigma^\star$ is a maximal entropy reconstruction of $\rho_A$ from local data. Denoting the disk of radius $r$ centered at $x_0$ by $D(r,x_0)$, a typical choice for the $A_j$ might be all regions of the form $A \cap D(r,x_0)$ for all $x_0$ and some fixed small $r$. The linear size, $R$, of $A$ will be much larger than $r$ in our constructions.

Why is this formalism useful? Note that $\rho_A$ is locally consistent with itself, so it is a candidate for the maximum entropy state and
\beq
S(\rho_A) \leq S(\sigma^\star).
\eeq
Furthermore, $\sigma^\star$ is by construction a local Gibbs state, so it is easier to manipulate than $\rho_A$.

\textbf{Localized excitations}

Let $\sigma_A$ be the maximum entropy state consistent with local data on patches of linear size $r \ll R$. Write
\beq
\sigma_A = e^{-\tilde{H}_A}
\eeq
and define $H_A$ to be the Hamiltonian $H$ restricted to terms having support just in $A$. We have just shown that $\tilde{H}_A$ is a sum of local operators supported on patches of linear size $r$. But what does $\tilde{H}_A$ look like? It turns out that $\sigma_A$ is close to being a ground state of $H_A$ as we now show.

The Hamiltonian of the whole system is $H = \sum_x H_x$ where, without loss of generality, we assume each term $H_x \geq 0$. Let the ground state energy of $H_A$ be $E_{g,A}$ and let the ground state projector be $P_{g,A}$. To control the energy of $\sigma_A$ we first bound the expectation value of $H_A$ in the state $\rho_A$ as follows \cite{2009PhRvA..80e2104M}.

Separate the Hamiltonian into three pieces, and $H = H_A + H_{\bar{A}} + H_{\partial A}$, where the terms act within $A$, $\bar{A}$, and at the boundary of $A$ respectively.  We use the positivity of $H_{\partial A}$ to bound $\langle H_A + H_{\bar{A}}\rangle$,
\beq
\langle H_A + H_{\bar{A}} \rangle_g \leq \langle H_A + H_{\bar{A}} + H_{\partial A}  \rangle_g = E_g,
\eeq
where we have taken expectation values in a global ground state $|g\rangle$. Now we bound the ground state energy using the variational principle and the trial state $\rho_{\text{factor}} = \frac{P_{g,A}}{\tr(P_{g,A})} \tr_A(|g\rangle \langle g|)$. We obtain
\beq
\langle H_A + H_{\bar{A}} + H_{\partial A}  \rangle_g \leq \tr(\rho_{\text{factor}}(H_A + H_{\bar{A}} + H_{\partial A} )) \leq E_{g,A} + \langle H_{\bar{A}} \rangle_g + \max_{x \in \partial A}(\| H_x \|) |\partial A|.
\eeq
Combined with the first inequality we have for $\langle H_A \rangle_g = \tr(\rho_A H_A) = \langle H_A \rangle_{\rho_A}$ the result
\beq
E_{g,A} \leq \langle H_A\rangle_{\rho_A} \leq E_{g,A} + J_\partial |\partial A|
\eeq
where $J_\partial = \max_{x \in \partial A}(\| H_x \|)$.

Because the correct expectation values of the terms in $H_A$ are included in the local data defining $\sigma_A$ (assuming $r$ is bigger than the range of the terms $H_x$), we also have the bound
\beq
E_{g,A} \leq \langle H_A\rangle_{\sigma_A} \leq E_{g,A} + J_\partial |\partial A|.
\eeq
Since $\tilde{H}_A$ a sum of local operators and since the average excitation energy of $\sigma_A$ is non-extensive, it must be that the entropy coming from excitations is non-extensive and scales like the average excitation energy.

For example, if we restrict the local data defining $\sigma_A$ to just the terms $H_x$ contained in $H_A$, then $\tilde{H}_A$ has the form
\beq
\tilde{H}_A = \sum_x g_x H_x
\eeq
where we have renamed the Lagrange multipliers $\lambda_i \rightarrow g_x$. The local effective temperature, $1/g_x$, must go to zero away from $\partial A$ in order for the excitation energy of $H_A$ to be proportional to $|\partial A|$. Equivalently, $\sigma_A$ reproduces ground state correlations of $H_A$ away from $\partial A$, so the local temperature must go to zero or equivalently the local gap must diverge away from $\partial A$.

\textbf{Local gap and local thermodynamics}

To justify the notion of a local gap, we first appeal to the stability of the phase. It is trivially true that $H$ and $g H$ give the same ground states for all $g > 0$.  Now consider the Hamiltonian $H[g] = \sum_x g_x H_x$.  We expect that the stability of the phase implies that the couplings in $H$ can be modulated slowly in space without closing the gap.  Suppose the variation in $g_x$ is bounded and small. Then since $H[g]-H[1]$ is a sum of bounded local operators, the gap must be preserved if the perturbation is small enough since the phase is stable.
To build up larger changes in $g_x$ we can consider as a basic building block bump configurations of $g_x$. These bump configurations, $g_x = g_0 + (g_1 - g_0)\chi_x(A)$, are smooth functions $g_x$ which approach $g_0$ outside region $A$ and $g_1$ inside region $A$.  If the region $A$ is sufficiently large, then inside $A$ the Hamiltonian is effectively indistinguishable from $g_1 H[1]$. Since $H[1]$ and $g_1 H[1]$ trivially have the same ground state, the difference in the ground states of $H[1]$ and $H[g_0 + (g_1-g_0)\chi_x(A)]$ is actually localized near $\partial A$.

Finally, since the ground state properties deep inside $A$ are indistinguishable from those far outside of $A$, the stability of the uniform Hamiltonian implies that we may make further changes deep inside $A$. Hence we may repeat the argument by adding another bump function localized deep inside $A$ which further increases $g_x$. To argue for stability to arbitrary smooth variations $g_x$, we first approximate $g_x$ by collection of bump functions, then we use the stability of the bump function Hamiltonian itself to smooth out the bump functions and produce $g_x$. This is possible because the difference between $g_x$ and its bump function approximation is a sum of bounded local operators and hence obeys the criterion of stability for sufficiently slow variations.

To be quantitative suppose we have some disk $B(l,x_0)$ of radius $l$ centered at $x_0$ where the Hamiltonian locally looks like
\beq
H = \sum_{x\in B} g_{x_0} H_x + (\partial g)_{x_0}(x-x_0) H_x + ...,
\eeq
where $...$ includes terms outside $B$ and higher derivative corrections.  These local terms are smoothly patched together to form the entire Hamiltonian, and as long as the local gap is larger than the perturbation, the phase should be stable.  If $\Delta$ is the bulk gap when $g_x = 1$ then the local gap is roughly $g(x_0)\Delta$, and the strength of the perturbation, assuming the norm of the Hamiltonian terms is of the same order as the gap, should be roughly $\xi (\partial g)_{x_0} \Delta$ where $\xi$ is the correlation length.  Hence stability only requires that $g$ vary slowly,
\beq \label{eq:slowg}
\xi (\partial g)_x \ll g_x .
\eeq
Stability for all slowly varying $g_x$ and hence the persistence of short-range correlations justifies the notion of a local gap.

\textbf{Infinite bulk gap}

We now make quantitative the idea that the local gap must diverge away from $\partial A$ sufficiently fast to bound the entropy of excitations by $|\partial A|$. This property of [Infinite Bulk Gap] implies that entropies may be bounded by the logarithm of the relevant ground state degeneracy plus a term of order $|\partial A|$.

Call $\varsigma_A$ the maximum entropy state consistent just with the expectation values of the local terms in $H_A$. This state has the form
\beq
\varsigma_A = \frac{\exp\left(- \sum_{x\in A} g_x H_x \right)}{Z_\varsigma}
\eeq
and satisfies the inequality $S(\varsigma_A) \geq S(\sigma_A)$ (because $\varsigma$ satisfies
fewer constraints than $\sigma$). Since the average excitation energy of this state is proportional to $|\partial A|$, local thermodynamics implies that the entropy of excitations is similarly bounded.

The condition for stability \eqref{eq:slowg} when interpreted as a statement about local temperatures is the condition for local thermodynamics to be valid,
\beq
\xi \frac{|\partial T|}{T} \ll 1,
\eeq
where $T=T(x)\sim 1/g_x$ is a position-dependent temperature. Given the validity of local thermodynamics, we can estimate the free energy of $\varsigma_A$ using the formula \eqref{eq:Fgap}, \beq
F = - T \log(G) + F_{\text{excited}}.
\eeq
In particular, we assumed $F_{\text{excited}}$ was extensive, so the excited state free energy \textit{per unit volume} goes like $ \sim e^{-\Delta/T}$ where $\Delta $ and $T$ are the local gap and temperature. Of course, there is an ambiguity in splitting the ratio $\Delta/T$ into a gap and a temperature; all that really matters is the ratio.

Suppose \eqref{eq:slowg} is not obeyed by $g_x$. Then $g_x$ increases at least exponentially fast away from $\partial A$ and clearly the entropy of excitations will be bounded by $|\partial A|$. Thus suppose $g_x$ does obey \eqref{eq:slowg} so that local thermodynamics is applicable. The free energy of excitations may be estimated as
\beq
F_{\text{excited}} \sim \sum_x e^{- \Delta g_x}
\eeq
where $\Delta$ is the gap when $g_x = 1$. The danger is this: it might be possible for $g_x$ to decay in such a way that the energy of excitations is bounded by $|\partial A|$ while the entropy of excitations scales less favorably with the size of $A$.

To put this danger to rest, consider the generalized free energy
\beq
\CF(T) = \CF_0 \sum_x e^{- \Delta g_x/T}
\eeq
which reduces to $F_{\text{excited}}$ when $T=1$. Conventional thermodynamics relates the energy, $E(T)$, and the entropy, $S(T)$, of excitations to $\CF$:
\beq
S(T)= - \partial_T \CF(T)
\eeq
and
\beq
E(T) = \CF(T)+T S(T).
\eeq
The entropy is thus
\beq
S(T) = \CF_0 \sum_x \frac{\Delta g_x}{T^2} e^{- \Delta g_x/T}
\eeq
while the energy is
\beq
E(T) = \CF_0 \sum_x \(1 + \frac{\Delta g_x}{T} \) e^{- \Delta g_x/T}.
\eeq
It is thus clear that since $g_x$ is increasing, the entropy of excitations cannot outgrow the energy of excitations. The entropy of excitations is bounded by $|\partial A|$ and $g_x$ must increase sufficiently rapidly away from $\partial A$ to guarantee these bounds.

As an aside, if one is uncomfortable with the idea of using local thermodynamics near zero temperature, another way to phrase the above results is in terms of the density of states. Since the magnitude of the local terms in $\sum_x g_x H_x$ is slowly increasing, it must be the density of states is also thinning relative to the density of states of $\sum_x H_x$.  With the rather mild assumption that the density of states thins relatively locally with $g_x$, the above claims about entropy again follow.

\subsection{Main argument for area law}
\label{subsec:mainarealaw}

Using the ideas just established plus the $s$ source framework, we now give our main argument for the area law. $A$ is a subregion with state $\rho_A$ inside a large gapped phase in a ground state of the global Hamiltonian $H$ .

Let $\sigma_A$ be the state of maximal entropy locally consistent with $\rho_A$. $\sigma_A$ has the form
\beq
\sigma_A = \frac{e^{-\tilde{H}_A}}{Z}.
\eeq
Since $\rho_A$ is locally consistent with itself, we have $S(\rho_A) \leq S(\sigma_A)$.

Recall that we may restrict to phases without protected edge states. $\tilde{H}_A$ is locally gapped away from $\partial A$, but $\tilde{H}_A$ may have accidental edge states. Repair these with a perturbation $V$ which is localized near $\partial A$.

Let the thermal state of the fully gapped Hamiltonian $\tilde{H}_A + V$ be
\beq
\sigma'_A = \frac{e^{-(\tilde{H}_A+V)}}{Z}.
\eeq
$\sigma'_A$ minimizes its own free energy, so we have (temporarily dropping the $A$ subscript)
\beq
\langle (\tilde{H}_A+V)\rangle_\sigma - S(\sigma) \geq \langle (\tilde{H}_A + V)\rangle_{\sigma'} - S(\sigma').
\eeq
Rearranging terms gives
\beq
S(\sigma) \leq S(\sigma') + \left[\langle (\tilde{H}_A+V)\rangle_\sigma - \langle (\tilde{H}_A + V)\rangle_{\sigma'}\right].
\eeq

The terms in $[...]$, whatever they may be, are proportional to $|\partial A|$ because $V$ is localized near $\partial A$ and $\sigma$ and $\sigma'$ give approximately the same expectation values for local terms $H_x$ in $\tilde{H}_A$ far from $\partial A$. In fact, the convergence of the local terms is exponentially fast since the system has a finite correlation length and the perturbation $V$ is localized near $\partial A$.

This gives
\beq
S(\sigma_A) \leq S(\sigma'_A) + \mathcal{O}(|\partial A|).
\eeq
$\tilde{H}_A+V$ is in the same phase as a gapped local Hamiltonian on $A$ and has a diverging local gap away from $\partial A$ by [Infinite Bulk Gap]. Thus the entropy of $\sigma'$ is bounded by the ground state degeneracy of $\tilde{H}_A+V$, which is the universal value associated with the phase on this open geometry, plus a term proportional to $|\partial A|$,
\beq \label{eq:GSbound}
S(\rho_A) \leq S(\sigma_A) \leq \log(G(\tilde{H}_A+V)) + \mathcal{O}(|\partial A|).
\eeq

If $S(\rho_A) \sim R^{d-1} f(R)$ with $f$ a growing function of $R$, then the number of ground states $G(\tilde{H}_A+V)$ must also grow faster than $e^{ c R^{d-1}}$. However, if we build up the open boundary system defined on $A$ using our RG procedure (which we can do if the system has no protected edge states), then the ground state degeneracy on $A$ will obey the same recursion relation as the torus ground state degeneracy. Violating the area law with a logarithmic correction requires $s=2^{d-1}$, but $s=2^{d-1}$ gives a ground state degeneracy growing only like $\log(G) \sim R^{d-1}$. This growth violates the lower bound on $\log(G)$ in \eqref{eq:GSbound}: we simply don't have enough ground states to account for the anomalous entropy. Hence $S(\rho_A) \leq S(\sigma_A)$ must obey the area law even with $s=2^{d-1}$. If we further assume that the entropy recursion relation is saturated, then there are no gapped phases with $s=2^{d-1}$.  Either way, we obtain a general area law for gapped phases.

Even without the $s$ source framework, the bound \eqref{eq:GSbound} implies the weak area law. Of course, the wormhole array argument in \S\ref{subsec:wal} gives more information than just the area law, but the present argument provides a useful independent check.

\subsection{Frustration free Hamiltonians}

Frustration free local Hamiltonians provide a very general setting in which the above argument can be made more rigorous. Suppose that the global ground states are frustration free ground states of the local Hamiltonian $H = \sum_x P_x$ which is assumed to be a sum of projectors $P_x$ (not necessarily commuting, the easier commuting case is discussed in Appendix~\ref{sec:commuting}). Frustration free means that every term $P_x$ independently annihilates the ground state.  The truncated Hamiltonian $H_A$ is still a sum of projectors and the state $\rho_A$ lies entirely within the ground state manifold of $H_A$.  This is because $\rho_A$ is annihilated by every projector in $H_A$, so we have $\tr(\rho_A H_A) = 0 $ which is the minimal energy of $H_A$.

The following lemma bounds the entropy of $\rho_A$ in terms of the number of ground states of $H_A$.
\begin{lem}[Frustration Free Entropy Bound]
Let $H$ be a frustration free Hamiltonian (meaning its ground states are frustration free) and let $\rho$ be a ground state of $H$, $\tr(\rho H)=0$.  Then the entropy $S(\rho)$ obeys $S(\rho) \leq \log(G(H))$ where $G(H)$ is the ground state degeneracy of $H$.
\end{lem}
Proof: (trivial) Since $\rho$ is a ground state, it cannot have more entropy than the maximum entropy ground state.  The maximum entropy ground state is the equal weight mixture of all $G(H)$ ground states and its entropy is $\log(G(H))$.

Applied to the case of $H_A$ and $\rho_A$, this lemma bounds the entropy of $\rho_A$ as desired.  In particular, if $H_A$ descends from another frustration free Hamiltonian $H'_A$ which differs from $H_A$ only at the boundary $\partial A$ and has fully gapped edge states, then one expects that deleting the boundary terms in $H'_A$ within $\ell$ of $\partial A$ can only add $e^{c |\partial A| \ell}$ additional ground states associated with the edge.  Then the entropy of $\rho_A$ would be bounded by the logarithm of the bulk ground state degeneracy of $H'_A$ (assumed to have no edge states) plus an area term.

\begin{thm}[Limited Growth of Ground State Manifold]
Let $\check{H}_A$ be a gapped, stable, and frustration free Hamiltonian written as a sum of positive operators with strictly bounded support on an open region $A$ with $G(\check{H}_A)$ locally indistinguishable zero energy ground states and let $H_A$ be obtained from $\check{H}_A$ by deleting operators within $\ell$ of $\partial A$.  Then we have $\log(G(H_A)) \leq \log(G(\check{H}_A)) + c |\partial A |\ell$.
\end{thm}

First Proof: Let $V$ denote $\check{H}_A - H_A$, i.e.~the edge terms which gap out the accidental edge states of $H_A$.  Then consider the Hamiltonian $H(\lambda) = \lambda H_A + V$ and let its thermal state at temperature $T$ be $\sigma(\lambda,T)$,
\beq
\sigma(\lambda,T) = \frac{e^{-H(\lambda)/T}}{Z(\lambda,T)}.
\eeq
Because $\sigma(\lambda,T)$ minimizes its own free energy, we have the bound
\beq
\tr\left((\lambda H_A + V) \frac{P_{g,A}}{G(H_A)}\right) - T \log(G(H_A)) \geq \tr\left((\lambda H_A + V) \sigma(\lambda,T)\right) - T S(\sigma(\lambda,T)).
\eeq
Rearranging terms, using the positivity of the Hamiltonian, using that $\tr(P_{g,A} H_A) = 0$, we find
\beq
T \log(G(H_A)) \leq \tr\left(V \frac{P_{g,A}}{G(H_A)}\right) + T S(\sigma(\lambda,T)).
\eeq

The above bound holds for all $\lambda$, so send $\lambda \rightarrow \infty$.  Since the ``bulk temperature" $T/\lambda$ is now zero, we would like to argue that $S(\sigma(\infty,T))$ is scales like $|\partial A|$.  To do this we compute the heat capacity, $C(T) = T \partial_T S(T)$, using the formula $C(T) = \partial_T E(T)$.  By definition, $E(T)$ is given by
\beq
E(T) = \lim_{\lambda \rightarrow \infty} \tr\left((\lambda H_A + V) \sigma(\lambda,T)\right).
\eeq
As $\lambda$ goes to infinity, the average of all the terms in $H_A$ are set to zero.  Indeed, every term in $H_A$ is positive definite, so if any term had a non-zero value, the energy would tend to infinity in the Boltzmann weight thus giving zero contribution.  Hence the contribution from $\lambda H_A$ must be zero as $\lambda$ goes to infinity.

This intuition may be proven by noting that, given two positive operators $P_1$ and $P_2$ with a common null space, the partition function $Z(\lambda_1, \lambda_2) = \tr\left(e^{-\lambda_1 P_1 - \lambda_2 P_2} \right)$ is a monotonically decreasing function of $\lambda_1$.  Indeed, we have $- \partial_{\lambda_1} \log(Z(\lambda_1,\lambda_2) = \langle P_1 \rangle \geq 0$.  Integrating both sides with respect to $\lambda_1$, we obtain the formula $-\log(Z(\infty,\lambda_2))+\log(Z(1,\lambda_2)) = \int_1^\infty \langle P_1 \rangle(\lambda_1) d\lambda_1$.  Since the left hand side is finite, it must be the case that $\langle P_1\rangle $ vanishes faster than $1/\lambda_1$ as $\lambda_1$ goes to infinity.  Hence $\lim_{\lambda_1 \rightarrow \infty} \lambda_1 \langle P_1 \rangle =0$.

Now since $V$ is explicitly localized near $\partial A$, it follows that $E(T)$ and $C(T)$ are bounded by $|\partial A|$.  Integrating the heat capacity to produce the entropy, we find that $S(T) - S(0)$ is also bounded by $|\partial A|$.  Hence we bound $\log(G(H_A))$:
\beq
\log(G(H_A)) \leq \mathcal{O}(|\partial A|) + \log(G(\lambda H_A + V)) =  \mathcal{O}(|\partial A|) + \log(G(\check{H}_A)).
\eeq
The second equality follows because $ \log(G(\lambda H_A + V)) = \log(G(\check{H}_A))$ since every ground state of a frustration free Hamiltonain $\sum_x H_x$ is also a ground state of $\sum_x g_x H_x$ for all $g_x > 0$ and vice versa.

Second (Restricted) Proof: We give another proof of a weakened version of the theorem. Let a Hamiltonian $H_A$ defined on an open region $A$ be called $\ell$ {\it bulk stable} if the ground state manifold is stable to all perturbations a distance greater than $\ell$ from $\partial A$.  Then if $\check{H}_A$ is stable and $H_A$ is at least bulk stable, we can again prove that $\log(G(H_A)) \leq \log(G(\check{H}_A)) + c |\partial A |\ell$.

Let $|g_A\rangle$ be a ground state of $H_A$ which is not a ground state of $\check{H}_A$ and let $|\check{g}_A\rangle$ be a ground state of $\check{H}_A$.  If $|g_A\rangle$ is locally indistinguishable from $|\check{g}_A\rangle$ then $|g_A\rangle$ is a ground state of $\check{H}_A$.  Thus $|g_A\rangle$ must be locally distinguishable from $|\check{g}_A\rangle$.  However, $|g_A\rangle $ cannot be distinguishable from $|\check{g}_A\rangle$ in the bulk because $H_A$ is bulk stable.  If $|g_A\rangle$ could be distinguished from $|\check{g}_A\rangle$ by a bulk operator then we could partially lift the degeneracy of $H_A$ by a bulk perturbation contradicting bulk stability.

Thus $|g_A\rangle$ must be distinguishable from $|\check{g}_A\rangle$ only near the boundary $\partial A$.  The number of states distinguishable from $|\check{g}_A\rangle$ only by operators within $\ell$ of $\partial A$ is bounded by $e^{c |\partial A|\ell}$ for some constant $c$.  Hence the ground state degeneracy of $H_A$ obeys $\log(G(H_A)) \leq \log(G(\check{H}_A)) + c |\partial A |\ell$ as claimed.

Two annoying features of the second (restricted) proof are the requirement of strict locality of the Hamiltonian terms and the extra assumption of bulk stability for $\check{H}_A$.  Does bulk stability not follow from the stability of $H_A$?  Indeed, it does but at the cost of relaxing strict locality.
\begin{lem}[Bulk Stability from Stability]
Let $\check{H}_A$ be a gapped, stable, and frustration free Hamiltonian written as a sum of positive operators with quasi-local support on an open region $A$ with $G(\check{H}_A)$ locally indistinguishable zero energy ground states and let $H_A$ be Hamiltonian obtained from $\check{H}_A$ by deleting up to $ c|\partial A|\ell$ operators localized near $\partial A$. Then $H_A$ can be taken to be bulk stable.
\end{lem}

Proof: Observe that every ground state of $\check{H}_A$ is a ground state of $H_A$.  Let $|g_A\rangle$ be a ground state of $H_A$ which is not a ground state of $\check{H}_A$ and assume $|g_A\rangle$ is distinguishable from a ground state $|\check{g}_A\rangle$ of $\check{H}_A$ in the bulk.  Then there exists a quasi-local bulk operator $O$ such that $O |g_A\rangle \neq 0$ but $O |\check{g}_A \rangle =0$.  Add to $\check{H}_A$ a bulk term $O^\dagger O$. The resulting bulk Hamiltonian has the same ground state manifold since $O|\check{g}_A \rangle = 0$, and the bulk Hamiltonian is of the same form as assumed in the theorem statement.  However, the resulting edge deleted Hamiltonian no longer has as a ground state the state $|\check{g}_A\rangle$ since there exist zero energy states, e.g., $|\check{g}_A\rangle$, but the energy of $|g_A\rangle$ is non-zero.  Hence we may assume that $H_A$ is bulk stable.

The technical subtlety is that while we can guarantee that a quasi-local $O$ exists, we cannot guarantee that a strictly local $O$ exists. If $O$ does have some quasi-local tail, then we must not delete $O^\dagger O$ from the bulk Hamiltonian when removing boundary terms.  We can truncate these tails with $\frac{1}{\text{poly}(R)}$ error by taking $\ell \sim \log(R)$, but this leads to weakened (and not useful for the area law) bound of the form $\log(G(H_A)) \leq \log(G(\check{H}_A)) + c |\partial A |\log(R)$.  This technical point leads us naturally to the general case.

\subsection{Reducing general gapped phases to frustration free phases?}

What about general gapped ground states $|g \rangle$ which may not be frustration free?  If it were possible to approximately reduce any gapped phase to a frustration free phase, then the logic of the previous section might be sufficient. On very general RG grounds, one expects that in the extreme long wavelength limit of a gapped phase, the ground state can be specified by local constraints. Unfortunately it is difficult to prove this intuition, although some progress is possible (see also the MERA discussion \S\ref{sec:mera}).

Suppose we have a general Hamiltonian $H = \sum_x H_x$ where each term satisfies $\langle g | H_x | g \rangle = 0$ and has bounded norm. Then we can construct new operators $\hat{H}_x$ that annihilate the ground state \cite{2006AnPhy.321....2K}.  Let the gap of $H$ be $\Delta$ and let $\tilde{f}(\omega)$ be smooth function satisfying $\tilde{f}(-\omega)=\tilde{f}^*(\omega)$, $\tilde{f}(0)=1$, and $\tilde{f}(\omega) = 0$ for $|\omega | \geq \Delta$.  The Fourier transform $f(t) = \int \frac{d\omega}{2\pi}e^{-i \omega t} \tilde{f}(\omega)$ decays faster than any power of $t$ and we can define a quasi-local $\hat{H}_x$ by
\beq
\hat{H}_x = \int dt f(t) e^{i H t} H_x e^{-i H t}.
\eeq
Each $\hat{H}_x$ then annihilates all ground states up to terms exponentially small in system size.  Indeed, since $\tilde{f}=0$ beyond the gap, the operators $\hat{H}_x$ keep us within the ground state manifold, and since the ground states are locally indistinguishable, the operators $\hat{H}_x$ don't connect different ground states.  Local operators that annihilate the ground state manifold are called local constraints.

Given local constraints, a simple local frustration free Hamiltonian with the same ground state manifold as $H$ can be defined.  Let $\hat{H}$ be
\beq
\hat{H} = \sum_x \hat{H}^2_x,
\eeq
so that every term is a positive operator and annihilates the ground states of $H$.  The issue is that $\hat{H}$ may not be gapped, although Kitaev has conjectured that a gapped Hamiltonian built from local constraints always exists.
\begin{conj}[Existence of local constraints (Kitaev \cite{2006AnPhy.321....2K})]
Every gapped phase with locally indistinguishable ground states admits a gapped Hamiltonian of the form $\hat{H} = \sum_x M_x^\dagger M_x$ where the $M_x$ are local operators that annihilate the ground state manifold.
\end{conj}
We may still allow exponentially small splittings of the ground state manifold and we have two versions of the conjecture depending on whether the constraints are assumed to be strictly local or only quasi-local.  For our purposes, Kitaev's conjecture with strict locality would certainly be sufficient to establish the required properties of [Universality] and [Infinite Bulk Gap].  Even the conjecture with quasi-locality may be sufficient, but it appears to require surmounting some technical obstacles.

\begin{wrapfigure}{R}{.5\textwidth}
  \centering
    \includegraphics[height=5cm, width=0.48\textwidth]{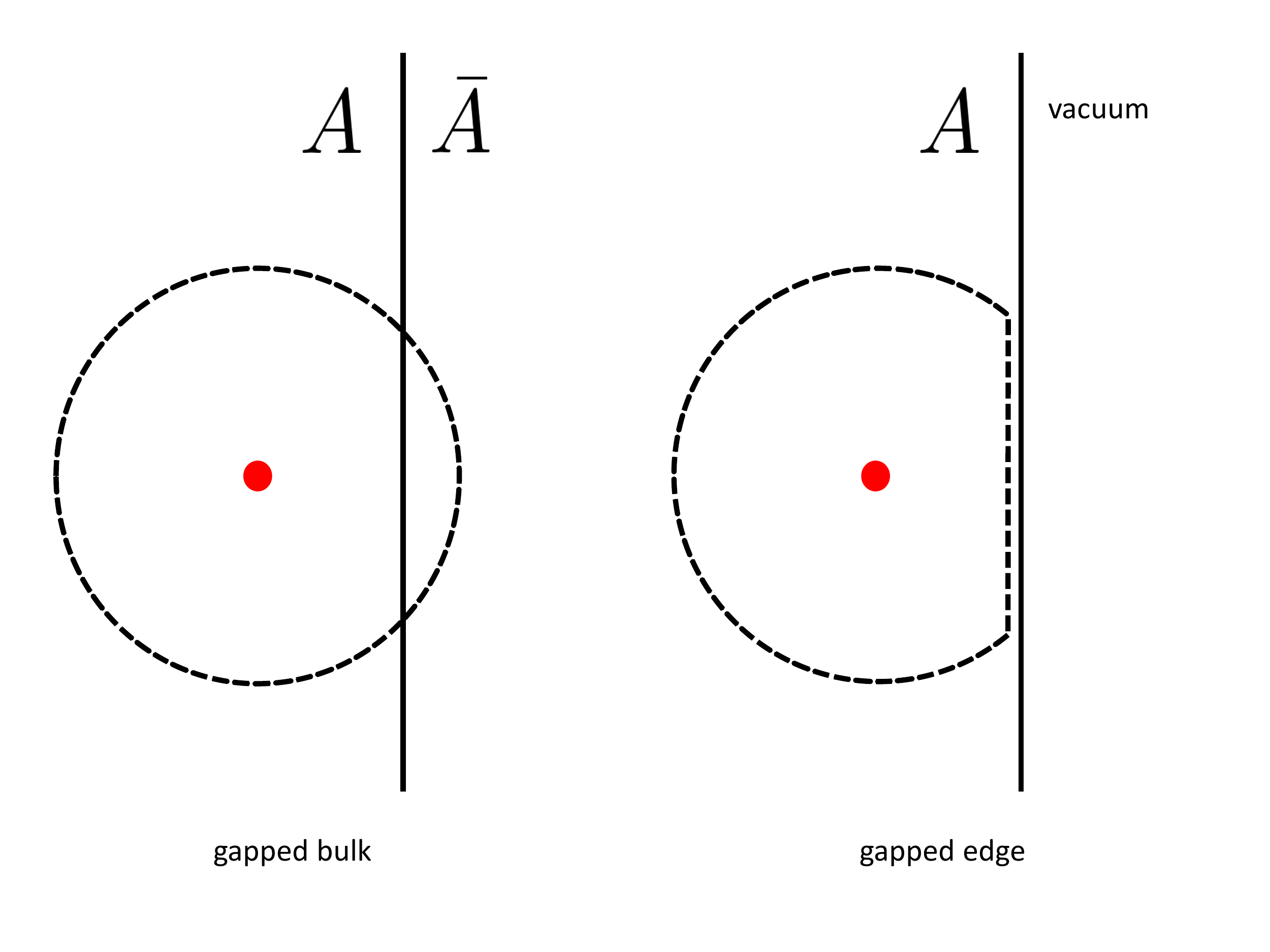}
  \caption{The red dot is local term in the Hamiltonian which is smeared into a quasi-local constraint.  The dashed circle is a cutoff where we truncate the qausi-local constraint to a strictly local constraint. Gappability of an edge suggests the constraint can be chosen to live strictly within $A$.}
  \label{deformedge}
\end{wrapfigure}

If we accept Kitaev's conjecture with strictly local operators, then the results of the previous section complete the argument. What about quasi-local constraints? A quasi-local constraint can always be truncated to a strictly local one of range $\ell$ with an error which decays faster than any power of $\ell$. To make the error smaller than $\frac{1}{\text{poly}(R)}$, take $\ell \sim \log^{1+\delta}(R)$ for any $\delta >0$. This extra $\sim \log(R)$ blowup seems dangerous to the strict area law, e.g., the effective width of the boundary region near $\partial$ may grow slightly with $R$.

However, suppose the phase does not have protected edge states. Then we have some intuition, illustrated in Fig.~\ref{deformedge}, that even quasi-local constraints may be sufficient to prove the that entropy is bounded by $\sim \log(G)$. As long as the number of constraints we must delete from $\hat{H}$ to obtain the Hamiltonian restricted to $A$ is bounded by $|\partial A|\ell$ (plus terms strictly in $\bar{A}$) for some system-size-independent $\ell$, then the arguments of the previous section would be sufficient. So the dangerous constraints are those that are further from $\ell$ from the boundary but closer than $\log(L)$ so that they cannot be truncated without further analysis. Given one of these dangerous distant but not too distant constraints, the idea is that if the edge of the system can be gapped, then there is a different quasi-local constraint, shown on the right in Fig.~\ref{deformedge}, which lives strictly within $A$ and which does the same job (e.g., we smear the local Hamiltonian term with the gapped Hamiltonian with edge).

Pick a region $A$ and suppose that all quasi-local constraints further than some system-size-independent $\ell$ but less than $\sim \log(L)$ from $\partial A$ can be deformed as in Fig.~\ref{deformedge} to live strictly within $A$.  Then we have a frustration free Hamiltonian $\hat{H}_{\text{deformed}}$ which has the property that when restricting the Hamiltonian to region $A$, the number of terms we must delete is bounded by $|\partial A|\ell$.  Since we already assumed the edge can be gapped, it follows that $\hat{H}_{\text{deformed},A}$ (the restriction of $\hat{H}_{\text{deformed}}$ to $A$) can be completed to a gapped frustration free Hamiltonian with a perturbation $V$ which consists of a boundary's worth of operators. Then the analysis of the previous section implies that the ground state degeneracy of $\hat{H}_{\text{deformed},A}$ is bounded by ground state degeneracy of $\hat{H}_{\text{deformed},A}+V$ times a factor of the order of $e^{c |\partial A|\ell}$ (the exponential of the number of operators in $V$).

It should be said that the above intuition about squeezing quasi-local constraints using gapped boundaries suggests that phases without protected edge states can be described by strictly local constraints or perhaps even commuting projector Hamiltonians.
This would be a converse to the result of
\cite{Lin-Levin-generalized}.

In this section we have given a general argument for the bound $S(\rho_A)\leq \mathcal{O}(|\partial A|) + \log(G(H_A))$.  It should be emphasized that we have not proven that the entanglement Hamiltonian (named in \cite{PhysRevLett.101.010504}), $\log(\rho_A)$, is local (although we believe this is probably true). Instead, we worked with the maximal entropy state $\sigma_A$ consistent with local data which is provably the Gibbs state of a local Hamiltonian and which can be more easily controlled. The bound \eqref{eq:GSbound} sharply encodes our intuition that many ground states are required to violate the area law. Besides our general arguments, we have proven this bound in the context of frustration free Hamiltonians. Finally, we showed how the above bound, together with the $s$ source framework, leads to an argument for the area law.

\section{Relation to MERA}
\label{sec:mera}

We now show how to cast our results into the form of a MERA provided the quasi-local unitaries are generated by quasi-local operators. Quasi-locality will mean that the effective range of the generator is bounded by a rapidly decaying function $h(r)$ which we may take to be, for example, $h(r) \sim e^{-r^{1-\delta}}$ or $h(r) \sim e^{-r/\log^2(r)}$. The basic idea of the construction is then to truncate quasi-local tails when they reach size $\frac{1}{\text{poly}(L)}$; this requires us to take $h(r_{\text{trunc}}) \sim \frac{1}{\text{poly}(L)}$ and hence $r_{\text{trunc}} \sim \log^{1+\delta}(L)$ or $r_{\text{trunc}} \sim \log(L)\log(\log(L))$. We then group $r_{\text{trunc}}^d$ sites into a single supersite and show that the quasi-local unitary may be approximated by a strictly bounded width circuit acting on these supersites.

We restrict our discussion here to MERA representations for $s=1$, although our techniques should also provide approximate branching MERA representations for $s>1$ states. We leave the details of these branching constructions to future work. Note that MERA has been applied to models which probably host $s=1$ fixed points \cite{PhysRevLett.104.187203}. Finally, although the bond dimensions we achieve are comparable to those recently obtained in the PEPS context using a very different method \cite{2014arXiv1406.2973M}, the MERA construction has the advantage that it is contractible in time polynomial in the bond dimension.  This gives an exponential speedup in the contractibility of the network in the worst case. Our results show that, given the MERA network (which may still be hard to find), it is possible to calculate properties of even complicated topological quantum liquids in time almost polynomial in system size.

The MERA construction also sheds light on the question of the existence of frustration free Hamiltonians for gapped states. In \S\ref{sec:fieldtheory} we will show how to construct MERAs for all TQLs by studying gapped field theories in an expanding universe.

\subsection{Truncating time evolutions with exponentially decaying interactions}

Given a quasi-local generator $K$, we may truncate the generator to a strictly finite range generator $K_\ell$ by setting to zero all interactions acting beyond range $\ell$ ($\ell$ is what we called $r_{\text{trunc}}$ just above).  The neglected terms have size of order $h(\ell)$.  We may determine the error in time evolution introduced by this truncation by studying the evolution under $K - K_\ell$. To be precise, we must compute the average of $e^{iK}e^{-i K_\ell}$ to determine the error due to evolving with $K_\ell$ instead of $K$, and this exponential can be processed using Baker-Campbell-Hausdorff to give $e^{i(K-K_\ell) + \frac{1}{2}[iK,iK_\ell] + ...}$ where $...$ denotes further commutators.  Since the commutator $[K,K_\ell$] is bounded by $h(\ell)$ and of the same order as $K-K_\ell$, it suffices to consider $K-K_\ell$ to get the scaling structure.

We compute the probability $p(t)$ to remain in the state $|\psi\rangle$ under time evolution by $\delta K = K-K_\ell$ in perturbation theory.  By definition we have
\beq
p(t) = |\langle \psi | e^{-i \delta K t} |\psi\rangle|^2,
\eeq
and expanding to first non-trivial order we obtain
\beq
p(t) \sim 1 - \frac{t^2}{2} \langle (\delta K)^2 \rangle.
\eeq
Suppose $\delta K$ is the sum of an extensive number of terms, $\delta K=\sum_x \delta K_x$, each of magnitude $J h(\ell)$ or less.  We then compute
\beq
\langle (\delta K)^2 \rangle  = \sum_{x, y} \langle \delta K_x \delta K_y \rangle \leq J^2  h^2(\ell) \sum_{x,y} e^{-|x-y|/\xi},
\eeq
where we have used the exponential decay of connected correlations and have assumed (without loss of generality) that $\langle \delta K_x \rangle =0$.

Evolving for a time of order $1/J$ we find
\beq
p(t\sim 1/J) \sim 1 - h^2(\ell) L^d \xi^d.
\eeq
Demanding that this probability be close to one, so that the perturbative calculation is valid, we must have
\beq
h(\ell) \sim \frac{1}{L^{\frac{d+q}{2}}}
\eeq
with $q > 0$. Then we are guaranteed that $p(t\sim 1/J) \sim 1 - L^{-q}$ which converges to one in the thermodynamic limit $L \rightarrow \infty$.

\subsection{Conversion of an $s=1$ fixed point to a MERA}
\label{subsec:mera}

\begin{wrapfigure}{R}{.5\textwidth}
\vskip-20pt
  \centering
    \includegraphics[height=5cm, width=0.48\textwidth]{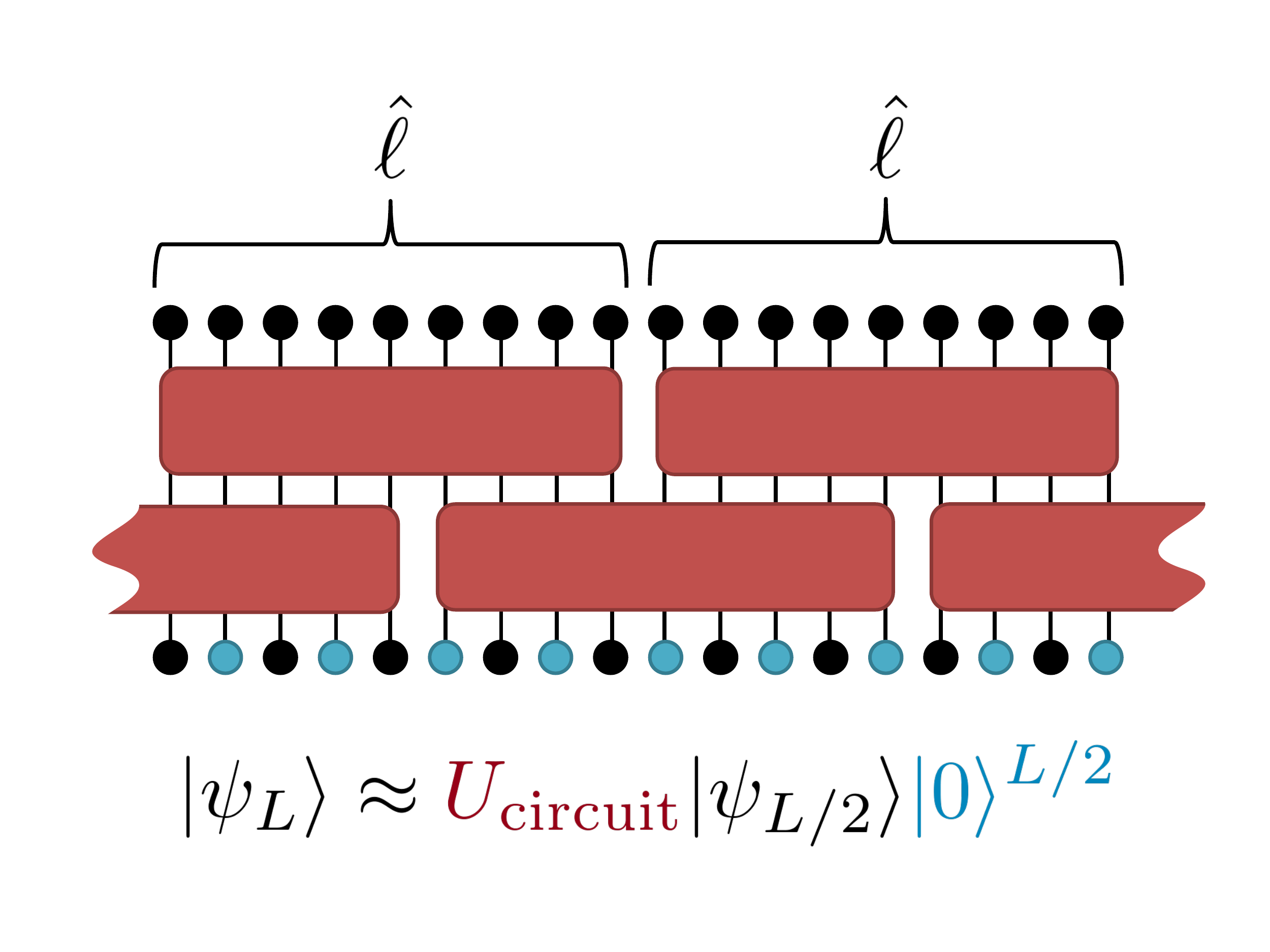}
  \caption{The staggered circuit composed of blocks of size $\hat{\ell}$ which approximates the action of the quasi-local unitary mapping $|\psi_{L}\rangle $ to $|\psi_{L/2}\rangle |0\rangle^{L/2}$ in $d=1$ for an $s=1$ fixed point. The colors of the circuit elements are coordinated with the colors of the terms in the equation in the figure.}
  \label{1dblock}
  \vskip-5pt
\end{wrapfigure}

We have just argued that to have the evolution under $K - K_\ell$ preserve the state in the thermodynamic limit, we must take $\ell \sim \log^{1+\delta}(L)$.  This cost is modest given the global accuracy since we are only required to coarse-grain chunks of $\ell^d \sim \log^{d(1+\delta)}(L)$ sites into supersites of total Hilbert space of dimension $e^{c \log^{d(1+\delta)}(L)}$ to have a local generator acting only on neighboring supersites. We now show that the unitary generated by $K$ can also be truncated to a strictly bounded-causal-width circuit acting only on neighboring supersites with local Hilbert space scaling in the same way with $L$. This circuit then constitutes one layer of a MERA of bond dimension $e^{c \log^{d(1+\delta)}(L)}$.

Note that the contraction of a MERA with $e^{c \log^{d(1+\delta)}(L)}$ bond dimension is almost polynomial in system size, and since a MERA is contractible in time polynomial in the bond dimension, it follows that physical properties of $s=1$ fixed points may be computed in time $e^{c \log^{d(1+\delta)}(L)}$ given the MERA circuit (which may be hard to find).  Furthermore, while this large a bond dimension may be prohibitive in practice, our result provides strong support for the conjecture that universal properties can be computed to high accuracy with a system-size-independent bond dimension,
as we discuss in \S\ref{subsec:bounded-bond-dim}.

To show that a $e^{c \log^{d(1+\delta)}(L)}$ bond dimension MERA exists, we must take the strictly local unitary evolution generated by the local operator $K_\ell$ and turn it into a quantum circuit with strictly bounded causal width.  In this case, we can again appeal to a coarse-graining argument.

\begin{wrapfigure}{R}{.5\textwidth}
\vskip-20pt
  \centering
    \includegraphics[height=5cm, width=0.48\textwidth]{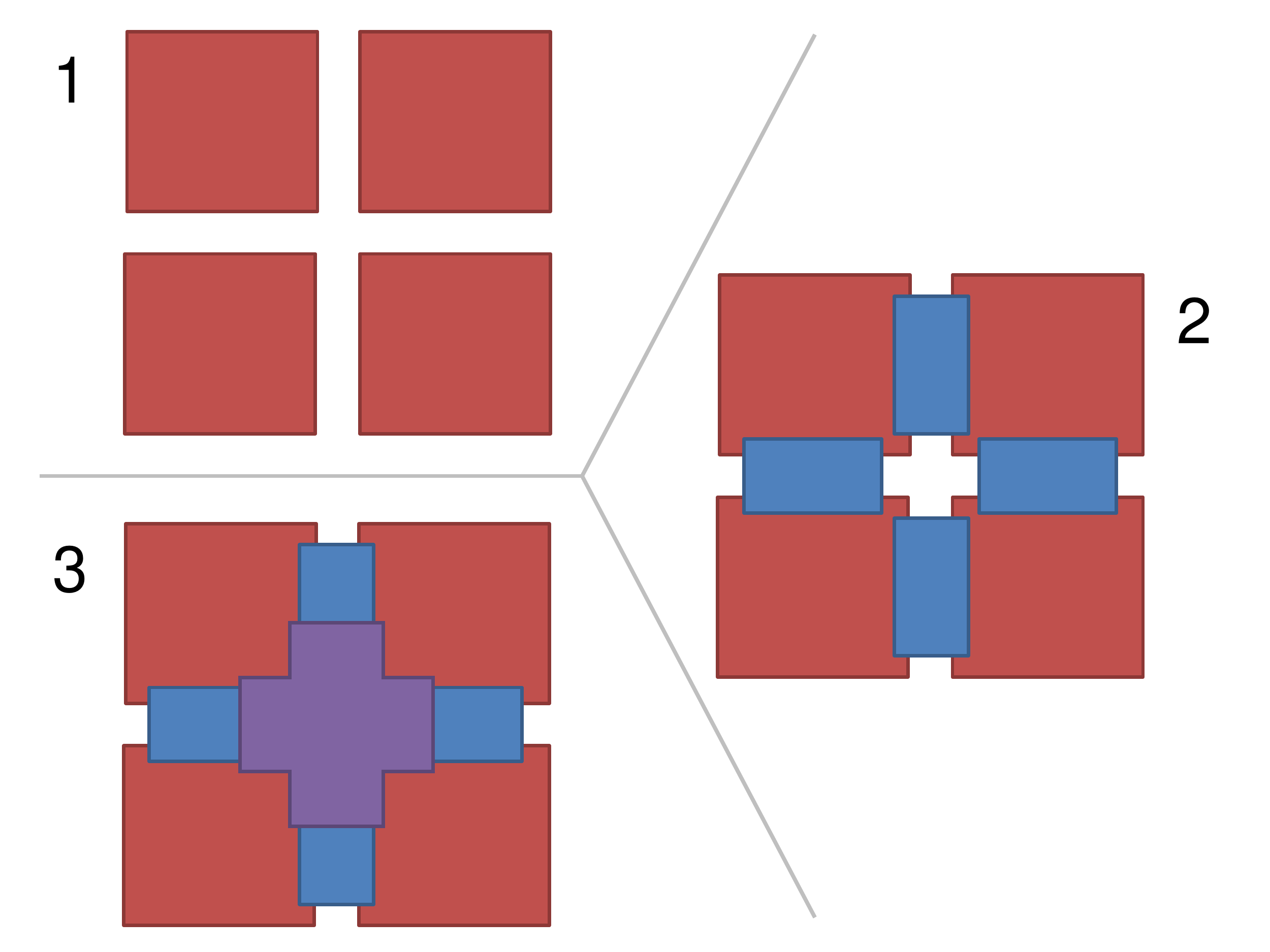}
  \caption{The three layers of an $d=2$ circuit approximation of the quasi-local unitary transformation. In layer 1 we apply $K$ in the red boxes to leave a qausi-one-dimensional network which is dealt with in layers 2 and 3 using the blue and purple unitaries similar to Fig.~\ref{1dblock}. We have $|\psi_L \rangle \approx \color{palatinatepurple} U_3 \color{black} \color{blue} U_2 \color{black}\color{red} U_1 \color{black}|\psi_{L/2}\rangle |0\rangle^{3L^2/4}$. The colors of circuit elements in the figure are coordinated with the colors of terms in the previous equation.}
  \label{2dblock}
  \vskip-5pt
\end{wrapfigure}

Consider first the case of $d=1$.  Suppose we are given a range $\ell$ two body Hamiltonian $K_\ell$ acting on qubits.  Group neighboring sets of $\hat{\ell}$ sites into supersites of Hilbert space dimension $2^{\hat{\ell}}$ as shown in Fig.~\ref{1dblock}.  By acting with one layer of unitaries on the supersites and one layer of unitaries between neighboring supersites (say between $\hat{\ell}/2$ on the left and $\hat{\ell}/2$ on the right) we obtain a causal width of $2\hat{\ell}$. To accuracy $\epsilon$ one can replicate the action of the local time evolution generated by $K_\ell$ by taking $\hat{\ell} \sim \ell +  v_{LR} t + \log(\epsilon)$ where $\ell$ is the interaction range, $v_{LR}$ is the Lieb-Robinson velocity \cite{lieb1972,2006CMaPh.265..119N}, and $t$ is the evolution time.

A crisp way to make the argument is to use the interaction picture with respect to the generator restricted to the size $\hat{\ell}$ blocks.  The remaining coupling terms between blocks get effectively smeared out by an amount much less than $\hat{\ell}$ by the Lieb-Robinson bound \cite{lieb1972,2006CMaPh.265..119N}.  Then take the resulting time evolution with these smeared generators which couple neighboring supersites and truncate the exponential tails beyond size $\hat{\ell}/2$ on either side of the interface. We have a two layer circuit consisting of staggered unitaries acting on blocks of linear size $\hat{\ell}$; this is the bounded-causal-width quantum circuit discussed above. Since $\ell$ already scales like $\log^{1+\delta}(L)$, it follows that (even with $\epsilon \sim L^{-q}$) $\hat{\ell}$ does as well. In fact, Lemma 1 of \cite{2012arXiv1206.6900M} can be adapted to rigorously prove that the above construction provides an excellent approximation to the time evolution; see also \cite{2006PhRvL..97o7202O,2007PhRvA..75c2321O} for earlier independent work along the same lines.

When $d>1$ a very similar construction may be used.  First, we block the system into blocks of linear size $\hat{\ell} \sim \log^{1+\delta}(L)$ as shown in Fig.~\ref{2dblock}.  Then we apply a unitary generated by $K$ restricted to have support completely within the blocks. Each block unitary commutes with every other block unitary by construction.  Next, we switch to the interaction representation of the block restricted $K$.  The remaining terms in $K$ will be smeared in the process, but provided we take $\hat{\ell}$ large enough, these interaction terms will be confined to thin regions near the boundaries of the blocks.  In $d=2$, for example, we would be left with a thin network of terms along the boundaries of the blocks.  These terms are now essentially one dimensional and the arguments in the previous paragraph can be used to deal with them. For example, by applying the blue and purple unitaries in Fig.~\ref{2dblock} we approximate the remaining quasi-local unitary acting on the quasi-one-dimensional network with a circuit. The only difference from the setup in Fig.~\ref{1dblock} is that we have junctions in the quasi-one-dimensional network, but the purple junction unitaries (which play the role of the second staggered layer of unitaries in Fig.~\ref{1dblock}) handle this overlap.

\begin{wrapfigure}{R}{.5\textwidth}
\vskip-20pt
  \centering
    \includegraphics[height=5cm, width=0.48\textwidth]{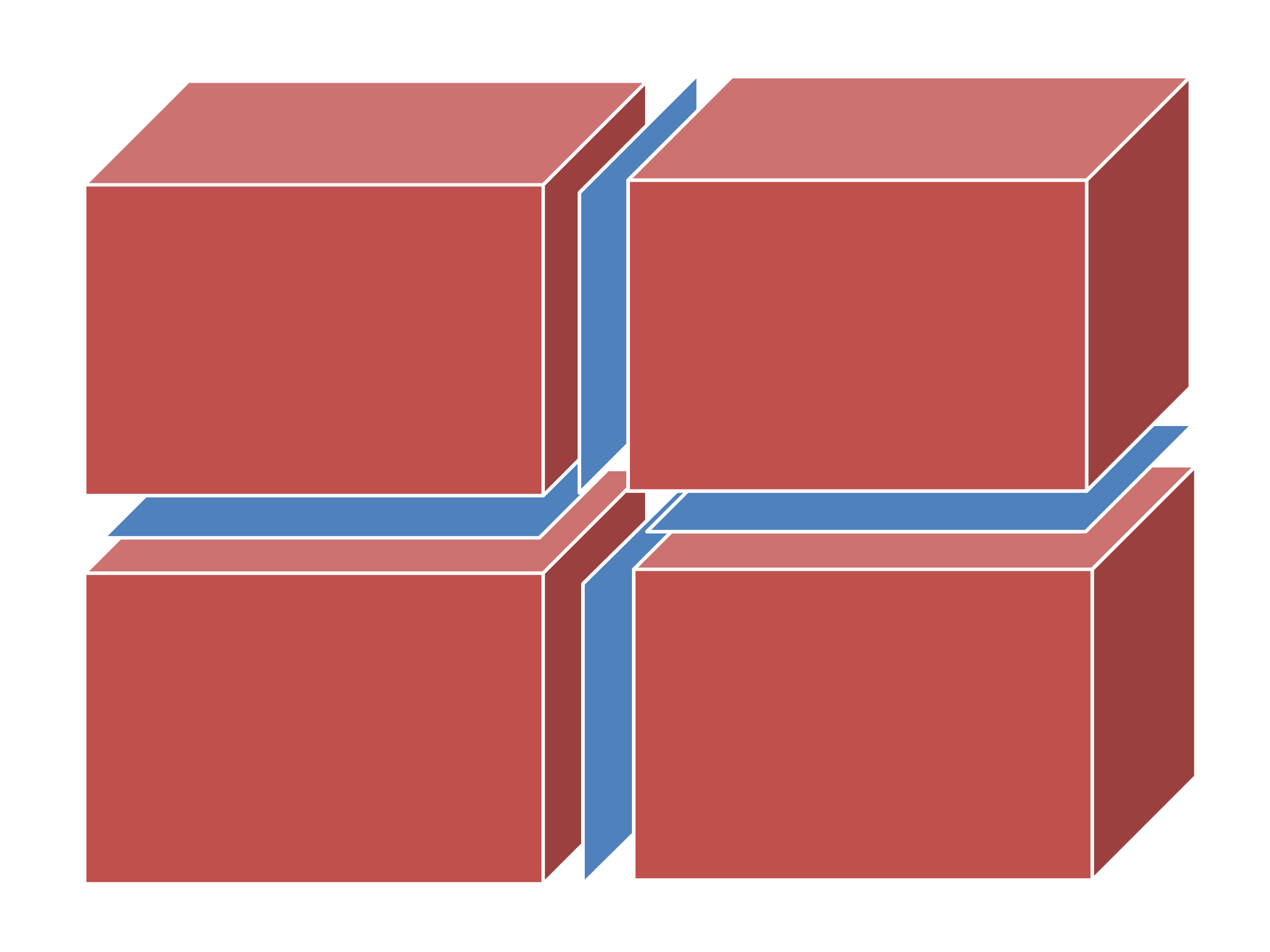}
  \caption{The blocking scheme in $d=3$. First, we deal with the red blocks. Then we deal with the blue faces. Finally, we are left with a quasi-one-dimensional network where the blue faces intersect.}
  \label{3dblock}
\end{wrapfigure}

In $d=3$ we would first block the system into cubic blocks (red blocks in Fig.~\ref{3dblock}) and apply the unitary generated by $K$ restricted to those blocks.  Then we would switch to the interaction representation of the blocks and apply a unitary generated by the terms in $K$ restricted to the faces between neighboring blocks (blue faces in Fig.~\ref{3dblock}. Then we would again be left with a quasi-one-dimensional network of unaccounted-for terms in $K$, and the one dimensional construction may be brought to bear.  In general $d$ dimensions, we recursively deal with the $d$-blocks, then the $d-1$-blocks between neighboring $d$-blocks, then the $(d-2)$-blocks between neighboring $d-1$-blocks, and so on until we reach the one dimensional limit.  In this way, a general quasi-local evolution may be blocked into a quantum circuit acting on up to $\sim \log^{d(1+\delta)}(L)$ degrees of freedom at a time.  This is a generalized MERA with bond dimension $\sim e^{c \log^{d(1+\delta)}(L)}$ which has unit overlap with the ground state in the thermodynamic limit.

\subsection{Polynomial bond dimension MERA?}

One reason to be optimistic that a polynomial bond dimension MERA exists is that the quasi-local tails which obstructed our construction above can perhaps be partially incorporated by modifying the tensors at smaller scales. Indeed, a MERA with bounded bond dimension can accommodate power-law decaying correlations, so at least in terms of raw expressive power, a polynomial bond dimension MERA should easily be able to accommodate exponential tails. We also know that many kinds of topological phases without edge states have, at a certain point in their phase diagram, an exact MERA representation with system-size-independent bond dimension.

We encode our speculations as a conjecture:
\begin{conj}[Polynomial Bond Dimension MERA]
Every $s=1$ RG fixed point has a MERA representation with poly$(L)$ bond dimension which achieves high overlap with the ground state in the thermodynamic limit.
\end{conj}

Note that the conjecture implies a strong result, that $s=1$ fixed points are well approximated in the thermodynamic limit by states with Schmidt rank bounded by $e^{c |\partial A|\log(L)}$ for any bi-partition $A\bar{A}$.  Such a result has already been proven in the context of regulated field theories in \cite{2013arXiv1304.6402S} which showed that truncating the reduced density matrix of a size $R$ region to its $e^{c|\partial A|+\delta}$ largest eigenvalues left a state which was still $\epsilon$ close to the correct reduced density matrix with $\delta \sim -\log(\epsilon)$. As a rough estimate, to produce a state with error $\epsilon \sim \frac{1}{\text{poly}(L)}$ for all $R$, we must take a bond dimension of order poly$(L)$.  With such a bond dimension, the Schmidt rank of any region will be $e^{c |\partial A| \log(L)}$ which is sufficient to produce small error. Note that our construction above gives a Schmidt rank going like $e^{c |\partial A| \log^{d(1+\delta)}(L)}$ for a $\epsilon \sim L^{-q}$ approximation to the true state, but there are subtleties in this analysis; see Appendix~\ref{sec:renyi} for a further discussion.

It should be further noted that ground states of frustration free Hamiltonians have Schmidt rank bounded by $G(H_A)$, the ground state degeneracy of the Hamiltonian truncated to region $A$. Since $G(H_A)$ obeys an area law for $s=1$ fixed points, it follows that these ground states have strictly area law Schmidt rank.  Combined with the existence of the quasi-local unitary mapping size $L$ to size $2L$, surely a polynomial bond dimension MERA exists.

We proceed to set up some definitions to reduce the above conjecture to a sharp technical statement.  To be concrete, we mostly consider $d=1$ and briefly remark about what changes in $d>1$.

Call $K_\ell$ a range $\ell$ quasi-local generator if it is a sum of local terms which decay faster than any power of distance beyond distance $\ell$.  Call $U_\ell$ a range $\ell$ quasi-local unitary if it is generated by a quasi-local generator of range $\ell'$ evolving for a time $t$ with $\ell = \ell'+t$ (we have put the Lieb-Robinson velocity to one).  It follows that the effects of a range $\ell$ quasi-local unitary decay faster than any power of distance beyond size $\ell$.  Finally, call a range $\ell$ quasi-local unitary acting on $L$ sites $(\ell',\ell_{\text{loc}},\epsilon)$ recursively localizable if its action on $|0\rangle^{L}$ can be reproduced up to error $\epsilon$ in norm by a two layer quantum circuit of staggered unitaries of strictly bounded range $\ell_{\text{loc}}$ times another quasi-local unitary of range $\ell'$ acting only on every other site (more generally acting only on the sites surviving at size $L/2$),
\beq
\left\| U_\ell |0\rangle^{L} - U_{\text{circuit}} U_{\ell'} |0\rangle^{L}\right\| < \epsilon.
\eeq

Since $|0\rangle^{ L}$ is the ground state of a local gapped Hamiltonian, our construction of the $e^{c \log^{d(1+\delta)}(L)}$ bond dimension MERA shows that every quasi-local unitary is $(0,\log^{1+\delta}(L),L^{-q})$ recursively localizable with the quasi-local unitary acting on $L/2$ sites taken to be the identity. By considering non-trivial quasi-local unitaries acting on $L/2$ sites we can hope to improve the parameters.

Recursive localizability of unitaries acting on $|0\rangle^L$ is clearly equivalent to recursive localizability of unitaries acting on any state obtained from $|0\rangle^L$ by a quasi-local unitary $V$ acting on every other site.  Indeed, we may simply absorb this $L/2$ site quasi-local unitary into the unitary defining the recursive localization at the cost of increasing $\ell'$ by the range of $V$. If we define recursive localizability of unitaries acting on the state $|\psi_{L/2}\rangle |0\rangle^{L/2}$ with $|\psi_{L/2}\rangle$ not quasi-locally equivalent to $|0\rangle^{L/2}$, then we obtain a potentially different classification of unitaries. A particularly interesting classification is obtained if $|\psi\rangle$ is allowed to be a local gapped $s=1$ ground state on $L/2$ sites.

Call a quasi-local unitary $U_\ell$ acting on $L$ sites $(\ell',\ell_{\text{loc}},\epsilon)_{|\psi\rangle}$ recursively localizable if there exists a staggered circuit and another quasi-local unitary acting on $L/2$ sites such that
\beq
\left\| U_\ell |\psi_{L/2} \rangle |0\rangle^{L/2} - U_{\text{circuit}} U_{\ell'} |\psi_{L/2} \rangle |0\rangle^{L/2}\right\| < \epsilon.
\eeq

These definitions generalize in an obvious way to higher dimensions by mimicking the structure of the $s=1$ fixed point. They are purposefully set up to be recursive. Indeed, suppose every quasi-local unitary $W_\ell$ is $(\ell/2,\ell_{\text{loc}},L^{-q})_{|\psi\rangle}$ recursively localizable for all $s=1$ RG fixed points $|\psi\rangle$ with $\ell_{\text{loc}}\sim \log^\zeta(L)$ and $\zeta \leq \frac{1}{d}$. Then every $s=1$ RG fixed point has a MERA representation with poly$(L)$ bond dimension. Note that $\ell/2$ was chosen as the first parameter because we are coarse-graining by a factor of $2$, so the quasi-local unitary on $L/2$ sites has the same effective range as before (since those $L/2$ sites are twice as far apart as measured in the un-decimated lattice).

Proof: We work in $d=1$ then remark on the extension to $d>1$ at the end. By assumption, there is a quasi-local unitary $U_{L,\ell}$ acting on $L$ sites with range $\ell$ that
accomplishes the $s=1$-source RG step
$U_{L,\ell} |\psi_{L/2}\rangle |0...0\rangle = |\psi_L\rangle$.  We assume that the range $\ell$ can be taken to be the same at every scale $L$ and that all range $\ell$ quasi-local unitaries are $(\ell/2,\ell_{\text{loc}},\epsilon)_{|\psi_{L/2}\rangle}$ recursively localizable for all $L$ with $\epsilon \sim L^{-q}$ and $\ell_{\text{loc}} \sim \log^\zeta(L)$.

Then there exists a quasi-local unitary $V_{L/2,\ell/2}$ and a strictly local circuit $V_{L,c}$ such that
\beq
\left\| U_{L,\ell} |\psi_{L/2}\rangle |\underbrace{0...0}_{L/2}\rangle - V_{L,c} V_{L/2,\ell/2} |\psi_{L/2}\rangle |\underbrace{0...0}_{L/2}\rangle \right\| < \epsilon.
\eeq
Multiply the quasi-local unitary $V_{L/2,\ell/2}$ by the unitary $U_{L/2,\ell}$ to produce a new quasi-local unitary $\tilde{U}_{L/2, 3\ell/2}$.  Apply recursive localizability for the unitary $W = \tilde{U}_{L/2,3 \ell/2}$ to produce a new circuit $V_{L/2,c}$ and a new quasi-local unitary $V_{L/4,3 \ell/4}$ such that
\beq
\left\|  V_{L,c} (V_{L/2,\ell/2} U_{L/2,\ell}) |\psi_{L/4}\rangle |\underbrace{0...0}_{3L/4}\rangle - V_{L,c} V_{L/2,c} V_{L/4,3 \ell/4} |\psi_{L/4}\rangle |\underbrace{0...0}_{3L/4}\rangle \right\| < \epsilon.
\eeq

Repeat the entire process by absorbing $V_{L/4,3 \ell/4} $ into $U_{L/4,\ell}$ to produce $\tilde{U}_{L/4,7 \ell/4}$.  Notice that the range of $\tilde{U}$ will always be less than $2 \ell$ so there is no blowup of the range in the recursive process.  This is important because $\ell_{\text{loc}}$ depends in principle on the range $\ell$, so to avoid a blowup of $\ell_{\text{loc}}$ we must avoid a blowup of the range. For example, at the next stage we use recursive localizability of $W = V_{L/4,3 \ell/4}U_{L/4,\ell} =\tilde{U}_{L/4,7 \ell/4}$ to exhibit a new circuit $V_{L/4,c}$ and a new quasi-local unitary $V_{L/8,7\ell/8}$ such that
\beq
\left\|  V_{L,c} V_{L/2,c} V_{L/4,3 \ell/4} U_{L/4,\ell} |\psi_{L/8}\rangle |\underbrace{0...0}_{7L/8}\rangle -V_{L,c} V_{L/2,c} V_{L/4,c} V_{L/8,7\ell/8} |\psi_{L/8}\rangle |\underbrace{0...0}_{7L/8}\rangle \right\| < \epsilon.
\eeq

To complete the argument, we iterate $\log(L)$ times, add and subtract the intermediate states within the norm, and use the triangle inequality to show that
\beq
\left\| |\psi_L \rangle - V_{L,c} V_{L/2,c} ... |\underbrace{0...0}_{L}\rangle\right\| < \epsilon \log(L).
\eeq
Since $\epsilon \sim L^{-q}$ we have shown high overlap between $|\psi_L\rangle $ and a MERA-like sequence circuits of range $\ell_{\text{loc}}$ acting on $|0...0\rangle$.

Returning to general $d$, grouping $\ell_{\text{loc}}^d$ sites into one supersite and using $\ell_{\text{loc}} \sim \log^\zeta(L)$, we produce a MERA with $e^{c \ell_{\text{loc}}^d} \sim e^{c \log^{d \zeta}(L)}$ bond dimension. If $\zeta =1/d$ can be achieved, then we have a polynomial bond dimension MERA.  If $\zeta < 1/d$ is possible, the MERA actually has sub-linear bond dimension.  We doubt this is possible generically, but it may be achievable in some special cases.

It would thus be very interesting to make progress on the technical problem of recursive localizability of quasi-local unitaries. As an intermediate step, we might conjecture that phases which have exact MERAs at some point in their phase diagam have at worst a poly$(L)$ bond dimension MERA throughout the entire phase.

\subsection{Universal properties from bounded bond dimension MERA?}
\label{subsec:bounded-bond-dim}

We have stated that our results support the idea that universal properties can be obtained with a bounded bond dimension MERA. We now sketch an argument for this conclusion, but first we must clarify what is meant by universal properties. It is difficult to give a general list of universal properties, but typically one means quantities that depend only the phase of matter and not on the particular realization (particular Hamiltonian) of that phase. Examples from two dimensional topological phases include the statistics of anyons, topological entanglement entropy, and the chiral central charge.

Because it is difficult to define these universal properties in complete generality not to mention rigorously prove that they are invariant under adiabatic deformations, we adopt a simpler approach. Having already shown that MERA captures the correct global structure of topological quantum liquids, we now argue that local properties may be obtained to high accuracy with bounded bond dimension. Good local properties plus the correct global (RG) structure of the network, when taken together, strongly suggest that universal physical properties can be obtained from a bounded bond dimension MERA. Indeed, it should always be kept in mind that demanding high overlap with the wavefunction in the thermodynamic limit is an absurd requirement from the point of view of most experimental settings where imperfectly known Hamiltonians, neglected degrees of freedom, dirt, etc. essentially always guarantee that a model wavefunction has tiny overlap with the physical state.

To argue for good local properties we appeal to the idea that adiabatic evolution for a finite time, while failing to preserve the global ground state, will still generate a controlled density of excitations. Alternatively, taking the quasi-adiabatic generator and truncating it to finite range (independent of system size) will again introduce a controlled density of excitations (while failing to preserve the global ground state). Some additional local error is also incurred in the truncation of the resulting local unitary evolution to a strictly bounded-causal-width circuit. We expect that both types of error can be made roughly exponentially small (at least decaying faster than any power) in the relevant cutoff length- or time-scale.

To make an estimate we suppose that approximating the exact quasi-local unitary with a strictly bounded-causal-width (independent of system size) quantum circuit produces a finite density of excitations. Let the induced energy density of excitations be $\delta \mathcal{E}$. As discussed above, we expect that $\delta \mathcal{E} \sim e^{-(\Delta \tau)^{1-\delta}}$ for a finite evolution time $\tau$ and gap $\Delta$; similarly, we expect the $\delta \mathcal{E} \sim e^{-\ell^{1-\delta}}$ where $\ell$ is the causal width of the truncated circuit approximating the full quasi-local unitary.

We estimate the energy density $\mathcal{E}_{2L}$ at scale $2L$ as follows. Recall that ($s=1$ fixed points) to obtain the state at scale $2L$ we take the state at scale $L$, add $(2^d - 1)L^d$ product states, and act with a quasi-local unitary. Thus given the energy density $\mathcal{E}_L$ at scale $L$, we first dilute it (since the product states are in their exact ground state) to obtain an energy density $\mathcal{E}_L/2^d$. Then we act with the approximate circuit which increases the density of excitations by $\delta \mathcal{E}$. The final energy density is thus
\beq
\mathcal{E}_{2L} = \frac{\mathcal{E}_L}{2^d} + \delta \mathcal{E}.
\eeq
Iterating this recursive equation then gives
\beq
\mathcal{E}_L \sim \sum_{n=0}^{\log(L)} \frac{\delta \mathcal{E} }{(2^d)^n} \sim \frac{\delta \mathcal{E}}{1 - 2^{-d}} + \mathcal{O}(L^{-d}).
\eeq
Thus the density of excitations at scale $L$ is essentially just given by $\delta \mathcal{E}$, so by choosing large but system-size-independent parameters $\tau$ and $\ell$ we may achieve a small density of excitations. In fact, the convergence appears to be almost exponentially fast.

Finally, why should universal properties be captured correctly by such an approximate state? One line of thought proceeds as follows. Once the energy density of the approximate state is sufficiently close to zero, there should exist another Hamiltonian $H'$ which is a perturbation of $H$ (whose exact ground state we are approximating) for which the approximate state is the correct ground state. Furthermore, because the energy density relative to $H$ is close to zero, the necessary perturbation to reach $H'$ should be small, hence the stability of the phase implies that $H$ and $H'$ are in the same phase and thus have the same universal properties. A candidate for the Hamiltonian $H'$ (which turns out to be frustration free) is a sum of projectors onto the null spaces of the local reduced density matrices of the approximate state.

This final point, that $H'$ is frustration free, is interesting. If $H'$ is also gapped, then this answers in the affirmative (for $s=1$ fixed points) Kitaev's conjecture about the existence of local constraints (strictly local case). It is hard to imagine that $H'$ is not gapped for sufficiently large (but still bounded) bond dimension, but we do not prove that here.

\subsection{Comments on algorithms}

In addition to showing that a MERA with modest resources exists for $s=1$ fixed points, we have given a novel procedure to construct such a MERA.  Start with the exact ground state on some small cluster; this data forms the initial condition of the MERA network (the ``top" tensor).  Then we take any path in Hamiltonian space that connects size $L$ to size $2L$ without closing the gap and form the quasi-local unitary that maps the ground states.  This can be converted into a layer of the MERA network as discussed above. Then repeat. When finished, we have the top tensor and all the layers of the network and at no point have we done a variational calculation for a large system.

Now of course it may be that finding such a gapped path in Hamiltonian space is hard, and it may be that constructing the quasi-adiabatic generator is hard.  On the other hand, for some problems of interest we may have a plausible guess for a path, or we may even be able to provably find such a path without knowing the ground state.  Furthermore, although constructing the quasi-adiabatic generator requires simulating time evolution, the effective time under which we evolve is of order one, so the quasi-local unitary should be open to efficient approximation. Alternatively, we could use the adiabatic approach instead of the quasi-adiabatic approach if we are interested only in local properties.

The point of this discussion is not that we have a provably superior algorithm, but simply to observe that our procedure provides a rather different approach to constructing a MERA.  In particular, we are never faced with the problem of an explicit variational calculation on a large system, so we might hope to avoid the problem of local minima in some cases.  Of course, such local minima may manifest in other ways, for example, as a small gap at some intermediate stage of the quasi-adiabatic evolution.  In any event, the present construction is close in spirit to the core motivation for the MERA construction where one has a picture of removing local entanglement scale by scale, a motivation that is to some extent obscured by the variational approach.

\section{Field theory construction}
\label{sec:fieldtheory}

In this section we consider what may be gained by studying topological quantum liquids in the continuum limit.  As discussed above, the continuum limit necessitates the consideration of a topological quantum liquid. Let us simply assume that the system has some conventional field theoretic representation where we may even impose Lorentz invariance if we wish.  We would like to implement the mapping from size $L$ to size $2L$ in this context. We show that this can be done by placing the system into a background geometry consisting of an expanding universe.  This construction shows that all massive field theories are $s\leq 1$ RG fixed points. We also give an explicit example with free fermions.

Imagine we have a field theory with some mass gap $m$ playing the role of the gap $\Delta$ above.  For example, we could consider a Chern-Simons theory, a massive Dirac fermion, a gapped non-linear sigma model, a gapped discrete gauge theory, or even fermions with a Fermi surface
gapped out by a superconducting order parameter.  We place the system into an expanding universe with metric
\beq \label{expand}
ds^2 = -dt^2 + a^2(t) d\vec{x}^2.
\eeq
Where necessary, we can compactify the spatial directions into a torus of coordinate size $L_0$; more generally, we could take the spatial geometry to be any closed $d$-dimensional manifold.  There may also be ambiguities in defining the field theory on such a curved spacetime geometry, but we may resolve these ambiguities any way we like provided the mass gap is preserved, e.g.~non-minimal couplings to the background gravitational field are allowed provided the gap is not closed.

In (\ref{expand}) the proper distance corresponding to a coordinate distance of $|\vec{x}|$ is $a(t) |\vec{x}|$.  Thus letting $a(t)$ run from $a_0$ at $t=t_0$ to $2 a_0$ at $t=t_0+\tau$ effectively doubles the linear size of the system.  Furthermore, if $\tau $ is long compared to $m^{-1}$, then we are in the adiabatic limit and the instantaneous ground state will be a good local approximation to the true state of the system at all times.  The most useful aspect of the field theory approach is that it dispenses with the lattice scale details and gives us a universal recipe for implementing our RG transformation.  Hence a very large class of topological theories, regarded in a continuum approximation, indeed have a quasi-adiabatic transformation which maps from $L$ to $2L$ and MERA representatives with the basic features outlined above.

Now it must be said that to be truly globally close to the ground state (i.e.~finite overlap
as $L\to \infty$), we must, as before, either use the quasi-adiabatic generator or perform an adiabatic evolution for a time poly-logarithmic in system size.  For variety, let us first analyze the adiabatic approach.  Assuming the function $a(t)$ is smooth and constant outside the interval $[t_0,t_0+\tau]$, the Fourier transform $\tilde{a}(\omega)$ can be made to decay faster than any power of $\omega$ for $|\omega|> \tau^{-1}$. First order perturbation theory then gives, for the probability to create an excitation, a quantity of order $|\tilde{a}(m)|^2$. An achievable decay of $\tilde{a}(\omega)$ is
\beq
\tilde{a}(\omega) \sim e^{-(\omega \tau)^{1-\delta}}
\eeq
for any $\delta >0$, hence by choosing
\beq
\tau \sim m^{-1} \log^{1+\delta}( L)
\eeq
we may assure that the probability to create an excitation is bounded by $L^{-q}$ where $L = a(t) L_0$ is the proper size of the system.  Furthermore, because the system is in finite volume perturbation theory converges.

A comment about regulators is in order.  If for example we impose a hard cutoff $\Lambda_0$ on momenta defined with respect to the coordinate distance $|\vec{x}|$, then as space expands the physical cutoff, $\Lambda = \Lambda_0/a$, decreases with time.  Without changing the cutoff $\Lambda_0$ the Hilbert space remains the same at all scales (unlike in our lattice constructions above).  In keeping with the lattice construction, it is better to keep the physical cutoff $\Lambda$ the same before and after space expands.  One way to accomplish this is to add to the system auxiliary heavy spectator fields.  Then as space expands some of the high energy states from these spectator fields can be incorporated into the ``low energy" (but still gapped) field theory of interest to keep the physical cutoff invariant.  In other words, we can always safely steal states from trivial field theories at very high energies (in fact, this is in a sense the only non-trivial part of the construction).  It may also be necessary to truncate some unbounded operators to apply our results for bounded strength interactions.  This should always be possible.  Hence we claim that any regularizable massive field theory obeys the area law and has a MERA representation with modest bond dimension.

\subsection{Example: Dirac fermion, $d=2$}

We will now work through the example of a massive Dirac fermion $\psi$ in $d=2$ evolving in a time dependent background.  This case is interesting because the system exhibits the quantized Hall effect determined by the sign of mass $m$, so our analysis will show this theory is an example of an $s=1$ fixed point.  The background geometry is
\beq
ds^2 = - dt^2 + a^2(t)( dx^2 + dy^2)
\eeq
which we cast in the form $g_{\mu \nu} = e^a_\mu e^b_\nu \eta_{ab}$ where $\eta$ is the flat metric and $e$ is the vierbein.  We read off the values of $e$ from the metric and find that
\beq
e^{\hat{x}}_x = e^{\hat{y}}_y = a, \,\,\, e^{\hat{t}}_t = 1.
\eeq
The spin connection $\omega$ is defined as
\beq
d e^a + \omega^a_b e^b = 0,
\eeq
and we find
\beq
\omega^{\hat{x}}_{\hat{t}} = \frac{\dot{a}}{a} e^{\hat{x}}, \,\,\, \omega^{\hat{y}}_{\hat{t}} = \frac{\dot{a}}{a} e^{\hat{y}},
\eeq
and all others zero.

For flat space $\gamma$ matrices we take $\gamma^{\hat{t}} = iZ$, $\gamma^{\hat{x}} = X$, and $\gamma^{\hat{y}}= Y$ which satisfy $\{\gamma^a,\gamma^b\} = 2 \eta^{ab}$.  Curved spacetime $\Gamma$ matrices may then be defined as $\Gamma^\mu = e^\mu_a \gamma^a$.  The Dirac action (with $\bar{\psi} = \psi^\dagger \Gamma^0$) is then
\beq
\mathcal{S}_D[\psi] = \int dt dx dy \,a^2 \left[\bar{\psi} \,\Gamma^\mu \left(i \partial_\mu - \frac{i}{2}\omega_{\mu ab}\sigma^{ab} \right)\psi - m\bar{\psi}\psi \right]
\eeq
where $\sigma^{ab} = \frac{i}{4} [\gamma^a, \gamma^b]$ are the Lorentz generators.

The necessary components of $\sigma$ are $\sigma^{\hat{t}\hat{x}} = -i Y/2$ and $\sigma^{\hat{t}\hat{y}} = i X/2$.  If we also switch to Fourier modes $\psi(x) = \sum_k e^{i \vec{k}\cdot \vec{x}} \psi_k$ then the resulting action is $S_D[\psi] = \sum_k \mathcal{S}_{D,k}[\psi_k]$
and the action for a given $k$ mode is
\bea
\mathcal{S}_{D,k}[\psi_k] = \int dt \, a^2(t)    & \left[
 \bar{\psi}_k \left(\frac{X}{a}\right) \left(- k_x + \frac{\dot{a}}{2} Y \right)\psi_k + \bar{\psi}_k \left(\frac{Y}{a}\right) \left(- k_y - \frac{\dot{a}}{2} X \right)\psi_k \right.
\cr \cr
&+ \left.   \bar{\psi}_k i Z i \partial_t \psi_k - m \bar{\psi}_k \psi_k \right].
\eea
Observe that the two terms from the spin connection both combine to give $2 i Z \frac{\dot{a}}{a}$. Performing a time dependent rephasing $\psi_k = a^{i/2} \Phi_k$ removes the spin connection term. The details are not ultimately important; what is important is that we have a Hamiltonian system of finite dimension which is changing adiabatically.

As reviewed above, we may compute the probability $p_k$ for each $k$ mode to remain in its ground state using perturbation theory. We have $p_k \geq 1 - c e^{-(\sqrt{k^2+m^2}\tau)^{1-\delta}}$, where $\tau$ is the evolution time. This perturbation theory converges for any finite $\tau \gg m^{-1}$ since each $k$ mode is finite dimensional.

Multiplying over all $k$ modes, the total probability to remain in the ground state is
\beq
p \sim \prod_k p_k \sim \exp\left( -\sum_k c e^{-(\sqrt{k^2+m^2}\tau)^{1-\delta}}\right),
\eeq
where the second estimate follows if $\tau \gg m^{-1}$.  Replacing the sum over $k$ with an integral we obtain
\beq
p \sim \exp\left(1 - c L^d \int \frac{d^d k}{(2\pi)^d} e^{-(\sqrt{k^2+m^2}\tau)^{1-\delta}}\right)
\eeq
which can be made to approach one as $1 - L^{-q}$ if $\tau \sim m^{-1} \log^{1+\delta}(L)$ for some $\delta >0$.  Crucially, the upper cutoff on $k$ does not enter because the integral converges rapidly.  Hence in this case the formal cutoff may be sent to infinity and no heavy spectator fields are required; the expansion of space smoothly brings down higher momentum modes to continually fill the growing number of long wavelength modes.

\subsection{Black holes and dS/CFT}

The preceding discussion of continuum field theory in expanding universe,
in particular of bringing in product states at the UV cutoff,
can be recognized
as a regulated description of the ``Unruh vacuum" for quantum fields in curved spacetime.
Its defining properties are ``reasonable at short distances"
-- that is, the large-$k$ modes are in their groundstates --
plus no particles initially.
The procedure we have described is just what is
done to compute density perturbations in inflation
and also Hawking radiation \cite{Mukhanov:2007zz},
and in particular is the resolution, in practice, of the so-called
`trans-Planckian problem' raised by large gravitational blue-shifts.

Such a connection between renormalization group
evolution and the physics of an expanding universe
also appears in the `dS/CFT correspondence'
for the case of de Sitter space
\cite{Strominger:2001pn} and for
more general FRW spacetimes \cite{Strominger:2001gp}.

This connection between entanglement renormalization
and gravitational physics is different from the one
proposed in \cite{Swingle:2009bg, Swingle:2012wq} (see also the further developments \cite{2011JSP...145..891E,PhysRevLett.110.100402,2012JHEP...10..193N})
in that here the evolution produced by the quantum circuit
is really timelike;
such a Wick rotated picture has been advocated in \cite{Hartman:2013qma} (see also the sketch in \cite{2013NJPh...15b3020B}).
An explicit calculation of the entanglement entropy of subregions
of an expanding universe
for free field theory was made in \cite{2013JHEP...02..038M}.

These previous analogies between FRW cosmology and the RG
were motivated by hopes of learning something about quantum
gravity and cosmology, while in the bulk of this paper, we are using this idea in the other direction.

The restriction to $\log(G) < cL^{d-1}$, when interpreted as a statement about an entropy, is temptingly reminiscent of the black hole entropy bound. One way to attempt to make a connection is to consider collapsing a shell of matter to form a black hole in a space which already supports such a highly entangled state. Now because the system is gapped and because the curvature is weak at the event horizon, one might imagine that the highly entangled ground state survives (at least away from the singularity). Further assuming that the entanglement entropy of the matter across the horizon contributes to the black hole entropy, we may be able to violate the Bekenstein area bound if we had a gapped phase that violated the area law. If so, the coupling to gravity would forbid violations of the area law in gapped ground states.
Notice that the indistinguishability
of the groundstates is important to ensure that
the state outside the horizon is not perturbed by the gravitational collapse.
At present, however, this argument is speculative.

Nevertheless, the coupling to gravity does provide constraints on the behavior of any putative topological field theory. Consider a topological quantum field theory $\CQ$. Its Euclidean path integral $Z_\CQ$ on $\Sigma^d \times S^1$ ($\Sigma^d$ is some closed $d$-manifold) computes $\tr\left(e^{-\beta H_{\CQ}(\Sigma^d)}\right)$ where $H_{\CQ}(\Sigma^d)$ is the Hamiltonian of the topological theory on space $\Sigma^d$ and $\beta$ is the length of the $S^1$ factor. Since in the topological limit the gap to excitations is infinite, the trace reduces to counting the number of ground states of $H_{\CQ}(\Sigma^d)$, that is
\beq
Z_{\CQ}[\Sigma^d \times S^1] = G(H_{\CQ}(\Sigma^d)).
\eeq

Without invoking the topological nature of $\CQ$ we must allow $Z_{\CQ}$ to depend on the metric $g^\Sigma_{ij}$ on $\Sigma^d$, but with the assumption that $\CQ$ is topological we can rule out interesting dependence on $g^\Sigma$.  Let $g$ be the metric of spacetime; assuming $\CQ$ couples minimally to gravity we have
\beq
Z_{\CQ}[g+\delta g] = Z[g] \exp\left( \frac{1}{2}\int_{\Sigma^d \times S^1} d^{d+1} x \sqrt{g} \delta g_{\mu \nu } T^{\mu\nu}_{\CQ}\right)
\eeq
where $T^{\mu\nu}_\CQ$ is the stress tensor of $\CQ$. But $T_\CQ = 0$ since $\CQ$ is topological, so $Z_\CQ[g]$ is independent of small deformations of $g$. Note also that the coupling to $T_\CQ$ is a small perturbation, so the stability of the phase guarantees that the gap does not collapse. This together implies that $Z_{\CQ}[\Sigma^d \times S^1]$ is independent of the size of $\Sigma^d$ and hence so is the ground state degeneracy.

This argument does not rule out systems with ground state degeneracy depending on the ``size" of the space, but it does imply that they must couple to gravity differently. For example, suppose we realized a phase with $G \sim e^{c L}$ in $d=3$ in the lab by constructing an array of coupled localized objects, e.g., a lattice of cold atoms. Now suppose that a gravitational wave passes through the system. What happens is that the distance between the different potential wells, say, is changed, but the number of wells is not modified. Hence the coupling to gravity modulates the couplings between different localized objects, but does not change the ``size" (number of localized objects) of the system. Said differently, there is extra data in the path integral $Z_{\CQ}$ on which the ground state degeneracy does depend and which is not sensitive to weak gravitational perturbations (because the phase is stable).

\subsection{Lorentz invariant entanglement Hamiltonian}

As a final application of the field theory construction, we may explicitly verify the claimed properties of the maximum entropy locally consistent state $\sigma_A$. For simplicity we analyze the case where region $A$ is a half-space, but we expect that the lessons generalize to all regions because of the gap.

As shown in \cite{bwtheorem,1976PhRvD..14..870U,1973PhRvD...7.2850F,1975JPhA....8..609D,PhysRevD.29.1656}, the entanglement Hamiltonian for a half-space in any Lorentz invariant quantum field theory can be related to a generator of boosts $M_{xt}$.  To be precise, suppose $A$ is a half-space given by $A = \{ \vec{x} | x \geq 0, x_\perp \in \mathbb{R}^{d-1} \}$. Associated to region $A$ we have the causal development $\CC(A)$ which is given by all $(t,x,x_\perp)$ with $(x,x_\perp) \in A$ and $|t| < x$. The causal development or ``Rindler wedge" $\CC(A)$ is mapped into itself by the flow generated by the boost generator
\beq
M_{xt} = \int_A d^{d-1}x_\perp dx \, x T_{00},
\eeq
where $T_{00}$ is the energy density. Then by constructing a path integral for $\rho_A$ in which the Euclidean angle in the $x-t$ plane is used as time, \cite{PhysRevD.29.1656} showed that
\beq
\rho_A = \frac{e^{-2\pi M_{xt}}}{\tr\(e^{-2\pi M_{xt}}\)}.
\eeq
In other words, the entanglement Hamiltonian $-\log(\rho_A)$ is local.

Since the entanglement Hamiltonian $2\pi M_{xt}$ is local, it follows that $\sigma_A = \rho_A$. Thus the maximum entropy locally consistent state explicitly has the form argued for in \S\ref{sec:arealaw} and in particular has the property of [Infinite Bulk Gap]. We may then pursue the kind of general local thermodynamic arguments given in \S\ref{sec:arealaw}. Alternatively, we may explicitly compute the spectrum of $M_{xt}$ in simple cases and verify that the entropy obeys an area law.

\section{Discussion and speculation}

In this paper we have argued for an area law for gapped phases, and we have shown how to produce tensor network representations of interesting phases. We introduced the idea of an $s$ source RG fixed point. Assuming all gapped phases are $s$ source fixed points, we argued that only phases with ground state degeneracy scaling like $G(L) \sim e^{c L^{d-1}}$ or faster could violate the area law. We also used ideas about local reconstruction of quantum states to argue for the bound $S(\rho_A) \leq \mathcal{O}(|\partial A|) + \log(G(H_A))$ which gave another proof of the claim that
a stable hamiltonian requires $G(L) \sim e^{c L^{d-1}}$ or greater to violate the area law. Combining the two approaches, we showed that even with $G(L) \sim e^{c L^{d-1}}$, we could not support the suggested logarithmic violation of the area law. More extreme violations of the area law were ruled out with weak spectral assumptions about the low temperature thermal free energy.

Some of our results are rigorous, including the proof of the area law for topological quantum liquids, the MERA construction, and the bound $S(\rho_A) \leq \mathcal{O}(|\partial A|) + \log(G(H_A))$ for ground states of frustration free Hamiltonians. Nevertheless, our overall argument for the area law rests on non-trivial physical assumptions and is not rigorous.  On the other hand, we see no immediate obstacle to making much of the general argument more rigorous. More interesting, in our opinion, is our claim that if a phase does violate the area law, then it must be a rather strange beast. For example, if it is an $s$ source fixed point and obeys the free energy condition, it seems that the phase must violate our reconstruction arguments in \S\ref{sec:arealaw}.  If a frustration free gapped phase violates the area law, then it must have a very large ground state degeneracy.  If the phase is not an $s$ source fixed point, then it is peculiarly disconnected from its peers at smaller and larger scales. So while it would be very interesting to exhibit such a peculiar beast, we hope to have convinced the reader that the area law holds for a huge class of systems including essentially all models of current physical relevance.

There are numerous directions for future work. We have not tried to optimize the analytic parts of the arguments to achieve the best possible bounds, so it should possible to do better than our simple estimates, e.g., in the MERA construction. Making progress on the question of recursive localizability or otherwise exhibiting a poly($L$) bond dimension MERA would be very interesting. Providing further arguments for the $s$ source framework (or counterexamples) is highly desirable. The inclusion of symmetry in the analysis is a logical next step. Another possible direction would be to explore the consequences of the $s$ source framework for defects, e.g., interfaces between phases. It would also be interesting to study the precise quantitative relationship between the gap and the entanglement entropy. Finally, of particular interest is the extension of our results to gapless systems.

A very natural speculation is that conventional conformal field theory (CFT) fixed points with gapless degrees of freedom match our definition of $s=1$ fixed points. One may object that we have only studied gapped phases in this work, but this objection has significantly less force than one might imagine. Various kinds of topological states in $d>1$ have just as much entanglement in their ground state as CFTs, so the amount and structure of entanglement is not obviously at issue. Furthermore, long range correlations can easily be included in the MERA network, so this too does not seem to be a real objection. We also only require the state to global accuracy $L^{-q}$; this is consistent if very high dimension operators are truncated from the spectrum (because they only contribute very rapidly decaying power law corrections which are well within our error threshold). The field theory constructions are also very promising. Non-local tensor networks that exactly represent gapless phases have been exhibited \cite{PhysRevLett.113.010401} and \cite{2013arXiv1304.6402S} has argued that even ground states of gapless regulated field theories can be approximated by states with limited Schmidt rank. It is also amusing to note that the structure of correlations in strongly coupled large $N$ gauge theories described by holographic duals is not so different from a gapped phase, e.g., short-ranged mutual information to leading order in $N$. Taken together, this evidence suggests that the conjecture that conventional field theory fixed points are also $s=1$ RG fixed points is quite reasonable. Of course, even if this conjecture is true, it remains to construct the required quasi-local unitary. We plan to address these points in a forthcoming companion paper.

\vskip.2in
{\bf Acknowledgements.}
We thank L. Huijse, K. Van Acoleyen, M. Mari\"{e}n, and M. Zaletel for helpful conversations on related topics. We thank D. Chowdhury and I. Kim for helpful comments on the manuscript.

BGS is supported by a Simons Fellowship through Harvard University.
This work was supported in part by
funds provided by the U.S. Department of Energy
(D.O.E.) under cooperative research agreement DE-FG0205ER41360,
in part by the Alfred P. Sloan Foundation.
BGS and JAM acknowledge the hospitality of the Perimeter Institute for Theoretical Physics during
the workshop ``Low Energy Challenges for High Energy Physicists".
Research at Perimeter Institute is supported by the Government of Canada through Industry Canada and by the Province of Ontario through the Ministry of Economic Development \& Innovation.

\appendix
\renewcommand{\theequation}{\Alph{section}.\arabic{equation}}

\section{What is a phase?}
\label{sec:phase}

In this appendix we briefly discuss some of the properties we expect of decent quantum phases of matter (clearly this will be a somewhat personal perspective; for a somewhat complementary discussion,
see \cite{2010arXiv1008.5137H}). The starting point is typically what we call a \textit{Hamiltonian motif} which is a function that maps a set $\mathcal{G}$ of graphs (or more generally a cell complex) to a set $\mathcal{H}$ of Hamiltonians defined on those graphs. The set of graphs often has some restrictions, e.g., to $d$-dimensional graphs, to trivalent graphs, to planar graphs, or to graphs with an even number of sites (e.g., in spin-1/2 systems). Crucially, the set of admissible graphs must include a sequence of graphs with size going to infinity to define a thermodynamic limit. For the present paper we always restrict to local graphs which can be understood as living in $d$ dimensions. The word motif is appropriate because typically the way the function works is to assign terms to the Hamiltonian based on local features or patterns in the graph, e.g. a term for every vertex, link, or plaquette. So when we speak of a phase of matter we are really considering an equivalence class of Hamiltonian motifs where two motifs are equivalent if they give the same global properties. In particular, a gapped phase refers at least to a family of Hamiltonians defined on systems of various sizes all having a system-size-independent gap (or lower bound on the gap).

However, not just any function from graphs to Hamiltonians can be a representative of a gapped phase of matter.  A Hamiltonian motif must obey certain rules to represent a gapped phase.  We do not attempt to give a completely rigorous definition of a gapped phase, but instead enumerate the most important rules that a gapped phase must obey. Indeed, there is some subtlety here. For example, the ground state manifold of Haah's code at size $L$ cannot typically be adiabatically connected to the ground state manifold of Haah's code at size $L+1$, so by some definitions these two systems are in different phases.  However, because they descend from the same Hamiltonian motif and because they share many properties, one might like to think of them as representing the same phase.  It is not clear to us which view-point is superior.

[Stability]: A phase of matter has the property of stability with respect to small changes in the Hamiltonian motif.  We may assign slightly different Hamiltonians to a given graph without encountering any change in the global properties of the system.  Indeed, there should be an open set in local Hamiltonian space around the Hamiltonian on any graph within which the global properties are unchanged.  In the case of gapped phases, one convenient way to encode this criterion is to demand that there be a family $H(\eta)$ of gapped Hamiltonians interpolating between the initial and final Hamiltonians.  Then ground states of the initial Hamiltonian may be mapped to ground states of the final Hamiltonian using a quasi-local unitary.

[Local indistinguishability]: Stability of the Hamiltonian implies that the number of ground states cannot depend on small local perturbations.  This leads one to the idea of local indistinguishability.  Truly stable gapped quantum phases must have the property that all ground states are approximately locally indistinguishable.  This ensures that no local perturbation can split the ground state manifold except possibly by an amount exponentially small in system size.  We will always assume that ground states are locally indistinguishable unless otherwise specified.

[Insensitivity to Boundary Conditions]: Related to the idea of local indistinguishability is the idea of insensitivity to boundary conditions. Given some region $A$ in a $d$ dimensional graph $G$ and given two gapped Hamiltonians $H_1$ and $H_2$ representing the same phase which differ only far away from $A$, it should be the case that the state of $A$ is approximately the same in any ground state of either Hamiltonian. Note, however, that this notion is subtle. For example, in an integer quantum Hall state on a torus, inserting flux through the cycles of the torus, which is a global operation, does lead to a non-trivial Berry phase, so we are not claiming that boundary conditions are totally irrelevant, far from it. Still, we will assume that local data is indeed insensitive to boundary conditions. Because this final assumption is not so straightforward as local indistinguishability and stability (since it requires the notion of a phase), we spend a little time discussing it.

The starting point for any discussion of insensitivity to boundary conditions should begin with the decay of correlations.  In any gapped phase of matter it can be proven that all connected correlations decay exponentially. In other words, although the system may have long-range entanglement, correlations of local operators always fall off rapidly with distance. As a necessary tool to prove the decay of correlations, one should also mention the Lieb-Robinson bound \cite{lieb1972,2006CMaPh.265..119N} which states that causal influences propagate with a finite velocity up to exponentially decaying tails. Causality, in the form of the Lieb-Robinson bound, is another important primitive in the discussion about insensitivity to boundary conditions.

Now suppose we have two Hamiltonians $H_1$ and $H_2$ differing only far from region $A$ such that there is a gapped Hamiltonian path $H(\eta)$ from $H_1$ to $H_2$ also differing only far from $A$. Then by constructing the quasi-adiabatic generator $K(\eta)$ and its associated quasi-local unitary, we can map ground states of $H_1$ to ground states of $H_2$.  Since $\partial_\eta H$ is only non-zero far from $A$, it follows that the quasi-local unitary generated by $K(\eta)$ has an effect on $A$ which is smaller than any power of the separation between $A$ and the region where $\partial_\eta H(\eta)$ is non-zero.

This result is nice, but it relied on the existence of a gap.  We want an even stronger notion of insensitivity to boundary conditions.  For example, we might introduce a boundary to the system which hosts gapless edge states, but we would still expect that regions far from the boundary are in approximately the same state as before the boundary was introduced.  There is thus a notion of a local gap which protects regions even from the effects of gapless degrees of freedom provided those degrees of freedom are localized far from the region of interest.

We can try to make this idea of a local gap sharper using the generator of quasi-local evolution defined as
\beq
-i K(\eta) = \int_{-\infty}^\infty dt F(t) e^{i H(\eta) t} \partial_\eta H(\eta) e^{-i H(\eta) t}.
\eeq
Suppose that all members of the family $H(\eta)$ have gapless edge states near some boundary, but we demand that $H(\eta)$ is only changing far from these edge states.  As usual, we choose the filter function $F$ such that its Fourier transform vanishes for energies less than $\Delta$.  Then if we had a bulk gap, we could take $\Delta$ to be the gap, but the presence of gapless edge states makes that impossible.  On the other hand, if the matrix elements of $K(\eta)$ between states of energy less than $\Delta$ are exponentially small, e.g. because such low energy states are localized far away from where $\partial_\eta H(\eta)$ is non-zero, then we still approximately map ground states to ground states.  This is one example of what we mean by a local gap and insensitivity to boundary conditions.

As the strongest notion of insensitivity to boundary conditions, we might demand that even if we delete entirely some part of the system, the state of distant regions remains approximately the same.  This situation can be viewed as an extreme version of the gapless edge state situation where we take an entire region of size $R$ through a phase transition into a trivial gapped phase (product state ground state).  The gap of the entire system will typically go to zero as $R^{-p}$, but we still expect that the state of distant regions will be little modified.  However, it should be noted that the ground state manifold can change in this process.  New ground states with splitting at most $e^{- R^\alpha}$ can come down into the ground state manifold during the phase transition. We expect all these new ground states to be locally indistinguishable far from the region which experienced the phase transition.

In some cases, this expectation of strong insensitivity to boundary conditions can be explicitly verified.  Suppose we wish to take a large region $A$ through a phase transition into a trivial phase.  Let us further suppose that there is a Hamiltonian $H(\eta)$ which interpolates between the initial and final Hamiltonians and which is gapped throughout the phase transition.  Only a non-local (but still few body) Hamiltonian could possibly maintain a gap throughout the phase transition, but if the non-locality can be approximately confined within $A$, then we may still prove a strong result.  Evolving for a finite time with the quasi-adiabatic generator $K(\eta)$ still generates a unitary which maps ground states to ground states, but now this unitary will be non-local within region $A$.  However, outside of region $A$ the unitary will again be quasi-local, so if $\partial_\eta H(\eta)$ is confined near region $A$, then we can prove that the state of regions far from $A$ are approximately preserved by the evolution.

One obstruction to the existence of such a gapped non-local Hamiltonian interpolating between two gapped local Hamiltonians is if the initial and final ground state degeneracies are different.  This is expected to be a concern if we are effectively changing the topology or changing the system size in an $s>1$ fixed point with ground state degeneracy which depends on system size. In the examples we understand, e.g., the layer construction and Haah's code, the boundary conditions far away are indeed provably irrelevant.  In the layer construction this is trivial while in Haah's code it follows because the Hamiltonian consists of commuting projectors.

There may be other obstructions and we do not give a general prescription for finding such a non-local Hamiltonian.  However, one idea is to force all pairs of local operators to have their correction expectation values, e.g. $H_{\text{non-local}} \sim \sum_{x,y,\alpha,\beta} (O_{x,\alpha} O_{y,\beta} - \langle O_{x,\alpha} O_{y,\beta}\rangle)^2$.

An example where the limited non-local approach does work is in the gluing together of two disks of integer quantum Hall fluid.  The difficult step is to exhibit a non-local gapped Hamiltonian whose ground state is a $d=1$ Fermi gas.  Consider fermions at half-filling on a one dimensional lattice of length $L$.  It is convenient to work in momentum space with states labelled by $k \in [-\pi,\pi)$.  The desired Hamiltonian can be constructed by demanding a single particle energy spectrum $\epsilon(k)$ which is given by
\beq
\epsilon(k) = \begin{cases} \Delta/2, & k \in [-\pi,-\pi/2) \cup [\pi/2,\pi) \\ -\Delta/2, & k\in[-\pi/2,\pi/2) \end{cases}.
\eeq
Then the free fermion ground state with states $k\in [-\pi/2,\pi/2]$ filled is an exact ground state and the Hamiltonian is gapped.  The real space hopping amplitudes which produce such a single particle spectrum may be found by Fourier transform and decay as one over distance.  We can further modify this Hamiltonian to adiabatically continue it to a local insulating Hamiltonian thus producing a gapped path from a product state to the fermion gas ground state.

We use the various physical properties just reviewed throughout the paper. For example, we assume some ability to place phases on different types of geometries.  If the phase can be represented as a Hamiltonian motif which only assigns terms to links on a graph, then we can place such a phase on any type of geometry.  More generally, at least some freedom is required to proceed with our results, e.g. we need tori and open regions of various sizes.  We also use the ideas of stability, local indistinguishability, and insensitivity to boundary conditions repeatedly. An important statement following from insensitivity to boundary conditions is that the entanglement entropy $S(R)$ of a region of linear size $R$ is independent of $L$ for $R \ll L$. However, when we give theorems we endeavor to state the mathematically precise assumptions.

\section{Adiabatic perturbation theory}
\label{sec:apt}

Suppose we have a Hamiltonian $H(t)$ which depends on time.  Let the instantaneous energy eigenstates and energies be given as
\beq
H(t) |n,t\rangle = E_n(t) |n,t\rangle.
\eeq
We start evolving at $t=0$ from $|\psi(0)\rangle = \sum_n c_n(0) |n,0\rangle$ and expand the time dependent state as
\beq
|\psi(t) \rangle = \sum_n c_n(t) e^{-i \int_0^t E_n(t') dt'} |n,t\rangle.
\eeq
$|\psi(t)\rangle $ obeys the Schrodinger equation $i \partial_t |\psi(t)\rangle = H(t)|\psi(t)\rangle $ which we want to convert into an equation for the $c_n$.

Taking the time derivative of $|\psi(t)\rangle$ we obtain three terms:
\bea
&& i \partial_t |\psi(t)\rangle = \sum_n \left( E_n(t) c_n e^{-i \int_0^t E_n(t') dt'} |n,t\rangle \right) + \nonumber \cr \\
&& \sum_n \left( (i \partial_t c_n)  e^{-i \int_0^t E_n(t') dt'} |n,t\rangle + c_n e^{-i \int_0^t E_n(t') dt'} i\partial_t |n,t\rangle \right).
\eea
The first term containing $E_n$ cancels with $H(t)|\psi(t)\rangle$, so we have
\beq
0 = \sum_n \left( i \partial_t c_n  e^{-i \int_0^t E_n(t') dt'} |n,t\rangle + c_n e^{-i \int_0^t E_n(t') dt'} i\partial_t |n,t\rangle \right).
\eeq

We take a derivative of the eigenvalue equation for $|n,t\rangle$ to find an equation for $\partial_t |n,t\rangle$.  First, since $\langle n,t|n,t\rangle = 1$ it follows that $\langle n,t|\partial_t |n,t\rangle =0$.  Then we obtain for $\partial_t |n,t\rangle $ the equation
\beq
i\partial_t |n,t\rangle = -i(H-E_n)^{-1}(\partial_t H) |n,t\rangle
\eeq
where it is understood that the singular term in the inverse is omitted. Expanding the time derivative of $|\psi(t)\rangle $ in the $|n,t\rangle$ basis we find (with some relabelling of $n$ and $m$)
\beq
(i \partial_t c_n ) e^{-i \int_0^t E_n(t') dt'} = \sum_{m\neq n} \frac{ie^{-i \int_0^t E_m(t') dt'}}{E_n -E_m} \langle n,t| \partial_t H |m,t\rangle.
\eeq
We can simply this equation to
\beq
\partial_t c_n = \sum_{m\neq n} \frac{e^{-i \int_0^t (E_m(t')-E_n(t')) dt'}}{E_n -E_m} \langle n,t| \partial_t H |m,t\rangle.
\eeq
See \cite{2008PhRvA..78e2508R,2010LNP...802...75D} for a recent general analysis of this formula and \cite{nenciu1993} for rigorous results; our needs are simpler.

Suppose we have a single unique ground state separated at all times by a gap of at least $\Delta$ from the rest of the spectrum.  We wish to estimate the probability to remain in the ground state using perturbation theory assuming that $\partial_t H(t)$ is a smooth function which vanishes for $t$ outside $[0,\tau]$. We have
\beq
c_0(\tau) - c_0(0) = \int_0^\tau dt \sum_{m\neq 0} \frac{e^{-i \int_0^t (E_m(t')-E_0(t')) dt'}}{E_0 -E_m} \langle 0,t| \partial_t H |m,t\rangle,
\eeq
and upon taking absolute values and using the triangle inequality we obtain
\beq
|c_0(\tau) - c_0(0)| \leq \sum_{m\neq 0} \left|\int_0^\tau dt \frac{e^{-i \int_0^t (E_m(t')-E_0(t')) dt'}}{E_0 -E_m} \langle 0,t| \partial_t H |m,t\rangle \right|.
\eeq
This expression is a sum of Fourier transforms of the matrix elements of $\partial_t H$ times a function of the energy differences.

To complete the analysis define $\delta E_n(t) = E_n(t) - E_n(0)$ and note that $E_m - E_0 \geq \Delta$ for all $m$.  Then we may write
\beq
|c_0(\tau) - c_0(0)| \leq \sum_{m\neq 0} \left|\int_0^\tau dt \frac{e^{-i(E_m -E_0)t}}{\Delta}\left[ e^{-i \int_0^t (\delta E_m(t')- \delta E_0(t')) dt'} \langle 0,t| \partial_t H |m,t\rangle\right] \right|.
\eeq
The function in brackets is smooth and has rapidly vanishing Fourier transform; call the Fourier transform $\mathfrak{H}_m(\omega)$. Then we have the bound
\beq
|c_0(\tau) - c_0(0)| \leq \frac{1}{\Delta}\sum_{m\neq 0} | \mathfrak{H}_m(E_m-E_0)|.
\eeq
Assuming $\mathfrak{H}_m(\omega)$ decays like $J e^{-(\omega \tau)^{1-\delta}}$ and assuming the number of non-vanishing matrix elements of $\partial_t H$ between excited states and the ground state is not too large, we find a bound like
\beq
|c_0(\tau) - c_0(0)| \leq \frac{J}{\Delta} e^{-(\Delta \tau)^{1-\delta}}.
\eeq
If we are considering a Hilbert space of bounded dimension then this bound follows immediately, and if the Hilbert space dimension is large, then we need a bound on the number of matrix elements going like poly$(\log(\mathcal{D}))$ for a Hilbert space of dimension $\mathcal{D}$.

The probability for the groundstate to decay is
$P_\text{decay} =  1 - |c_0(\tau)|/^2 $ (with the initial condition $ c_n(0) = \delta_{n,0}$).
The above bound implies that $ |c_0(\tau)| \geq 1 - \frac{J}{\Delta} e^{-(\Delta \tau)^{1-\delta}} $
and hence
$$ P_\text{decay} \leq 2\frac{J}{\Delta} e^{-(\Delta \tau)^{1-\delta}}.$$

\section{Controlling the Renyi entropy}
\label{sec:renyi}

A unitary $U$ acting on a Hilbert space $\mathcal{V}_1 \otimes \mathcal{V}_2$ of dimension $\mathcal{D}^2$ (assume for simplicity that $\mathcal{D}_1 =  \mathcal{D}_2 =\mathcal{D}$) can only increase the Schmidt rank of a state by a factor of $\mathcal{D}^2$. This may be proven by noting that $U$ may always be decomposed as
\beq
U = \sum_{i=1}^{\mathcal{D}^2} O_{1i} O_{2i}
\eeq
since $U$ is a vector in the space $(\mathcal{V}_1 \otimes \mathcal{V}^*_1) \otimes (\mathcal{V}_2 \otimes \mathcal{V}^*_2)$. If $\mathcal{V}_1$ and $\mathcal{V}_2$ are parts of larger systems, $\mathcal{V}_{1E_1} = \mathcal{V}_1 \otimes \mathcal{V}_{E_1}$ and $\mathcal{V}_{2E_2} = \mathcal{V}_2 \otimes \mathcal{V}_{E_2}$, then this bound remains true. In fact, the bound may be saturated by applying a swap operator which exchanges $1$ and $2$ to an initial state in which $1$ is maximally entangled with $E_1$ and $2$ is maximally entangled with $E_2$.

Applying this simple fact to the case where $1E_1 = A$ and $2E_2 = \bar{A}$ with $1$ and $2$ small regions neighboring $\partial A$, the Schmidt rank of $\rho_A$ can change by at most a factor of  $\min(\mathcal{D}^2_1,\mathcal{D}_2^2)$. Having approximated the sequence of quasi-local unitaries with a sequence of circuits acting on $\ell^d \sim \log^{d(1+\delta)}(L)$ degrees of freedom at a time, the total Schmidt rank of a region $A$ in $d>1$ can bounded by estimating the number of such circuit chunks acting across $\partial A$. A simple counting argument shows that this number is $N_{\text{chunks}} \sim |\partial A|/\ell^{d-1}$ in $d>1$. Since the Hilbert space of a block of size $\ell^d$ has dimension of order $e^{c \ell^d}$, it follows that the state built from the sequence of circuits has Schmidt rank across $\partial A$ bounded by $e^{N_{\text{chunks}}\ell^d} = e^{c |\partial A| \ell} \sim e^{c |\partial A| \log^{1+\delta}(L)}$. This bound is independent of $d$ and provides a better bound than PEPS constructions. Recall that the resulting state is also within $\epsilon \sim L^{-q}$ of the true ground state. Thus there is a approximation to $|\psi_L\rangle$ with limited Schimdt rank for any region $A$.

However, this does not imply that $|\psi_L\rangle$ has limited Schmidt rank.  Indeed, the Schmidt rank is badly discontinuous. Furthermore, all Renyi entropies $S_n$ with $n<1$ are only barely continuous. The Renyi entropy is defined as
\beq
S_n(\rho_A) = \frac{1}{1-n}\log\left(\tr(\rho_A^n)\right),
\eeq
where $S_1 = - \tr(\rho_A \log(\rho_A))$ is the usual entanglement entropy. For $n=1$ we have the Fannes-Audenaert inequality \cite{1751-8121-40-28-S18,fannes}: if $\frac{1}{2} \|\rho-\sigma\|_1 = T \leq 1$ is the trace distance and if $\rho$ and $\sigma$ are defined on a space with dimension $\mathcal{D}$, then
\beq
|S_1(\rho) - S_1(\sigma)| \leq T \log(\mathcal{D}-1) -T \log(T) - (1-T) \log(1-T).
\eeq
The inequality is saturated for \beq
\rho = \text{diag}(1,\underbrace{0, ...}_{\mathcal{D}-1})
\eeq
and
\beq
\sigma = \text{diag}(1-T,\underbrace{T/(\mathcal{D}-1),...}_{\mathcal{D}-1})
\eeq
with $S_1(\rho)=0$ and $S_1(\sigma) = T \log(\mathcal{D}-1) -T \log(T) - (1-T) \log(1-T)$.

Given the same two states $\rho$ and $\sigma$, an elementary exercise gives $S_n(\rho)=0$ and
\beq
S_n(\sigma) = \frac{1}{1-n}\log\left((1-T)^n + (\mathcal{D}-1)^{1-n} T^n\right).
\eeq
To have $S_n(\sigma)$ of order $\epsilon$, we must take $T \sim \epsilon^{1/n} \mathcal{D}^{-\frac{1-n}{n}}$ which is much smaller than the $T \sim \frac{\epsilon}{\log(\mathcal{D})}$ needed for $n=1$. Since $\mathcal{D}$ grows exponentially with system size, we need states to be exponentially close to bound the Renyi entropy for $n<1$, and hence the Renyi entropy is effectively discontinuous.

We still conjecture that the Renyi entropy of $s=1$ fixed points $|\psi_{L}\rangle$ obeys an area law in keeping with the analysis of \cite{2013arXiv1304.6402S}, but our results here are insufficient to prove this.  We have shown that there is an approximate state with Renyi entropy which can at most modestly violate the area law.

\section{Dilute array of non-abelian anyons}
\label{sec:anyon}

Suppose we have an array of $N$ non-abelian anyons $a$ in $d=2$ dimensions with quantum dimension $d_a >1$.  Associated with these anyons is a non-local fusion space $\mathcal{V}$ of dimension $\text{dim}(\mathcal{V}) \sim d_a^N$.  If we distribute the anyons roughly equidistant from each other (with pinning potentials, say), then the spacing between anyons will be roughly $n^{-1/2}$ where $n=\frac{N}{L^2}$ is the anyon density.  Since the underlying topological phase is gapped with correlation length $\xi$, the states in $\mathcal{V}$ are locally coupled with strength $J_a \sim e^{-n^{-1/2}/\xi}$.  The total spectral width of the anyon Hamiltonian is then of order $N J_a$ and hence if $n^{-1/2}$ increases as $L^\alpha$ then the all $d_a^N$ states are essentially only exponentially split.

Given a finite region of size $R$, the number of anyons contained within it is $n R^2$, so unless $n$ approaches a constant in the thermodynamic limit $L \rightarrow \infty$, the number of anyons in a finite region approaches zero. Then even if we imagine sitting in a highly entangled state in $\mathcal{V}$, the extra entanglement in a region of size $R$ will be negligible as $L\rightarrow \infty$. This conclusion is slightly delicate since the states in $\mathcal{V}$ are not strictly labeled by local data, but if no anyons are present in a region, then the state of the system will be the same as in the ground state which obeys the area law.

Thus while this is an interesting case (and clearly permits highly entangled states to be formed, e.g., as in a topological quantum computation), there are states in $\mathcal{V}$ which are lightly entangled. In any event, the setup violates our assumptions.

\section{Topological entanglement entropy is RG invariant}
\label{sec:TEE}

In this appendix we
use the s-sourcery to
give an argument
that the topological entanglement entropy (TEE) is a well-defined property of an $s=1$ fixed point,
that is, it is preserved under the $s=1$ RG step we have defined.
This argument is complementary to an argument for universality given in
\cite{Kitaev:2005dm, PhysRevLett.96.110405} and provides a check on our methods.

\begin{wrapfigure}{R}{.5\textwidth}
  \centering
    \includegraphics[width=0.48\textwidth]{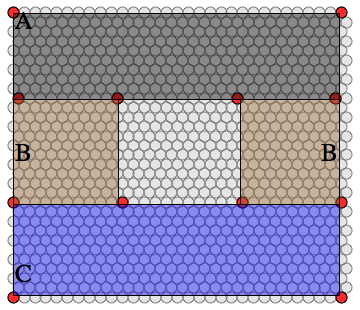}
  \caption{$A,B,C$ label regions used in the definition of the topological entanglement entropy.
  The ancillas which are unentangled before the action of the quasilocal unitary
  are not pictured.
  The grey disks represent regions of linear size $\hat \ell$,
on which a single layer of the staggered circuit representations of the quasilocal unitary
has support, as in Fig.~\ref{1dblock}.
  }
  \label{fig:TEE}
\end{wrapfigure}
The TEE can be
defined \cite{Kitaev:2005dm, PhysRevLett.96.110405} as $\gamma$ in
$$ 2 \gamma \equiv S_{AB}+ S_{BC} - S_B - S_{ABC} $$
with regions $A, B, C$ as in the figure.
We assume $A, B, C$ have linear size much larger than $\ell$,
the range of the quasilocal unitary
(small disks in Fig.~\ref{fig:TEE}).
We will restrict the discussion to $d=2$,
but believe that the argument extends to the
generalization to arbitrary dimensions given in \cite{Grover:2011fa}.

The $s=1$ RG step acts on a copy of the system
tensored with a collection of
decoupled ancillas;
the subspace labelled $A$ includes both
the system Hilbert space associated to region $A$
and the accompanying ancillas which will be intercalated
by the RG step.
We need to show that
a quasilocal unitary of range $\ell$,
acting on the system at size $L$
times these ancillas
preserves the combination $\gamma$,
up to corrections polynomial in $1/L$.

For any region $R$,
the change in its entanglement entropy
produced by such a quasilocal unitary can be approximated
as
\beq\label{eq:quasi-local-entropy-change}
 \Delta S_R =  \int_{\partial R}d\sigma s(\sigma) + \sum_{\text{corners}, \alpha} c(\theta_\alpha)
 \eeq
where $s$ is a smooth geometric function localized to the boundary of $R$,
and $c(\theta_\alpha)$ is the contribution from a corner of $\partial R$
which makes an angle $\theta_\alpha$.
This formula is similar in spirit to the formula of
\cite{Grover:2011fa} for the whole entanglement entropy
for regions of topological quantum liquids.
The precise $\Delta S_R$ is a Riemann-sum approximation
to such an integral, with error determined by $\hat \ell$.

To accomplish this, approximate the quasilocal unitary by a
staggered circuit as in
\S\ref{subsec:mera} and in particular
Fig.~\ref{1dblock}
(the support of one layer of the circuit is depicted by the gray disks in Fig.~\ref{fig:TEE}).
The error in the entanglement entropy from this circuit approximation
is usefully bounded by using the Fannes-Audenaert inequality again;
a useful approximation requires $ \epsilon \sim L^{-q}$ as before.
The contributions to the change in entanglement entropy of any region $R$
from each layer of the circuit
comes only from disks which intersect the boundary.
Away from corners of the region (red disks in Fig.~\ref{fig:TEE}), these contributions
can be represented by a derivative expansion in $\ell$ times
local geometric functionals of the shape of the boundary, as in \cite{Grover:2011fa}:
$$ s(\sigma) = a_0 + a_2 \ell  \kappa^2  + a_3 \ell \partial_\sigma \kappa+... $$
where $ \kappa$ is the extrinsic curvature of the boundary
and the ellipsis represents terms suppressed by more powers of $\ell$.
Terms in this expansion which are odd under exchanging the inside and outside
of region $R$ vanish because the whole system is in a pure state \cite{Grover:2011fa}.

Corners, where the shape of the boundary is not smooth, even at the scale $\ell$, must be treated specially.  The only property of the corner contribution $c(\theta)$ we require is that
it depends only on the angle between the edges which enter and exit
the corner disk ($\theta = \pi$ is no corner).

Adding up the contributions in the form \eqref{eq:quasi-local-entropy-change} to $ \Delta \gamma $,
the area-law contributions proportional to $ a_0 L^1 $ cancel by design,
leaving behind terms proportional to $  {\ell \over L}  $.
(This step of the argument is identical to that of \cite{Grover:2011fa},
with $\ell$ here playing the role of the correlation length there.)
The corner contributions also directly cancel in pairs.
In the thermodynamic limit, therefore, we find $ \Delta \gamma = 0$.

\section{Commmuting projector Hamiltonians}
\label{sec:commuting}

Here we review prior work on commuting projector Hamiltonians as a simple illustration of the frustration free setting. Many workers have extensively developed this machinery (see e.g.~\cite{petz1986,2004CMaPh.246..359H,2010PhRvL.104e0503B,2011PhRvL.106h0403P,2012arXiv1206.0755B}).

Suppose $H = \sum_x P_x$ is a sum of commuting projectors with $P_x |g \rangle =0$ for all locally indistinguishable ground states $|g\rangle$. It has already been proven that the ground states of such Hamiltonians obey the area law.  Our aim is to use this case to illustrate our alternative approach.  However, let us first establish the area law using an argument similar to that in \cite{2006PhRvL..97e0401B}.

Consider the ground state projector $P_g$ which can be obtained thermodynamically as
\beq
P_g = \lim_{\beta \rightarrow 0} e^{-\beta H}.
\eeq
Because $H$ is a sum of commuting projectors, the thermal state of $H$ is a quantum Markov chain for all $\beta$ \cite{2012arXiv1206.0755B}.  For our purposes this means that the conditional mutual information, $I(A:C|B) = S(AB)+S(BC)-S(B)-S(ABC)$, vanishes whenever $B$ isolates $A$ from $C$. Since the Markov property holds for all $\beta$, it also holds for the normalized ground state projector.  Furthermore, provided we work locally, local indistinguishability implies that the conditional mutual information in the ground state projector is the same as in any particular ground state.

Hence we have that $I(A:C|B)=0$ whenever $B$ isolates $A$ from $C$ in every ground state. Let $A$ be any simply connected region of linear size $R$, let $B$ be a strip of width $W$ bordering $A$, and let $C$ be the rest of the system. Then we have
\beq
0 = I(A:C|B) = S(AB) + S(BC) - S(B) - S(ABC),
\eeq
but because the state of $ABC$ is pure we have $S(ABC)=0$, $S(AB)=S(C)$, $S(BC)=S(A)$, and $S(B)=S(AB)$. Then we also find that
\beq
0 = S(C) + S(A) - S(AC),
\eeq
which states that the mutual information $I(A,C)$ vanishes. Thus we have
\beq
S(A) = S(AC)-S(C) = S(B)-S(AB) \leq 2 S(B) - S(A)
\eeq
by the Araki-Lieb inequality \cite{araki1970}, $S(AB) \geq |S(A)-S(B)|$. Since the size of $B$ is bounded by $R^{d-1} W$ and since the mutual information vanishes once $W$ is greater than the range of the Hamiltonian, we immediately find
\beq
S(A) \leq W R^{d-1}
\eeq
which is the area law.

The Markov property also implies that we can reconstruct states of subregions using only local data \cite{petz1986,2004CMaPh.246..359H}. In terms of our previous variables, $\sigma_A = \rho_A$ for quantum Markov chains. Furthermore, $\sigma_A$ is given by
\beq
\sigma_A = \frac{P_{g,A}}{G(H_A)},
\eeq
where $P_{g,A}$ is the ground state projector for $H_A$, the Hamiltonian truncated to region $A$.
(The formula for $S_A$ used in \S\ref{sec:haah} is a consequence of this relation.) Because commuting projector Hamiltonians cannot support protected edge states, there is another commuting projector Hamiltonian $\check{H}_A$ which has a full gap except for locally indistinguishable ground states. These two Hamiltonians differ only in boundary terms localized near $A$. The ground state degeneracy of $H_A$ is then bounded by the ground state degeneracy of $\check{H}_A$ plus an area law piece. The ground state degeneracy of $\check{H}_A$ is something we can relate to $s$ using the RG framework, so we have precisely the situation discussed in the section \ref{sec:arealaw}.

\bibliographystyle{ut}

\bibliography{rg_area_law}

\end{document}